# A Robust Partial Correlation-based Screening Approach

Xiaochao Xia[1]

[1]*College of Mathematics and Statistics, Chongqing University, Chongqing, China*

**Abstract**

As a computationally fast and working efficient tool, sure independence screening has received much attention in solving ultrahigh dimensional problems. This paper contributes two robust sure screening approaches that simultaneously take into account heteroscedasticity, outliers, heavy-tailed distribution, continuous or discrete response, and confounding effect, from the perspective of model-free. First, we define a robust correlation measure only using two random indicators, and introduce a screener using that correlation. Second, we propose a robust partial correlation-based screening approach when an exposure variable is available. To remove the confounding effect of the exposure on both response and each covariate, we use a nonparametric regression with some specified loss function. More specifically, a robust correlation-based screening method (RC-SIS) and a robust partial correlation-based screening framework (RPC-SIS) including two concrete screeners: RPC-SIS(L2) and RPC-SIS(L1), are formed. Third, we establish sure screening properties of RC-SIS for which the response variable can be either continuous or discrete, as well as those of RPC-SIS(L2) and RPC-SIS(L1) under some regularity conditions. Our approaches are essentially nonparametric, and perform robustly for both the response and the covariates. Finally, extensive simulation studies and two applications are carried out to demonstrate the superiority of our proposed approaches.

## 1 Introduction

Ultrahigh dimensional data are nowadays frequently encountered in many areas, such as DNA sequencing, medical health, and traffic online monitoring. Sure independence screening (SIS) approach has been a powerful tool to handle ultrahigh dimensional problems since the seminal work of Fan and Lv (2008), who firstly established the sure screening property under the framework of Gaussian linear model. That is, with an overwhelming probability, the SIS can select a subset from original predictors that contain all truly associated predictors with the response when the number of predictors, $p_n$, is much bigger than the sample size, $n$. More precisely, $p_n$ and $n$ may satisfy the mathematical relationship: $p_n = O(\exp(n^a))$ for some $a \in (0, 1/2)$, which is a nonpolynomial (NP) order. Although the SIS serving as an initial step in reducing the dimensionality is rough and simple, it usually works efficiently in practice in terms of the issues: computational complexity and algorithm stability in optimization, and simultaneously safeguards the statistical accuracy for both estimation and prediction for NP problems (Fan and Song (2010)). On the other hand, the regularization-based variable selection approaches, such as the LASSO in Tibshirani (1996), the SCAD in Fan and Li (2001), and the MCP in Zhang (2010), suffer from above issues in ultrahigh dimensional setting. Owing to its simplicity and efficiency in ultrahigh dimensional statistical data analysis, the SIS has received an increasing attention for various successful applications over the last decade. Many variants were proposed to extend the range of its application. Some relevant literature can be referred to, such as Fan and Song (2010), Zhu et al. (2011), Li et al. (2012), Li, Zhong and Zhu (2012), Fan, Feng and Song (2011), He, Wang and Hong (2013), Chang, Tang and Wu (2013, 2016), Fan, Ma and Dai (2014), Liu, Li and Wu (2014), Li et al. (2016), Han (2019) and reference therein. It can be seen that most of these papers focus on a linear or nonlinear dependence between the response variable and each covariate with or without the framework of parametric or nonparametric models.

On one hand, for ultrahigh dimensional data, we may frequently encounter outliers, heteroscedasticity and heavy-tailed distribution along each dimension of the data, rendering any convenient variable screening approach that focuses only on the normal data become unstable and difficult. To deal with such types of data, many useful screeners were developed including the sure independent ranking and screening (SIRS, Zhu et al. (2011)), the distance correlation based screening (DC-SIS, Li, Zhong and Zhu (2012)), the Kendall's $\tau$-based screening (Kendall's $\tau$-SIS, Li et al. (2012)), the quantile-adaptive sure independence screening (QaSIS, He, Wang and Hong (2013)), and the survival impact index based screening (SII, Li et al. (2016)). However, almost all concentrate merely on the robustness against the response variable. This would result in undesirably finite-sample performance for these methods when predictor variables are neither light-tailed and symmetric nor normally distributed. A more likely situation is that both the response and predictor variables are heavy-tailed, asymmetric and with outliers. However, none of aforementioned variable screening procedures has investigated this issue. To tackle this problem, Xia and Li (2021)'s paper provides a good solution in which a copula-based correlation (CC) based variable screening approach (CC-SIS) is developed. Their numerical results show that the CC-SIS performs robustly for both response and each covariate.

However, it is only evaluated at a fixed pair of quantile levels, which may lead to a loss of efficiency in the performance. Moreover, less of aforementioned methods can be applied to discrete response.

On the other hand, when some extra information such as conditional or exposure variables are available in advance, marginal variable screening approach that ignores the joint effect of other predictors certainly produces unpromising screening results, owing to the existence of spurious correlation among the high-dimensional covariates, which may result inevitably in misleading conclusions. In order to improve the performance of the marginal screening methods, conditional correlation-based screening and partial correlation-based screening methods are developed in the literature, including the conditional Pearson's correlation-based screening of Liu, Li and Wu (2014), the conditional quantile correlation-based screening (CQC-SIS) of Xia, Li and Fu (2019), the method of Barut, Fan and Verhasselt (2016) for generalized linear models, the method of Chu, Li and Reimherr (2016) for time-varying coefficient models, the thresholded partial correlation-based screening by Li, Liu and Lou (2017) under elliptical linear regression models, the quantile partial correlation-based screening (QPC-SIS) by Ma, Li and Tsai (2017), and the copula-based partial correlation (CPC-SIS) screening by Xia and Li (2021), among others. It is worth pointing out most of the above methods are robust only for the response variable, and most are parametric methods. In particular, almost all the partial correlation-based approaches remove the confounding effect of conditional variables on the response variable and predictors through fitting parametric models. However, using parametric models to remove the confounding effect may be less flexible and inaccurate when the confounding effect on response and each predictor can be captured by a nonparametric model. This motivates us in part to develop a flexible and robust screener that can perform robustly for both response and each predictor and simultaneously is also flexible for controlling the effect of conditional variables.

It is worth mentioning that this work is not a trivial extension of CC-SIS by Xia and Li (2021). Several remarkable differences between CC-SIS and the method proposed in current paper can be summarized as follows.

- (i) In the idea of two papers, the CC introduced in Xia and Li (2021) is originally formulated in terms of the idea of the quantile correlation measure (Li, Li and Tsai (2015)), which is an asymmetric function of covariate $X$ and response $Y$ in their formula. Xia and Li (2021)'s work proposed a symmetric quantile correlation, named CC coefficient in their paper. Instead, the current paper directly focuses on the Pearson correlation of two random indicators.

- (ii) For variable screening, Xia and Li (2021) used the *absolute* value of CC as screening utility, which is evaluated at a fixed pair of quantiles levels, whereas the current paper utilizes a weighted integral of the *square* of the correlation over the range of sample observations to serve as screening utility.

- (iii) The definitions of screening utilities introduced in the two papers are different. For example, the definition of CC introduced in Xia and Li (2021) is $\varrho_{Y,X}(\tau,\iota) = \frac{F_{Y,X}(F_Y^{-1}(\tau), F_X^{-1}(\iota)) - \tau\iota}{\sqrt{\tau(1-\tau)\iota(1-\iota)}}$ for two chosen numbers $0 \leq \tau, \iota \leq 1$ and the screening utility is $u_j^{CC} = |\varrho_{Y,X_j}(\tau,\iota)|$. Whereas, the robust correlation (denoted as RC hereafter) defined in this paper is the formula (1) in Section 2.1, and the corresponding screening utility is $u_j^{RC} = E\{[\rho(Y, X_j)]^2\}$. We can see that $u_j^{RC}$ clearly differs from $u_j^{CC}$ since the denominator in CC is a constant, however, the denominator in RC is *not* constant and it actually varies with $x$ and $y$ via $F_X(x)$ and $F_Y(y)$.

- (iv) Another remarkable difference is that whether CC-SIS is applicable to analyzing the ultrahigh dimensional data with the discrete response is unknown at least in theory. But our RC-SIS works for the discrete response in ultrahigh dimension with a theoretical guarantee of sure screening property.

We can see that the proposed $u_j^{RC}$ takes the impact of marginal distribution of each covariate into account in variable screening while $u_j^{CC}$ does not. This partly reveals the advantage of $u_j^{RC}$ over $u_j^{CC}$ theoretically. Furthermore, when $X_j$ is associated with $Y$, the CC has a risk of being zero or closing to zero at a particularly chosen pair of quantiles, resulting in that $X_j$ can not be selected by CC-SIS due to a very small value of $u_j^{CC}$. Whereas, using a weighted integral of squared CC, i.e., $u_j^{RC}$, is capable of reducing this risk.

The contribution of current paper can be summarized as follows. First, we provide the definition of the RC between the response and each predictor. We then use the RC to serve as our screener utility for variable screening, denoted as RC-SIS accordingly. Second, we propose a robust partial correlation (RPC) that measures the association between the response and each predictor given an exposure variable, as an important extension of the RC. Our RPC can reveal the nonlinear effect of an exposure variable on both the response and each predictor, while the classical partial correlation only describes the linear effect of exposure variables. Specifically, a simple estimator of RPC is constructed by a sample RC of two residuals that are obtained through fitting two nonparametric functions in the framework of nonparametric regression with some specified loss function. Afterwards, a RPC-based variable screening procedure, denoted as RPC-SIS, is formed under two special losses ($L_2$ loss and $L_1$ loss), resulting in two concrete variable screening procedures: RPC-SIS(L2) and RPC-SIS(L1), respectively. In nature, our RC-SIS and RPC-SIS are totally nonparametric without requiring any parametric model assumption. Third, we show theoretically that both the proposed RC-SIS and RPC-SIS enjoy sure screening consistency under mild regularity conditions. In particular, when the response is categorical, our RC-SIS also works well and enjoys sure screening property.

The rest of the paper is organized as follows. In Section 2, we present the RC-SIS and RPC-SIS procedures. In Section 3, we investigate the large-sample properties including sure screening properties for RC-SIS and RPC-SIS (RPC-SIS(L2) and RPC-SIS(L1), respectively). In Section 4, simulation studies are carried out to evaluate the finite-sample performance of proposed methods. Two real-world data sets are analyzed in Section 5. Extensions and concluding remarks are given in Section 6. Some numerical results and all proofs of theorems are given in the supplementary file.

## 2 Methodology

### 2.1 Variable Screening Based on A Robust Correlation (RC-SIS)

Suppose that the random variables, $Y$ and $X$, have marginal cumulative distribution functions (CDF), $F_Y(x)$ and $F_Y(y)$, defined on the regions $\mathcal{Y}$ and $\mathcal{X}$, respectively. We define the RC for $Y$ and $X$ as follows. For any $(y, x) \in \mathcal{Y} \times \mathcal{X}$,

$$\rho(y, x) = \frac{\operatorname{cov}(I(Y \leq y), I(X \leq x))}{\sqrt{\operatorname{var}(I(Y \leq y))\operatorname{var}(I(X \leq x))}}, \qquad (1)$$

where $I(\cdot)$, $\operatorname{cov}(\cdot, \cdot)$ and $\operatorname{var}(\cdot)$ denote the indicator function, the covariance and the variance, respectively. $\rho(y, x)$ is simply the Pearson correlation of two random indicator variables $I(Y \leq y)$ and $I(X \leq x)$. Clearly, $\rho(y, x) \equiv 0$ holds for all $(y, x) \in \mathcal{Y} \times \mathcal{X}$ if and only if $X$ and $Y$ are mutually independent. Because indicators can capture the robustness of the association measure against outliers and extreme values, thus the RC should be robust also. Furthermore, applying any monotone transformation to both $Y$ and $X$ does not affect the value of $\rho(y, x)$.

Given the observations $\{(Y_i, X_i), i = 1, \ldots, n\}$, independent and identically distributed (i.i.d.) from the population $(Y, X)$, a simple estimate of $\rho(y, x)$ is given by

$$\widehat{\rho}(y, x) = \frac{\widehat{F}_{Y,X}(y, x) - \widehat{F}_Y(y)\widehat{F}_X(x)}{\sqrt{[\widehat{F}_Y(y) - \widehat{F}_Y^2(y)][\widehat{F}_X(x) - \widehat{F}_X^2(x)]}}, \qquad (2)$$

where $\widehat{F}_{Y,X}(y, x) = \frac{1}{n}\sum_{i=1}^{n} I(Y_i \leq y, X_i \leq x)$ is empirical estimate of joint distribution function $F_{Y,X}(y, x)$, and $\widehat{F}_Y(y) = \frac{1}{n}\sum_{i=1}^{n} I(Y_i \leq y)$ and $\widehat{F}_X(x) = \frac{1}{n}\sum_{i=1}^{n} I(X_i \leq x)$ are marginal empirical CDFs of $Y$ and $X$, respectively.

The above $\widehat{\rho}(y, x)$ can be used as an empirical screening index for feature screening for ultrahigh di-mensional data $\{(Y_i, \mathbf{X}_i), i = 1, \ldots, n\}$, where $\mathbf{X}_i \in \mathbb{R}^{p_n}$, and $p_n$ can be $o(\exp(n^a))$ for some $0 < a < 1/2$. Specifically, we use the following marginal utility for feature screening

$$u_j^{RC} = E\{\rho^2(Y, X_j)\}, 1 \leq j \leq p_n. \tag{3}$$

Its sample version of $u_j^{RC}$ can be obtained as

$$\widehat{u}_j^{RC} = \frac{1}{n} \sum_{i=1}^{n} \widehat{\rho}^2(Y_i, X_{ij}), 1 \leq j \leq p_n. \tag{4}$$

Thence, we can select a subset of predictors

$$\widehat{\mathcal{M}}_a = \{j : \widehat{u}_j^{RC} > \varsigma_n, 1 \leq j \leq p_n\}. \tag{5}$$

as an important set of relevant predictors that are believed to be dependent with the target variable $Y$, where $\varsigma_n$ is a screening tuning parameter and usually user-specified. We refer to this procedure as robust correlation-based sure independence screening, denoted as RC-SIS.

We remark that the copula-based correlation (CC) based screening (CC-SIS) introduced by Xia and Li (2021) is closely related to the RC-SIS as above. In fact, CC is a special case of our RC in the sense that $\rho(y, x)$ is evaluated at $(y, x) = (F_Y^{-1}(\tau), F_X^{-1}(\iota))$, where $F_Y^{-1}(\tau)$ and $F_X^{-1}(\iota)$ denote the $\tau$th and $\iota$th quantiles of $Y$ and $X$, respectively. Unlike the empirical utilities of CC-SIS in Xia and Li (2021), which needs to specify a fixed pair of quantiles from sample in advance. This per se depends prior knowledge on the data. However, our empirical utilities $\widehat{u}_j^{RC}$s in this paper are computed over entire sample observations. Hence, from this point of view, RC-SIS avoids the choice of quantile levels and makes use of full information from sample data than CC-SIS. As a result, the performance of RC-SIS is expected to be better than CC-SIS. Furthermore, it can be seen that $u^{RC}$ is a weighted Cramér-von Mises distance.

## 2.2 Variable Screening Based on A Robust Partial Correlation (RPC-SIS)

In this subsection, we introduce a flexible RPC for two continuous random variables $Y$ and $X$ while controlling an exposure variable or a conditional variable, $Z$. The idea of RPC is rooted in the definition of the classical partial correlation that first removes the effect of $Z$ on $X$ and $Y$ by fitting two regressions, respectively, and then uses the residuals from the regressions as new variables in the calculation of correlation. To be specific, we use $\varepsilon_{X|Z} = X - m_x(Z)$ and $\varepsilon_{Y|Z} = Y - m_y(Z)$ to denote two residuals, where $m_x(Z)$ and $m_y(Z)$ represent two regression functions that regress $X$ and $Y$ on $Z$, respectively. Let $\ell(u)$ denote some loss function in $u$, such as $L_2$ loss $\ell(u) = u^2$ and $L_1$ loss $\ell(u) = |u|$. Then, $m_y(Z)$ is defined as the minimizer of

$$\min_{g \in L^2(P)} E\{\ell(Y - g(Z))\},$$

where $L^2(P)$ denotes a class of square integrable functions under the measure $P$. Clearly, we have $m_y(Z) = E(Y|Z)$ for the $L_2$ loss and $m_y(Z) = \text{median}(Y|Z)$ for the $L_1$ loss. Besides, $m_x(Z)$ can be similarly defined.

Assume that $\varepsilon_{X|Z}$ and $\varepsilon_{Y|Z}$ are well defined on two regions $\mathcal{V}$ and $\mathcal{U}$, respectively. We now formally formulate RPC for $Y$ and $X$ given $Z$ as

$$\varrho(u,v) = \frac{\text{cov}(I(\varepsilon_{Y|Z} \leq u), I(\varepsilon_{X|Z} \leq v))}{\sqrt{\text{var}(I(\varepsilon_{Y|Z} \leq u))\text{var}(I(\varepsilon_{X|Z} \leq v))}}, \tag{6}$$

which is the correlation coefficient given in (1) for random variables $\varepsilon_{X|Z}$ and $\varepsilon_{Y|Z}$. When $Y$ and $X$ are conditionally independent given $Z$, we have $\varrho(u,v) = 0$ for all $(u,v) \in \mathcal{U} \times \mathcal{V}$.

We next provide an estimate of $\varrho(u,v)$. Given sample data $\{(Y_i, X_i, Z_i), i = 1, \ldots, n\}$, an estimate of $\varrho(u,v)$ can be given as

$$\widetilde{\varrho}(u,v) = \frac{\widehat{F}_{\varepsilon_{Y|Z}, \varepsilon_{X|Z}}(u,v) - \widehat{F}_{\varepsilon_{Y|Z}}(u)\widehat{F}_{\varepsilon_{X|Z}}(v)}{\sqrt{[\widehat{F}_{\varepsilon_{Y|Z}}(u) - \widehat{F}^2_{\varepsilon_{Y|Z}}(u)][\widehat{F}_{\varepsilon_{X|Z}}(v) - \widehat{F}^2_{\varepsilon_{X|Z}}(v)]}}, \tag{7}$$

where $\widehat{F}_{\varepsilon_{Y|Z}, \varepsilon_{X|Z}}(u,v), \widehat{F}_{\varepsilon_{Y|Z}}(u)$ and $\widehat{F}_{\varepsilon_{X|Z}}(v)$ are empirical estimates of $F_{\varepsilon_{Y|Z}, \varepsilon_{X|Z}}(u,v), F_{\varepsilon_{Y|Z}}(u)$ and $F_{\varepsilon_{X|Z}}(v)$, respectively. However, $\varepsilon_{Y|Z}$ and $\varepsilon_{X|Z}$ can not be obtained directly from the sample due to the presence of unknown $m_x(Z)$ and $m_y(Z)$. So they need to be estimated before use. Many popular nonparametric approaches, such as kernel-based local smoothing approaches, series expansion-based methods, and polynomial spline approximation approach (see Fan and Gijbels (1996); Stone (1982)) can be used for estimating $m_x(Z)$ and $m_y(Z)$. In this paper, because of the benefit from computational convenience of B-spline smoothing technique, we employ the B-spline approximation approach to estimate them although other nonparametric approaches can be applied as well.

We only present the details for the estimation of $m_y(z)$ below since the estimate of $m_x(z)$ can be obtained explicitly in the same manner. Specifically, we let $\{B_k(\cdot), k = 1, \cdots, L_n\}$ denote a normalized B-spline basis with $\|B_k\|_\infty \leq 1$, where $L_n$ is the sum of the polynomial degree and the number of knots and $\|\cdot\|_\infty$ is the sup norm. Let $\mathcal{G}_n$ be the space of polynomial splines of degree $l \geq 1$. By the theory of B-spline approximation (de Boor (2001)), under some smoothness conditions, $m_y(z)$ can be well approximated by functions in $\mathcal{G}_n$.

That is, there exists a vector $\boldsymbol{\gamma} \in \mathbb{R}^{L_n}$ such that $m_y(z) \approx \mathbf{B}(z)^T\boldsymbol{\gamma}$, where $\mathbf{B}(\cdot) = (B_1(\cdot), \cdots, B_{L_n}(\cdot))^T$. Given a sample of observations $\{(Z_i, Y_i), i = 1, \ldots, n\}$, the estimate of $m_y(z)$ is $\widehat{m}_y(z) = \mathbf{B}(z)^T\widehat{\boldsymbol{\gamma}}$, where $\widehat{\boldsymbol{\gamma}}$ is the estimate of $\boldsymbol{\gamma}$, which is defined as the minimizer of

$$\min_{\boldsymbol{\gamma} \in \mathbb{R}^{L_n}} \frac{1}{n} \sum_{i=1}^{n} \ell(Y_i - \mathbf{B}(Z_i)^T\boldsymbol{\gamma}). \tag{8}$$

If the $L_2$ loss is used in (8), then $\widehat{\boldsymbol{\gamma}}$ has a closed form, i.e., $\widehat{\boldsymbol{\gamma}} = (\mathbf{B}^T\mathbf{B})^{-1}\mathbf{B}^T\mathbb{Y}$, where $\mathbf{B} = (\mathbf{B}(Z_1), \cdots, \mathbf{B}(Z_n))^T$ and $\mathbb{Y} = (Y_1, \cdots, Y_n)^T$. Thus, $\widehat{m}_y(z) = \mathbf{B}(z)^T\widehat{\boldsymbol{\gamma}} = \mathbf{B}(z)^T(\mathbf{B}^T\mathbf{B})^{-1}\mathbf{B}^T\mathbb{Y}$. If the $L_1$ loss is used in (8), then $\widehat{\boldsymbol{\gamma}}$ becomes the least absolute deviation estimator without a closed form. Similarly, we can obtain the estimate of $m_x(z)$ as $\widehat{m}_x(z)$.

With the above, we can obtain the sample surrogates of $\varepsilon_{Y|Z}$ and $\varepsilon_{X|Z}$ by $\widehat{\varepsilon}_{i,Y|Z} = Y_i - \widehat{m}_y(Z_i)$ and $\widehat{\varepsilon}_{i,X|Z} = X_i - \widehat{m}_x(Z_i)$, respectively, for $i = 1, \ldots, n$. As a result, a practical estimate of $\varrho(u, v)$ can be obtained as

$$\widehat{\varrho}(u, v) = \frac{\widehat{F}_{\widehat{\varepsilon}_{Y|Z}, \widehat{\varepsilon}_{X|Z}}(u, v) - \widehat{F}_{\widehat{\varepsilon}_{Y|Z}}(u)\widehat{F}_{\widehat{\varepsilon}_{X|Z}}(v)}{\sqrt{[\widehat{F}_{\widehat{\varepsilon}_{Y|Z}}(u) - \widehat{F}^2_{\widehat{\varepsilon}_{Y|Z}}(u)][\widehat{F}_{\widehat{\varepsilon}_{X|Z}}(v) - \widehat{F}^2_{\widehat{\varepsilon}_{X|Z}}(v)]}}. \tag{9}$$

Suppose we have collected the ultrahigh-dimensional data $\{(Y_i, \mathbf{X}_i, Z_i), i = 1, \ldots, n\}$ with $\mathbf{X}_i \in \mathbb{R}^{p_n}$, where $p_n$ is ultra-high dimensional, and $Z$ is an exposure variable which is usually known in advance to have an effect on both the response $Y$ and covariates in the data analysis. For example, for the genetic data analysis, the response can be a clinical outcome such as the cure time of some disease, the ultrahigh dimensional covariates can be thousands of gene expression measurements, and the exposure variable can be the age of patients. In such data, an essential goal is to identify which genes are most influential ones to the outcome when taking into account the information of patient's age. The aforementioned RPC can be applied to serving as a screening index in variable screening for such a situation. To this end, we define a marginal utility for the $j$th predictor, $X_j$, on population level as

$$u_j^{RPC} = E\{\varrho^2(\varepsilon_{Y|Z}, \varepsilon_{X_j|Z})\}, 1 \leq j \leq p_n. \tag{10}$$

A simple sample estimator of $u_j^{RPC}$ can be formulated as

$$\widehat{u}_j^{RPC} = \frac{1}{n} \sum_{i=1}^{n} \widehat{\varrho}^2(\widehat{\varepsilon}_{i,Y|Z}, \widehat{\varepsilon}_{i,X_j|Z}), 1 \leq j \leq p_n, \tag{11}$$

where $\widehat{\varepsilon}_{i,Y|Z}$ and $\widehat{\varepsilon}_{i,X_j|Z}$ can be obtained using the previous method, and $\widehat{\varrho}$ is given in (9). Then, we select the following set of predictors

$$\widehat{\mathcal{M}}_b = \{j : \widehat{u}_j^{RPC} > \nu_n, 1 \leq j \leq p_n\}. \tag{12}$$

as an empirical active set when used for prediction, where $\nu_n$ is the user-specified threshold value. We refer to this procedure as the robust partial correlation-based sure independence screening (denoted as RPC-SIS).

It is worth noting that, when implementing the RPC-SIS as above, if one specifies the $L_2$ loss in (8), we refer to this procedure as RPC-SIS(L2); and if one specifies the $L_1$ loss in (8), we refer to this procedure as RPC-SIS(L1). Intuitively, since using $L_1$ loss can yield more stable residuals than using $L_2$ loss, so RPC-SIS(L1) would be expected to behave more robust than RPC-SIS(L2) in the presence of abnormal response and covariates. In what follows, we will systematically study the sure screening properties and finite-sample performance of RPC-SIS(L2) and RPC-SIS(L1), respectively.

## 3 Theoretical Results

### 3.1 Weak convergence of $\widehat{u}_j^{RC}$

**Proposition 3.1.** *Let $\ell^\infty(\mathcal{Y} \times \mathcal{X})$ denote the Skorohod space equipped with the uniform norm. Then,*

$$\sqrt{n}[\widehat{\rho}(y,x) - \rho(y,x)] \overset{w}{\rightsquigarrow} \phi'_{\boldsymbol{\theta}}(\mathbb{G})$$

*in $\ell^\infty(\mathcal{Y} \times \mathcal{X})$, where $\overset{w}{\rightsquigarrow}$ indicates "convergence weakly", and $\phi'_{\boldsymbol{\theta}}(\mathbb{G})$ is a Gaussian process with mean zero and covariance function $\Xi(x_1, y_1, x_2, y_2)$ given in the Appendix B.*

This proposition can be used for statistical inference on $\rho(y,x)$, such as testing if $\rho(y,x)$ equals zero or constructing a confidence interval for $\rho(y,x)$ at a specified $(y_0, x_0)$. For a fixed pair $(y,x)$, $\sqrt{n}[\widehat{\rho}(y,x) - \rho(y,x)] \overset{d}{\to} N(0, \Xi(x,y))$, where $\Xi(x,y)$ is the asymptotic variance given in the Appendix B. With this result, we have the following weak convergence for $\widehat{u}_j^{RC}$.

**Theorem 3.1.** *(i) If $X_j$ and $Y$ are independent, then $\widehat{u}_j^{RC} = 0$, and as $n \to \infty$,*

$$P(n\widehat{u}_j^{RC} < v\omega_j) - P(Q_j < v) \to 0, \text{ for } v \in \mathbb{R}^+,$$

where $\omega_j = E\{[\phi'_{\boldsymbol{\theta}_j}(\mathbb{G})(\widetilde{Y}, \widetilde{X}_j)]^2\} = E\{\Xi(\widetilde{Y}, \widetilde{X}_j, \widetilde{Y}, \widetilde{X}_j)\}$, in which $(\widetilde{Y}, \widetilde{X}_j)$ is an independent copy of $(Y, X_j)$, and $Q_j = \sum_{k=1}^{\infty} \lambda^*_{j,k} \chi^2_k(1)$, $\chi^2_k(1)$s are independent $\chi^2(1)$ random variables, and $\lambda^*_{j,k}$ are non-negative constants that depend on the joint distribution of $(X_j, Y)$, with sum equal to one.

(ii) If $X_j$ and $Y$ are not independent, then $\widehat{u}^{RC}_j > 0$, and as $n \to \infty$,

$$P(n^{1/2}(\widehat{u}^{RC}_j - u^{RC}_j) < t) - P(T_j < t) \to 0, \text{ for } t \in \mathbb{R},$$

where $T_j \sim N(0, \Delta_j)$ and $\Delta_j$ is given in the Appendix. Consequently, $n\widehat{u}^{RC}_j$ diverges to infinity as $n \to \infty$.

This theorem suggests that $\widehat{u}^{RC}_j$ is $n$-consistent if $X_j$ and $Y$ are independent, and is $\sqrt{n}$-consistent otherwise, which implies that the $\widehat{u}^{RC}_j$-based test can be used to test whether two random variables are independent, and has nontrivial power. To put the test into practice, we need to decide the critical values in the test since the weights $\lambda^*_{j,k}$ involved in the limiting distribution are unknown. We can use a wild bootstrap procedure as was adopted in Zhou et al. (2020). Specifically, we define a wild bootstrap sample $\{(X^*_{ij}, Y_i)\}$, where $X^*_{ij} = \bar{X}_j + \iota_i \epsilon_{ij}$ and $\epsilon_{ij} = X_{ij} - \bar{X}_j$, $\bar{X}_j$ is the sample mean of $\{X_{ij}, i = 1, \ldots, n\}$, and $\iota_i$ satisfies $P(\iota_i = 1) = 1/2$ and $P(\iota_i = -1) = 1/2$. We repeat the wild bootstrap sample $D$ times. Then on the basis of these $D$ bootstrapped sample sets, we can obtain $\widetilde{\widehat{u}}^{RC}_{j(d)}, d = 1, \ldots, D$, of which, we calculate the $(1-\alpha)$-th quantile, denoted by $u^*_{1-\alpha}$. The null hypothesis will be rejected if $\widehat{u}^{RC}_j$ is greater than $u^*_{1-\alpha}$, and not be rejected otherwise.

## 3.2 Sure Screening Property for RC-SIS

Denote the active set of predictors by $\mathcal{M}_{1*} = \{j: F_{Y|\mathbf{X}}(y|\cdot) \text{ functionally depends on } X_j \text{ for some } y \in \mathcal{Y}\}$. We need the following technical conditions to establish the sure screening property for RC-SIS.

(D1) For $j = 1, \ldots, p_n, i = 1, \ldots, n$, we denote $U_{i0} \triangleq F_Y(Y_i)$ and $U_{ij} \triangleq F_{X_j}(X_{ij})$. We use $F_{0j}(u_1, u_2)$ to denote the joint distribution function of $(U_{i0}, U_{ij})$. Assume $F_{0j}$ has continuous partial derivatives $\frac{\partial F_{0j}}{\partial u_1}$ and $\frac{\partial F_{0j}}{\partial u_2}$, and there exist uniform constants $M_1, M_2 > 0$ such that

$$\max_{1 \leq j \leq p_n} \sup_{(y, x_j) \in \mathcal{Y} \times \mathcal{X}_j} \left| \frac{\partial F_{0j}(u_1, u_2)}{\partial u_1} \big|_{(u_1, u_2) = (F_Y(y), F_{X_j}(x_j))} \right| \leq M_1,$$

$$\max_{1 \leq j \leq p_n} \sup_{(y, x_j) \in \mathcal{Y} \times \mathcal{X}_j} \left| \frac{\partial F_{0j}(u_1, u_2)}{\partial u_2} \big|_{(u_1, u_2) = (F_Y(y), F_{X_j}(x_j))} \right| \leq M_2.$$

(D2) There exist two uniform constants $K_1$ and $K_2$ such that

$$\inf_{y \in \mathcal{Y}} F_Y(y)[1 - F_Y(y)] \geq K_1 > 0,$$
$$\min_{1 \leq j \leq p_n} \inf_{x_j \in \mathcal{X}_j} F_{X_j}(x_j)[1 - F_{X_j}(x_j)] \geq K_2 > 0.$$

(D3) $\min_{j \in \mathcal{M}_{1*}} u_j^{RC} \geq 2Cn^{-\kappa}$ for a constant $C > 0$ and some $\kappa \in (0, 1/2)$.

*Remark 1* Conditions (D1)-(D3) are very mild and easily satisfied. Condition (D1) imposes the smoothing assumption on the joint distribution of $(X_j, Y)$. Condition (D2) requires that the variances of indicators related to all predictors and the response variable are bounded away from zero uniformly in $j$, which can be used to guarantee that the marginal signal of each active predictor does not vanish. Condition (D3) is a standard assumption in the literature of sure independence screening (Fan and Lv (2008)), which ensures that the strength of minimal marginal signal can be identified and decrease not very quickly as the sample size increases.

It is noteworthy that condition (D1) and condition (D7) given below are two technical requirements on first-order partial derivatives of copula functions. Although the forms involved seem a bit complex and restrictive, there exist some bivariate symmetric copula families whose partial derivatives can be bounded on the entire region. For example, for the Gaussian copula defined as $C(u_1, u_2; \rho) = \Phi_2(\Phi^{-1}(u_1), \Phi^{-1}(u_2), \rho)$, the first-order partial derivative is $\frac{\partial}{\partial u_2} C(u_1, u_2; \rho) = \Phi((\Phi^{-1}(u_1) - \rho\Phi^{-1}(u_2))/\sqrt{1-\rho^2})$, where $\Phi_2(\cdot, \cdot, \rho)$ is the joint distribution of two standard normal distributed random variables with correlation $\rho \in (-1, 1)$, $\Phi(\cdot)$ is CDF of $N(0, 1)$, and $\Phi^{-1}(\cdot)$ is the quantile function. Clearly, $\frac{\partial}{\partial u_2} C(u_1, u_2; \rho)$ is bounded by one, thus conditions (C1) and (C7) are fulfilled without requiring $\mathcal{X}_j$ and $\mathcal{Y}$ to be compact regions. More other examples related to heavy-tailed distributions can be found in Aas et al. (2009) and Schepsmeier and Stöber (2014). In addition, the constant bounds $M_1$ and $M_2$ can be further relaxed to $M\psi_n$, where $M$ is a finite constant and $\psi_n$ is a slowly diverging rate depending on $n$ such as $\log n$, the sure screening property for RC-SIS still holds but with a longer modification of the proof.

When the response $Y$ is a continuous variable, we denote $\boldsymbol{\pi}(y, \mathbf{x}) = (\pi_1(y, x_1), \ldots, \pi_j(y, x_{p_n}))^T$, where $\pi_j(y, x_j) = \text{cov}(I(Y \leq y), I(X_j \leq x_j))$ for $j = 1, \ldots, p_n$, and we have the following properties for RC-SIS.

**Theorem 3.2.** *(Consistency Property for RC) Suppose that conditions (D1) and (D2) are satisfied. Then, for any $\epsilon > 0$, if $n\epsilon \to \infty$ and $n\epsilon^2/\log n \to \infty$ as $n \to \infty$, there exists a uniform constant $\tilde{c}_1 > 0$ such that*

$$P\Big( \max_{1 \leq j \leq p_n} |\widehat{u}_j^{RC} - u_j^{RC}| > \epsilon \Big) \leq 36p_n \exp(-\tilde{c}_1 n\epsilon^2).$$

**Theorem 3.3.** *(Sure Screening Property for RC-SIS) Suppose that conditions (D1)-(D3) hold and by choosing $\varsigma_n = Cn^{-\kappa}$, then (i) we have*

$$P\big(\mathcal{M}_{1*} \subset \widehat{\mathcal{M}}_a\big) \geq 1 - 36s_n \exp(-\tilde{c}_2 n^{1-2\kappa})$$

*for some positive constant $\tilde{c}_2$ and all sufficiently large $n$; (ii) if $E\|\boldsymbol{\pi}(Y,\mathbf{X})\|^2 = O(n^\iota)$ for some $\iota > 0$, then*

$$P(|\widehat{\mathcal{M}}_a| \leq O(n^{\kappa+\iota})) \geq 1 - 36p_n \exp(-\tilde{c}_2 n^{1-2\kappa}),$$

*and (iii) furthermore, if $u_j^{RC} = 0$ for $j \notin \mathcal{M}_{1*}$ and $\log p_n = o(n^{1-2\kappa})$, then*

$$P(\widehat{\mathcal{M}}_a = \mathcal{M}_{1*}) \to 1, \ n \to \infty.$$

*Remark 2* Theorem 3.2 established the consistency of empirical screening utility, which is the key to the proof of Theorem 3.3. Theorem 3.3 says that our proposed RC-SIS can select, with probability approaching one, a set of predictors that contain all truly active predictors, i.e., the sure screening property (Fan and Lv (2008)), and also implies that the dimensionality handled by RC-SIS can be as high as $o(\exp(\tilde{c}_1 n^{1-2\kappa}))$. A natural question is how large the size of the selected set by RC-SIS is? The parts (ii) and (iii) of Theorem 3.3 provide an answer. The part (ii) means the selected model size can be controlled at a polynomial rate of $n$ under the assumption that $E\|\boldsymbol{\pi}(Y,\mathbf{X})\|^2 = O(n^\iota)$ for some $\iota > 0$, which is similar to the condition that $\sum_{j=1}^{p_n} |\text{qpcor}_\tau(Y, Xj|X_{S_j})| = O(n^\varsigma)$ for some $\varsigma > 0$ imposed in Ma, Li and Tsai (2017). The part (iii) implies that if the truly active and noisy predictors are well-separated, the proposed RC-SIS is selection consistent.

When the response $Y$ is a discrete variable, the following theorem says the sure screening property for RC-SIS still holds. To this end, assume that $Y$ satisfies $P(Y = y_k) = p_k$ for $k = 1, \ldots, J$, where $J$ can be allowed to diverge with $n$.

**Theorem 3.4.** *Under conditions (D2) and (D3), if $\frac{n^{1-2\kappa}}{J^2 \log(\max(n,J))} \to \infty$ and $\log p_n = o(n^{1-2\kappa}/J^2)$ as $n \to \infty$, then (i)*

$$P\big(\mathcal{M}_{1*} \subset \widehat{\mathcal{M}}_a\big) \to 1, \ as \ n \to \infty;$$

*(ii) if $E\|\boldsymbol{\pi}(Y,\mathbf{X})\|^2 = O(n^\iota)$ for some $\iota > 0$, then $P(|\widehat{\mathcal{M}}_a| \leq O(n^{\kappa+\iota})) \to 1$ as $n \to \infty$; and (iii) if further $u_j^{RC} = 0$ for $j \notin \mathcal{M}_{1*}$, then our RC-SIS is selection consistent.*

## 3.3 Sure Screening Property for RPC-SIS

### 3.3.1 Sure Screening Property for RPC-SIS(L2)

Throughout this subsection, we denote $m_0(z) = E(Y|Z = z)$ and $m_j(z) = E(X_j|Z = z)$ and write $\varepsilon_{i0} = Y_i - m_0(Z_i)$ and $\varepsilon_{ij} = X_{ij} - m_j(Z_i)$. In order to simplify the statements and without causing much confusion, we use $\varepsilon_j$ to indicate the population version of $\varepsilon_{ij}$ and denote by $\mathcal{E}_j$ the range of $\varepsilon_j$ for $j = 0, 1, \ldots, p_n$. We keep in mind that the case of $j = 0$ corresponds to the response $Y$ and $j > 0$ is associated with the $j$th predictor $X_j$. Furthermore, we denote $F_{\varepsilon_j}(u)$ by the CDF of $\varepsilon_j$ for $j = 0, 1, \ldots, p_n$, and $F_{\varepsilon_0, \varepsilon_j}(u, v)$ by the joint distribution function of $\varepsilon_0$ and $\varepsilon_j$ for any $j > 0$. We denote the active set of predictors by

$$\mathcal{M}_{2*} = \{j : F_{Y|\mathbf{X}, Z}(y|\cdot, \cdot) \text{ functionally depends on } X_j \text{ for some } y \in \mathcal{Y}\}.$$

The sure screening property for the RPC-SIS(L2) is established under the following conditions.

(D4) The functions $m_j, j = 0, 1, \ldots, p_n$ belong to a class of functions $\mathcal{B}$, where the $r$th derivative $m^{(r)}$ of any class member $m$ exists and is Lipschitz of order $\alpha$. That is,

$$\mathcal{B} = \{m(\cdot) : |m^{(r)}(s) - m^{(r)}(t)| \leq M|s-t|^\alpha \text{ for } s, t \in \mathcal{T}\},$$

for some positive constant $M$, $r$ a nonnegative integer, and $\alpha \in (0, 1]$ such that $d \equiv r + \alpha > 0.5$.

(D5) The support of index variable $Z$ is bounded, say $\mathcal{Z} = [a, b]$ with finite constants $a$ and $b$, with density $f_Z$ bounded away from zero and infinity. That is, there exist two positive constants $\bar{M}_1, \bar{M}_2$ such that $\bar{M}_1 \leq \inf_{z \in \mathcal{Z}} f_Z(z) \leq \sup_{z \in \mathcal{Z}} f_Z(z) \leq \bar{M}_2$.

(D6) For all $j = 0, 1, \ldots, p_n$, there exists a positive uniform constant $\bar{K}_1$ such that $P(|W_j| > x|Z = z) \leq \bar{K}_1 \exp(-\bar{K}_1^{-1} x)$ uniformly in $z$ and $j$, where $W_0 = Y$ and $W_j = X_j, j \geq 1$.

(D7) The density of $F_{\varepsilon_j}(u)$, $f_{\varepsilon_j}(u)$, satisfies $\max_{0 \leq j \leq p_n} \sup_u f_{\varepsilon_j}(u) \leq M_{f_\varepsilon}$ for some positive constant $M_{f_\varepsilon}$, and the first-order partial derivatives of $F_{\varepsilon_0, \varepsilon_j}(u, v)$ satisfy $\sup_{u,v} \left|\frac{\partial F_{\varepsilon_0, \varepsilon_j}(u,v)}{\partial u}\right| \leq M_{F_1}$ and $\sup_{u,v} \left|\frac{\partial F_{\varepsilon_0, \varepsilon_j}(u,v)}{\partial v}\right| \leq M_{F_2}$ uniformly in $j$. In addition, we use $\bar{F}_{0j}(u, v)$ to denote the joint distribution function of $(\bar{U}_{i0}, \bar{U}_{ij})$, where $\bar{U}_{ij} = F_{\varepsilon_j}(\varepsilon_{ij}), i = 1, \ldots, n$, and further assume that $\bar{F}_{0j}$ has continuous partial derivatives $\frac{\partial \bar{F}_{0j}}{\partial u}$ and $\frac{\partial \bar{F}_{0j}}{\partial v}$, and that there exist uniform constants $\bar{M}_3, \bar{M}_4 > 0$ such that

$$\max_{1 \leq j \leq p_n} \sup_{(u,v) \in \mathcal{E}_0 \times \mathcal{E}_j} \left|\frac{\partial \bar{F}_{0j}(u,v)}{\partial u}\right| \leq \bar{M}_3, \max_{1 \leq j \leq p_n} \sup_{(u,v) \in \mathcal{E}_0 \times \mathcal{E}_j} \left|\frac{\partial \bar{F}_{0j}(u,v)}{\partial v}\right| \leq \bar{M}_4.$$

(D8) There exist two uniform constants $K_3$ and $K_4$ such that

$$\inf_{u \in \mathcal{E}_0} F_{\varepsilon_0}(u)[1 - F_{\varepsilon_0}(u)] \geq K_3 > 0,$$

$$\min_{1 \leq j \leq p_n} \inf_{v \in \mathcal{E}_j} F_{\varepsilon_j}(v)[1 - F_{\varepsilon_j}(v)] \geq K_4 > 0.$$

(D9) $\min_{j \in \mathcal{M}_{2*}} u_j^{RPC} \geq 2\bar{C} n^{-\tau}$ for a constant $\bar{C} > 0$ and some $\tau \in (0, 1/2)$.

*Remark 3* Conditions (D4)-(D5) are regularity conditions for the smoothness of conditional expectation function, which facilitates us to employ the theory of B-spline approximation. Condition (D6) is equivalent to a conditional Cramér condition, which requires conditional tail probability to be sub-exponential and implies a bounded conditional moment of finite order. Similar to condition (C1), condition (D7) is imposed to ensure that the marginal (joint) distribution functions of $(\varepsilon_0, \varepsilon_j)$ are continuous and (partially) differentiable, and its partial derivatives are uniformly bounded. Condition (D8) requires that the variance of $I(\varepsilon_j \leq u)$ is bounded away from zero uniformly in $j$ and $u$. Such a condition can guarantee that the marginal utility of each of the active predictors does not vanish. Similar assumptions are also imposed in the literature, such as condition (1) in Fan, Ma and Dai (2014), condition (C4) in Liu, Li and Wu (2014) and condition (C5) in Xia, Li and Fu (2019). Condition (D9) is a constraint on the minimum signal of active predictors, which is similar to condition (D3).

**Theorem 3.5.** *(Sure Screening Property for RPC-SIS(L2)) Suppose that conditions (D4)-(D9) are satisfied and if $n^{1-2\tau} L_n^{-4} / \max(\log L_n, \log n) \to \infty$ and $L_n^{d-1/2} n^{-\tau} \to \infty$ as $n \to \infty$, then (i) by choosing $\nu_n = \bar{C} n^{-\tau}$, there exist positive constants $\bar{c}_1^*$ and $\bar{c}_2^*$ such that*

$$P(\mathcal{M}_{2*} \subset \widehat{\mathcal{M}}_b) \geq 1 - \bar{c}_1^* s_n \exp(-\bar{c}_2^* n^{1-2\tau} L_n^{-4})$$

*for sufficiently large $n$; (ii) if $\sum_{j=1}^{p_n} u_j^{RPC} = O(n^\iota)$ for some $\iota > 0$, then*

$$P(|\widehat{\mathcal{M}}_b| \leq O(n^{\kappa + \iota})) \geq 1 - \bar{c}_1^* p_n \exp(-\bar{c}_2^* n^{1-2\tau} L_n^{-4}),$$

*and (iii) if we further assume that $u_j^{RPC} = 0$ for every $j \notin \mathcal{M}_{2*}$ and $\log p_n = o(n^{1-2\tau} L_n^{-4})$, then RPC-SIS(L2) is selection consistent, i.e., $P(\widehat{\mathcal{M}}_b = \mathcal{M}_{2*}) \to 1$, as $n \to \infty$.*

*Remark 4* Compared with Theorem 3.3, we can see that the dimensionality handled by RPC-SIS can be of order $o(\exp(\bar{c}_2 n^{1-2\tau} L_n^{-4}))$. Such an order is slower than the one obtained by the proposed RC-SIS approach.

The difference between those two rates is the term $L_n^{-4}$ in the exponent. This term can be regarded as the price that we pay for sufficing the dimensionality to consistently obtain a nonparametric estimator of unknown conditional expectation function by B-spline approximation.

*Remark 5* The tail probability bound established in Theorem 3.5 is comparable to some existing results in the literature. For instance, if we replace condition (D9) with the assumption that the minimal signal satisfies $\min_{j \in \mathcal{M}_{2*}} u_j^{RPC} \geq 2\bar{C}L_n n^{-2\tau}$ for a constant $\bar{C} > 0$, then by using the same arguments and choosing $\nu_n > \bar{C}L_n n^{-2\tau}$, we can prove that there exist positive constants $\bar{c}_3$ and $\bar{c}_4$ such that $P(\mathcal{M}_{2*} \subset \widehat{\mathcal{M}}_b) \geq 1 - \bar{c}_3 s_n \exp(-\bar{c}_4 n^{1-4\tau} L_n^{-2})$ for sufficiently large $n$, provided that $n^{1-4\tau} L_n^{-2}/\log L_n \to \infty$ as $n \to \infty$. This implies that we can handle the NP dimensionality, $p_n = o(\exp(\bar{c}_4 n^{1-4\tau} L_n^{-2}))$, the same as that in Xia, Li and Fu (2019), a slightly higher order than those in Fan, Feng and Song (2011) and Fan, Ma and Dai (2014).

### 3.3.2 Sure Screening Property for RPC-SIS(L1)

In current subsection, we mean $m_0(z) = \text{median}(Y|Z = z)$ and $m_j(z) = \text{median}(X_j|Z = z)$, and other notations and conditions given in the previous subsection are adapted accordingly. To establish the sure screening property for RPC-SIS(L1), we need one more condition as follows.

(D10) For $j = 0, 1, \cdots, p_n$, conditional density function $f_{W_j|Z}(w|\cdot)$ is bounded away from zero and infinity in a neighborhood of $m_j(Z)$ uniformly in $j$, where $W_0 = Y$ and $W_j = X_j$, $j \geq 1$.

**Theorem 3.6.** *(Sure Screening Property for RPC-SIS(L1)) Suppose that conditions (D4), (D5) and (D7)-(D10) are satisfied and if $\max(n^{1-2\tau} L_n^{-3}, n^{1-4\tau} L_n^{-2})/\log n \to \infty$ and $L_n^{d/2} n^{-\tau} \to \infty$ as $n \to \infty$, then (i) by choosing $\nu_n = \bar{C} n^{-\tau}$, there exist positive constants $\bar{c}_3^*$, $\bar{c}_4^*$, $\bar{c}_5^*$ and $\bar{c}_6^*$ such that*

$$P(\mathcal{M}_{2*} \subset \widehat{\mathcal{M}}_b) \geq 1 - s_n\{\bar{c}_3^* \exp(-\bar{c}_4^* n^{1-2\tau} L_n^{-3}) + \bar{c}_5^* \exp(-\bar{c}_6^* n^{1-4\tau} L_n^{-2})\}$$

*for sufficiently large $n$; (ii) if $\sum_{j=1}^{p_n} u_j^{RPC} = O(n^\iota)$ for some $\iota > 0$, then*

$$P(|\widehat{\mathcal{M}}_b| \leq O(n^{\kappa+\iota})) \geq 1 - p_n\{\bar{c}_3^* \exp(-\bar{c}_4^* n^{1-2\tau} L_n^{-3}) + \bar{c}_5^* \exp(-\bar{c}_6^* n^{1-4\tau} L_n^{-2})\},$$

*and (iii) if we further assume that $u_j^{RPC} = 0$ for every $j \notin \mathcal{M}_{2*}$ and $p_n = o(\exp(-\bar{c}_4^* n^{1-2\tau} L_n^{-3}) + \bar{c}_5^* \exp(-\bar{c}_6^* n^{1-4\tau} L_n^{-2}))$, then RPC-SIS(L1) is selection consistent.*

*Remark 6* Condition (D10) is a standard condition in the literature on quantile related variable screening (He, Wang and Hong (2013), Ma, Li and Tsai (2017)). This result reveals that the dimensionality handled by RPC-SIS(L1) can be $p_n = o(\exp(\bar{c}_4^* n^{1-2\tau} L_n^{-3}) + \exp(\bar{c}_6^* n^{1-4\tau} L_n^{-2}))$, a higher order than the one obtained by RPC-SIS(L2). If taking the optimal rate for nonparametric B-spline regression, i.e., $L_n = O(n^{\frac{1}{2d+1}})$, then the sure screening property for RPC-SIS(L1) holds even if $\tau$ and $d$ satisfy the relationship that $\tau < \frac{2d-3}{2(2d+1)}$, which requires $d > 3/2$. Meanwhile, the the sure screening property for RPC-SIS(L1) is guaranteed if $\tau < \frac{d-1}{2d+1}$ with $1 < d < 3/2$ or $\tau < \frac{2d-1}{4(2d+1)}$ with $d \geq 3/2$.

## 4 Simulation Study

In this section, we conduct several simulation examples to investigate the finite-sample performance of our proposed variable screening procedures: RC-SIS and RPC-SIS, respectively. Examples 1 and 2 are designed for illustration of RC-SIS, and the remaining examples are conducted for RPC-SIS. For each of these two screening approaches, the screening threshold parameter involved is determined by selecting the first $d_n = \lfloor n/\log n \rfloor$ predictors that are mostly associated with the response $Y$ in terms of our proposed correlation coefficients: RC and RPC, where $\lfloor a \rfloor$ indicates the largest integer but not bigger than $a$. All of the simulation results are assessed via the following three criteria: (i) the minimum model size (MMS), the minimum number of the selected predictors that contain all the active predictors, and its robust standard deviation (RSD); (ii) the rank for each active covariates ($\mathcal{R}_j$); and (iii) the proportion of all the active predictors being selected ($\mathcal{P}$) with the screening threshold $d_n$ over $N = 200$ replications, where the medians of MMS and RSD are reported in the tables below. Notice that the approach that possesses the smallest values of MMS (RSD) and $\mathcal{R}_j$ and the largest value of $\mathcal{P}$ is considered to work most satisfactorily.

### 4.1 Performance of RC-SIS

In Examples 1 and 2, we have compared our RC-SIS with several popularly used variable screening approaches developed in the literature: SIS (Fan and Lv (2008)) which is based on Pearson's correlation coefficient, SIRS (Zhu et al. (2011)), DC-SIS (Li, Zhong and Zhu (2012)), Kendall's $\tau$-SIS (Li et al. (2012)), CC-SIS with $(\tau, \iota) = (0.5, 0.5)$ (Xia and Li (2021)), QC-SIS with $\tau = 0.5$ (sure independence screening based on the quantile correlation proposed by Li, Li and Tsai (2015).

*Example 1* We generate the covariates $\mathbf{X} = (X_1, \ldots, X_{p_n})^T$ from a mixture distribution of $\mathbf{X}_0$ with probability 0.8 and $\boldsymbol{\epsilon}$ with probability 0.2, i.e., $\mathbf{X} = 0.8\mathbf{X}_0 + 0.2\boldsymbol{\epsilon}$, where $\mathbf{X}_0 = (X_{01}, \ldots, X_{0p_n})^T$ is from a $p_n$-variate normal distribution $N(\mathbf{0}_{p_n}, \Sigma)$ with $\Sigma = (\sigma_{ij})_{1 \leq i,j \leq p_n}$ and $\sigma_{ij} = \rho_0^{|i-j|}$, and all elements of $\boldsymbol{\epsilon}$ are

i.i.d. and from a standard Cauchy distribution, i.e., $Cauchy(0, 1)$. The response variable $Y$ is simulated as

$$Y = 3X_{01} + 3X_{02} + 2X_{03} + 2X_{04} + 2X_{05} + \varepsilon,$$

where $\varepsilon$ is independent of both $\mathbf{X}$ and $\boldsymbol{\epsilon}$, and distributed as $Cauchy(0, 1)$. The corresponding results are reported in Table 1, from which, we can see that our proposed RC-SIS generally has best performance among aforementioned seven approaches, while the Kendall's $\tau$-SIS has a very competitive performance with RC-SIS, both of which are uniformly better than other popularly used screening approaches. Meanwhile, a closer look at Table 1 gives that the screening performance of RC-SIS can be improved by enhancing the correlation among covariates and by enlarging the sample size. Furthermore, it seems that, in general, the larger the dimensionality, the harder the method to do screening.

*Example 2* Similar to Example 1, the covariates $\mathbf{X}$ are simulated from the mixture distribution $w_0 \mathbf{X}_0 + (1 - w_0)\boldsymbol{\epsilon}$, where $\mathbf{X}_0 \sim N(0_{p_n}, \Sigma)$ with $\Sigma = (\sigma_{ij})_{1 \leq i,j \leq p_n}$ and $\sigma_{ij} = \rho_0^{|i-j|}$, and $\boldsymbol{\epsilon}$ has the same distribution as that in Example 1. Note that $w_0 = 1$ is corresponding to the case of $\mathbf{X} = \mathbf{X}_0$, that is, the covariates are all normally distributed. We generate the response from the following four models, respectively,

(b1): $Y = 5X_{01}I(X_{01} < 0) + 5X_{02}I(X_{02} > 0) + 5\sin(X_{010}) + \varepsilon,$

(b2): $Y = 5\beta_1 X_{01} + 5\beta_2 X_{02} + 5\beta_3 X_{03} + 5\beta_4 X_{04} + \varepsilon,$

(b3): $Y = 5\beta_1 X_{01}^2 + 5\beta_2 X_{02} X_{03} + 5\beta_3 I(X_{04} > 0) + \varepsilon,$

(b4): $Y = \exp(3\beta_1 \sin(X_{01})) + 2\beta_2 \exp(X_{02}) + 3\beta_3 I(X_{03} > 0) + \log(4|X_{04}| + 0.5) + \varepsilon,$

where $\varepsilon$ is distributed as $Cauchy(0, 1)$ for all models. Note that models (b1), (b3) and (b4) are nonlinear models. Under each setting, the dimensionality of $\mathbf{X}$ is taken as $p_n = 1000$. The corresponding results are given in Table 2. We can see that under various model settings, sample sizes and strengths of dependence of covariates, our proposed RC-SIS performs uniformly better than other competitive methods in terms of MMS (RSD) and selection proportion. When comparing the columns of $w_0 = 0.8$ and $w_0 = 1$, we can further observe that the normal covariates tend to be easier to be screened than the covariates containing outliers.

*Example 3* In this example, we simulate the response variable from the following nonparametric regression model:

$$Y = 6g_1(X_1) + 6g_2(X_2) + 3g_3(X_3) + 6g_4(X_4) + \varepsilon,$$

where $g_1(u) = u$, $g_2(u) = (3u - 1)^2$, $g_3(u) = 2\sin(2\pi u)/(2 - \sin(2\pi u))$, and $g_4(u) = 0.1\sin(2\pi u) + 0.2\cos(2\pi u) + 0.3\sin(2\pi u)^2 + 0.4\cos(2\pi u)^3 + 0.5\sin(2\pi u)^3$. We generate the covariates $\{X_j, 1 \leq j \leq p_n\}$ as $X_j = (T_j + tU)/(1 + t)$, where $T_j \sim_{i.i.d.} U(0,1)$ and $U \sim U(0,1)$ are independent. The value of $t$ is chosen such that $\mathrm{corr}(X_j, X_k) = \rho_0$ for $j \neq k$. For the model error $\varepsilon$, we consider four distributions: (Normal) $\varepsilon \sim \sqrt{1.74}N(0,1)$, (Student's $t$) $\varepsilon \sim t(3)$, (Cauchy) $\varepsilon \sim \frac{1}{3}Cauchy(0,1)$, and (Mixed Normal) $\varepsilon \sim 0.5N(-2,1) + 0.5N(2,1)$, respectively. We fix $p_n = 1000$ and $\rho_0 = 0.4$, and let $n = 200$ and $n = 400$, respectively, in this example. The simulation results are given in Table 3, from which, we can see that the proposed RC-SIS performs better than other competitive methods across different model errors.

Furthermore, when the response takes on discrete values, in order to see how the proposed RC-SIS behaves, we consider two cases (Bernoulli and Poisson models) to generate the response via four additional examples (S1-S4) given in the Appendix A in the supplementary file. The results are reported in Tables S1-S4 in the supplementary file, showing that our RC-SIS works well for both binary and Poisson responses. Therefore, these evidences reveal that the proposed RC-SIS method not only works for continuous response but also behaves well for discrete response. Besides, RC-SIS produces better finite-sample performance than some existing competitive approaches, and has robustness against many different distributions of covariates and response. Moreover, the performances of RC-SIS and Kendall's $\tau$-SIS are comparable in some settings for which we provide some discussion in the Appendix A2 in the supplementary file.

### 4.2 Performance of RPC-SIS

In this subsection, we investigate the performance of proposed RPC-SIS(L2) and RPC-SIS(L1) and compare it with some existing variable screening procedures: the quantile partial correlation based screening (QPC-SIS, Ma, Li and Tsai (2017)), the copula partial correlation based screening (CPC-SIS, Xia and Li (2021)), the conditional quantile correlation based screening (CQC-SIS, Xia, Li and Fu (2019)) and the nonparametric independence screening under varying coefficient model (NIS, Fan, Ma and Dai (2014)), where we specify the QPC-SIS with $\tau = 0.5$, CPC-SIS with $(\tau, \iota) = (0.5, 0.5)$, and the CQC-SIS with $\tau = 0.5$. Here, the number of B-spline basis functions involved in both CQC-SIS and NIS is set as $L_n = 4$ as suggested in the papers.

*Example 4* Let $\mathbf{X} \sim N(\mathbf{0}_{p_n}, \Sigma)$, where $\Sigma = (\sigma_{ij})_{1 \leq i,j \leq p_n}$ and $\sigma_{ij} = 0.8^{|i-j|}$, and $Z$ is uniformly distributed as $U(0,1)$. We generate the response variable $Y$ through

$$Y = \mu + \varepsilon \equiv \theta\{2\exp(Z)X_1 + 5(2Z - 1)^2\exp(X_2) + 3\sin(2\pi Z)X_3^2\} + \varepsilon,$$

where we consider two cases for the random error $\varepsilon$ as $\frac{1}{3}Cauchy(0,1)$ and student's $t(3)$, respectively. The value of $\theta$ determines the signal intensity of active covariates, and is chosen such that $R^2 = \frac{\text{var}(\mu)}{\text{var}(Y)}$ approximately equals $0.05, 0.1$ and $0.3$, respectively. In this example, we fix the dimensionality of $\mathbf{X}$ as $p_n = 1000$ and take the sample size $n = 200$. The simulation results regarding the above five approaches are displayed in Table 4. From this table, we can see that our proposed RPC-SIS(L2) and RPC-SIS(L1) have very similar performance, while both behave much better than other methods in terms of both MMS and selection proportion. In addition, we can see that QPC-SIS works also well relative to RPC-SIS. The reason is probably because the covariates $\mathbf{X}$ are considered as normal and the response are heavy-tailed in this example.

*Example 5* In this example, we consider three models to generate the response $Y$. That is,

$$(d1): \quad Y = 2ZX_{01} + 5(2Z-1)^2 X_{02} + 3\sin(2\pi Z)X_{03} + \varepsilon,$$

$$(d2): \quad \log(0.5\exp(1.25Y) - 1) = 2ZX_{01} + 5(2Z-1)^2 X_{02} + 3\sin(2\pi Z)X_{03} + \varepsilon,$$

$$(d3): \quad Y = 2I(Z > 0.4)X_{02} + (1+Z)X_{0100} + (2-3Z)^2 X_{0400} + \exp(Z/(1+Z))X_{0600} + \varepsilon.$$

In each of above models, the model error $\varepsilon$ is simulated from $\frac{1}{3}Cauchy(0,1)$ or student's $t(3)$, the covariates $\mathbf{X}$ are generated as a mixture of two multivariate distributions, $\mathbf{X} = w_0 \mathbf{X}_0 + (1-w_0)\boldsymbol{\epsilon}$, where $w_0$ is a probability weight taking value 0.5 or 0.8, $\mathbf{X}_0 \sim N(0_{p_n}, \Sigma)$ with $\Sigma = (0.8^{|i-j|})_{1 \leq i,j \leq p_n}$, and the conditional variable $Z$ is independent of covariates $\mathbf{X}$ and distributed as $U(0,1)$. For each of models (d1)-(d3), we fix the dimensionality $p_n = 1000$ and the sample size $n = 200$. The simulation results for two probabilities $w_0 = 0.5$ and $w_0 = 0.8$ are reported in Table 5. From the table, we can make a similar conclusion to that in Example 4. It can also be observed that the RPC-SIS(L1) slightly outperforms the RPC-SIS(L2) in the most cases.

*Example 6* In this example, we consider a challenging design, in which the response $Y$ is generated from the model

$$Y = 3X_{01} + 4\sqrt{Z + 1/2}X_{02} + 2\exp(Z)X_{03} + 6\sin(2\pi Z)X_{04}/(2 - \sin(2\pi Z)) + \varepsilon,$$

where $\varepsilon$ is distributed as $\frac{1}{3}Cauchy(0,1)$, the observed covariates follows the mixed distribution: $\mathbf{X} = w_0 \mathbf{X}_0 + (1-w_0)\boldsymbol{\epsilon}$, and the conditional variable $Z$ is independent of $\mathbf{X}$ and distributed as $U(0,1)$. We let $Z$ be correlated with $\mathbf{X}_0$ in the following way. Specifically, for every $1 \leq j \leq p_n$, we set $X_{0j} = (T_j + t_1 U_1)/(1+t_1)$ and $Z = (U_2 + t_2 U_1)/(1+t_2)$, where $U_1, U_2 \sim_{i.i.d.} U(0,1)$ and $T_j \sim_{i.i.d} N(0,1)$, which are independent, and we set the values of $t_1$ and $t_2$ such that $\text{corr}(X_{0j}, X_{0k}) = \rho_0$ for any $j \neq k$ and $\text{corr}(X_{0j}, Z) = 0.4$. We consider three settings for $(n, \rho_0) = (200, 0.4)$, $(200, 0.8)$ and $(400, 0.8)$, respectively. We fix the dimensionality as $p_n = 1000$. The simulation results under three settings with two different probabilities ($w_0 = 0.8$ and $w_0 = 1.0$) are reported in Table 6. From the table, although the covariate $X_4$ seems hard to rank in the top by all aforementioned method, however, we can see that the proposed RPC-SIS(L1) performs best in all the settings, which is closely followed by the RPC-SIS(L2) in general. Despite the relatively large values of MMS for all the methods, the RPC-SIS(L1) still produces the smallest values of MMS among these methods. Besides, the performance for each method can be improved by increasing the sample size.

## 5 Real Data Analysis

In this section, we illustrate the proposed approaches (RC-SIS and RPC-SIS) by three real-world applications. The first application is about the eye gene expression measurements on 120 12-week old male rats, which is used to illustrate our RC-SIS when continuous response variable is available. The details of the analysis and related results are given in the Appendix A1 of the supplementary file. The second application is for lung cancer data, which is used to examine the performance of RC-SIS in the case of discrete response. The third application is regarding a synthetic ultrahigh dimensional data based on a low-dimensional prostate specific antigen data, which is employed to evaluate the usefulness of our RPC-SIS.

### 5.1 Lung Cancer Data

Lung cancer begins in the lungs and may spread to lymph nodes or other organs in the body. It has become the leading cause of cancer deaths for both men and women in the world. Accurately early diagnosis for lung cancer becomes very crucial in order to receive timely treatment. The data set comes from a large retrospective, multi-site, blinded study (Shedden et al. (2008)) and contains 442 lung adenocarcinomas, a specific type of lung cancer that is increasing in prevalence. This set consists of expression measurements of 22,283 gene probes, which are generated by four different laboratories under a common protocol. This data were analyzed for different purposes in the literature, see Li et al. (2016) and Xia, Li and Fu (2019) for example. The data contain the gene information from a total of 442 patients, of which two are excluded because of missing measurements. We consider the patient's survival time as the response. In particular, we construct the binary response whose value is coded 1 if the patient can survive over 50 months, and coded 0 otherwise. The median follow-up time is 46.5 months. We choose 3000 genes whose expression measurements have the largest variance, which are treated as covariates and standardized to have mean zero and unit variance in the downstream analysis.

We now have the sample size $n = 440$ and the dimensionality $p_n = 3000$. Hence, our RC-SIS can apply to such a discrete response data with ultrahigh dimensional covariates. Our aim here is to identify a small proportion of gene probes that significantly affect the patients' survival time, and then use them to predict whether or not a patient can live longer than 50 months once detected positive. For this, we conduct a two-stage procedure for which the proposed RC-SIS method is firstly applied to this data in order to select the top $d_n = \lfloor n/\log n \rfloor = 72$ influential genes, and then a follow-up penalized logistic regression (denoted as penLR) model is carried out on the selected ones. Such a two-stage approach is abbreviated as RC-SIS+penLR. We also compare our RC-SIS+penLR with other two-stage methods for which the methods mentioned in the simulation study and the mean-variance based variable screening (MV-SIS) by Cui, Li and Zhong (2015) are accounted for in the first stage, and the same penLR method is implemented in the second stage.

To evaluate the out-of-sample performance, we randomly split the data into two parts: the training set that is used to build a fitted model, and the test set, over which a test error is computed using the fitted model, with a training-to-test sample size ratio being 5 : 1. To assess the accuracy of the final classification, we report the area under the ROC curve (AUC) as well. We repeat such a random partition 100 times and draw the boxplots of test errors and the AUCs over 100 random samples. The results are displayed in Figure 1. From the figure, it seems that our RC-SIS+penLR achieves the lowest test error and the highest accuracy in terms of AUC, although MV-SIS+penLR performs very comparably to the RC-SIS+penLR method. This indicates our RC-SIS works well for discrete response data.

## 5.2 Prostate Specific Antigen Data

The prostate specific antigen data were collected from a study by Stamey et al. (1989) to examine the association between prostate specific antigen (PSA) and several clinical variables that are potentially associated with PSA in men who were about to receive a radical prostatectomy. The original dataset contains 97 observations on 8 covariates (*lcavol*, log cancer volume; *lweight*, log prostate weight; *age*, the man's age; *lbph*, log of the amount of benign hyperplasia; *svi*, seminal vesicle invasion (1=Yes, 0=No); *lcp*, log of capsular penetration; *gleason*, gleason score; *pgg45*, percent of gleason scores 4 or 5) and the logarithm of PSA which serves as the response variable. The data set can be downloaded at https://web.stanford.edu/~hastie/ElemStatLearn/ or obtained from R package "*mgcv*". In the analysis, we use *age* as the exposure variable $Z$. Here, our goal is to examine how the value of log PSA changes with the man's age using the information from covariates. As a benchmark of comparison, we remove the binary variable *svi* from the data in the next analysis.

To evaluate the finite-sample performance of various variable screening methods in an ultrahigh dimensional setting, we augment the data set by adding some artificial predictors as follows. First, we use $X_j, j = 1, \ldots, 6$, to represent the six covariates: *lcavol*, *lweight*, *lbph*, *lcp*, *gleason*, and *pgg45*. For each covariate $X_j$, we generate 20 extra predictors $\{0.15X_j + c_0W_k, k = 1, \ldots, 20\}$ which are correlated with original variables, where $W_k$s are i.i.d. $N(0, 1)$ and we let $c_0$ vary from 0.5 to 2.5 to see the robustness of various methods. It can be seen that these 120 simulated predictors are spuriously associated with the response variable, resulting in various screening approaches applied to such a data set more difficult. Second, for each $X_j$, we further generate 179 independent auxiliary predictors by bootstrap, which ensures that the resulting auxiliary predictors share the same data structure as the original predictors. Hence, we have a total of 1200 predictors and one exposure variable for variable screening.

With the above augmented data, we randomly split the data into a training set with size 60 and a test set with size 37. For the training set, we apply various screening methods to select the top $d_n = 30$ predictors and compute the selection rate for each original covariate $X_j$, $j = 1, \ldots, 6$. Afterwards, a group SCAD regularization by Fan, Ma and Dai (2014) is applied to the $d_n$ selected predictors to further kill the irrelevant predictors. Thus, we obtain a refined subset of predictors by such a two-stage variable selection procedure, and then fit a varying-coefficient model with an intercept based on the training set and use this fitted model to compute prediction errors using the test set, where the mean square prediction error (PE) defined in Section 5.1 is obtained. We report the selection rate and the average of PEs through randomly generating the above data augmentation 500 times. The results with different $c_0$ are displayed in Figure 2.

It can be observed from Figure 2 that first of all, there is no one method that always outperforms others in terms of selection rate. While, almost all methods fail to identify the variable *lbph* due to very low selection rates. For other predictors, it seems that CPC-SIS(0.5, 0.5) works best because it can select two variables *pgg45* and *lweight* with highest probabilities, however, when picking up the variable *lcp*, CPC-SIS(0.5, 0.5) has only a half of chance, much lower than our RPC-SIS(L2) and RPC-SIS(L1). Even though NIS is able to select *lcp* and *lcavol* with high probabilities, it identifies *gleason*, *lweight* and *pgg45* with less than a quarter of chance. Moreover, CQC-SIS(0.5) seems not suitable to analyze this data because the probability of selecting all predictors is low. Overall, our proposed RPC-SIS(L1) has most satisfactory performance in terms of selection rates, and our RPC-SIS(L2) also behaves comparable to RPC-SIS(L1), and is sequently followed by CPC-SIS(0.5). In addition, in terms of prediction performance, our RPC-SIS(L1) is best since it is significantly superior to other methods although it is slightly better than RPC-SIS(L2). In particular, CQC-SIS(0.5) has prediction error larger than 10 over all $c_0$s, rendering the PE curve for CQC-SIS(0.5) undisplayed.

As a final touch of the analysis, we fit a varying-coefficient model with a constant for six original covariates and an index variable *age* over the entire data. The estimated varying-coefficient functions with 95% pointwise confidence region are displayed in Figure 3. It can be seen that all the fitted curves have an explicitly nonlinear trend in age, where the estimated coefficient functions for *lcavol*, *lpbh*, and *lcp* are positive.

## 6  Extensions and Concluding Remarks

In Section 2.2, we proposed a general variable screening approach based on a robust partial correlation, RPC-SIS. Two possible extensions can be as follows. First, we note that the exposure variable $Z$ considered in the RPC is univariate. It is convenient to extend it to the situation in which multiple exposures $Z = (Z_1, \ldots, Z_q)^T$ are encountered. For example, when some of $Z_j$'s, $\{Z_j, j \in I_d\}$, are discrete and the others, $\{Z_j, j \in I_c\}$, are continuous with two index sets $I_d$ and $I_c$, where $\{1, \ldots, q\} = I_d \cup I_c$, we can model $m_x(Z)$ (and $m_y(Z)$ similarly) in the definition of RPC as $\beta_0 + \sum_{j \in I_d} \beta_j Z_j + \sum_{l \in I_c} m_l(Z_l)$, where $\beta_j$s are unknown parameters and $E(m_l(Z_l)) = 0, l \in I_c$. This means that we can adjust the effect of $Z$ $X$ and $Y$ by fitting two partially linear additive models. When all of $Z_j$'s are discrete variables, we can remove the effect of $Z$ on $X$ and $Y$ by fitting two parametric models. When all of $Z_j$'s are continuous variables, we can remove the effect of $Z$ $X$ and $Y$ by fitting two nonparametric additive models. Second, we need to choose a loss function when implementing RPC-SIS. Different choices for loss function would result in different performance of RPC-SIS. Other robust loss, such as Huber's loss, or a more general loss can be studied further. For above two aspects, it also deserves to investigate the performance of RPC-SIS in both application and theory in the future. In addition, extending RPC-SIS to handling the censoring data in survival analysis would be another topic of interest.

Finally, we conclude this paper. In this paper, we proposed a robust correlation-based screening approach, RC-SIS, and then presented a robust partial correlation-based screening approach, RPC-SIS, as an important extension. Both RC-SIS and RPC-SIS can be used to analyze the ultrahigh dimensional data especially when both response variable and predictors are asymmetric, heavy-tailed and heteroscedastic. Besides, RPC-SIS can remove the confounding effect of an exposure or a conditional variable on both the response and covariates in a more flexible way. It is worth noting that RC-SIS not only can handle the continuous response, but also works for the discrete response. Theoretically, we established sure screening properties for two variable screening procedures under some mild technical conditions. Extensive simulation studies and real-world applications further confirmed the effectiveness and superiority of our proposals.

**Supplementary Materials** The supplementary materials contain some additional simulation results, a real data analysis, and all the proofs of theorems stated in the manuscript.

| $(n, p_n, \rho_0)$ | Method | $\mathcal{R}_1$ | $\mathcal{R}_2$ | $\mathcal{R}_3$ | $\mathcal{R}_4$ | $\mathcal{R}_5$ | MMS (RSD) | $\mathcal{P}$ |
|---|---|---|---|---|---|---|---|---|
| $(100, 1000, 0.4)$ | SIS | 227 | 221 | 198 | 230 | 266 | 782 (270) | 0.00 |
| | SIRS | 389 | 349 | 384 | 400 | 406 | 679 (160) | 0.00 |
| | DC-SIS | 56 | 32 | 65 | 68 | 227 | 791 (285) | 0.01 |
| | Kendall's $\tau$-SIS | 3 | 1 | 3 | 4 | 7 | 12 (25) | 0.63 |
| | CC-SIS(0.5,0.5) | 3 | 2 | 3 | 7 | 21 | 74 (104) | 0.22 |
| | QC-SIS(0.5) | 99 | 66 | 54 | 135 | 215 | 769 (240) | 0.02 |
| | RC-SIS | 3 | 1 | 3 | 4 | 6 | 10 (24) | 0.65 |
| $(100, 1000, 0.8)$ | SIS | 97 | 95 | 56 | 56 | 100 | 624 (371) | 0.03 |
| | SIRS | 319 | 306 | 286 | 325 | 348 | 576 (185) | 0.00 |
| | DC-SIS | 11 | 6 | 5 | 5 | 17 | 635 (416) | 0.10 |
| | Kendall's $\tau$-SIS | 3 | 2 | 2 | 3 | 5 | 5 (0) | 1.00 |
| | CC-SIS(0.5,0.5) | 3 | 1 | 2 | 4 | 5 | 5 (1) | 0.94 |
| | QC-SIS(0.5) | 30 | 14 | 10 | 18 | 32 | 696 (354) | 0.03 |
| | RC-SIS | 4 | 1 | 2 | 3 | 5 | 5 (0) | 1.00 |
| $(100, 2000, 0.4)$ | SIS | 380 | 357 | 432 | 573 | 597 | 1521 (553) | 0.01 |
| | SIRS | 703 | 733 | 791 | 778 | 842 | 1355 (312) | 0.00 |
| | DC-SIS | 249 | 58 | 69 | 275 | 276 | 1577 (504) | 0.01 |
| | Kendall's $\tau$-SIS | 3 | 2 | 3 | 4 | 10 | 17 (44) | 0.55 |
| | CC-SIS(0.5,0.5) | 4 | 2 | 3 | 10 | 34 | 129 (217) | 0.18 |
| | QC-SIS(0.5) | 236 | 113 | 144 | 275 | 282 | 1430 (553) | 0.01 |
| | RC-SIS | 3 | 1 | 3 | 4 | 9 | 16 (34) | 0.55 |
| $(200, 2000, 0.4)$ | SIS | 529 | 439 | 489 | 491 | 738 | 1595 (493) | 0.00 |
| | SIRS | 756 | 708 | 717 | 758 | 789 | 1316 (350) | 0.00 |
| | DC-SIS | 89 | 23 | 49 | 120 | 328 | 1498 (546) | 0.01 |
| | Kendall's $\tau$-SIS | 3 | 1 | 3 | 4 | 5 | 5 (1) | 0.95 |
| | CC-SIS(0.5,0.5) | 3 | 1 | 3 | 4 | 8 | 12 (18) | 0.79 |
| | QC-SIS(0.5) | 191 | 40 | 173 | 357 | 486 | 1470 (527) | 0.00 |
| | RC-SIS | 3 | 1 | 3 | 4 | 5 | 5 (1) | 0.96 |

Table 1: Screening results for Example 1, where $\mathcal{R}_j$ represents the rank for the $j$th active covariate, MMS represents the minimum number of the selected predictors that contain all the active predictors, its robust standard deviations (RSD) are given in the parenthesis, and $\mathcal{P}$ stands for the proportion of all the active predictors being selected with the screening threshold $d_n = \lfloor n/\log n \rfloor$.

| Model | $(n, \rho_0)$ | Method | $w_0 = 0.8$ | | $w_0 = 1$ | |
|---|---|---|---|---|---|---|
| | | | MMS (RSD) | $\mathcal{P}$ | MMS (RSD) | $\mathcal{P}$ |
| (b1) | $(100, 0.5)$ | SIS | 686 (309) | 0.01 | 95 (437) | 0.38 |
| | | SIRS | 586 (176) | 0.00 | 71 (81) | 0.20 |
| | | DC-SIS | 584 (451) | 0.07 | 3 (1) | 0.87 |
| | | Kendall's $\tau$-SIS | 4 (5) | 0.86 | 3 (0) | 1.00 |
| | | CC-SIS(0.5,0.5) | 16 (41) | 0.54 | 4 (6) | 0.87 |
| | | QC-SIS(0.5) | 672 (422) | 0.06 | 3 (1) | 0.96 |
| | | RC-SIS | 4 (5) | 0.88 | 3 (0) | 1.00 |
| (b2) | $(200, 0.8)$ | SIS | 781 (219) | 0.00 | 406 (555) | 0.31 |
| | | SIRS | 621 (214) | 0.00 | 30 (276) | 0.54 |
| | | DC-SIS | 734 (370) | 0.07 | 6 (205) | 0.60 |
| | | Kendall's $\tau$-SIS | 5 (92) | 0.69 | 5 (104) | 0.71 |
| | | CC-SIS(0.5,0.5) | 7 (163) | 0.63 | 6 (145) | 0.67 |
| | | QC-SIS(0.5) | 729 (344) | 0.04 | 5 (124) | 0.68 |
| | | RC-SIS | 5 (88) | 0.69 | 5 (97) | 0.69 |
| (b3) | $(200, 0.8)$ | SIS | 813 (205) | 0.00 | 736 (341) | 0.06 |
| | | SIRS | 800 (191) | 0.00 | 596 (471) | 0.07 |
| | | DC-SIS | 777 (250) | 0.01 | 37 (146) | 0.51 |
| | | Kendall's $\tau$-SIS | 377 (498) | 0.16 | 204 (533) | 0.31 |
| | | CC-SIS(0.5,0.5) | 454 (455) | 0.15 | 221 (472) | 0.29 |
| | | QC-SIS(0.5) | 787 (194) | 0.00 | 313 (566) | 0.26 |
| | | RC-SIS | 51 (191) | 0.45 | 8 (40) | 0.69 |
| (b4) | $(200, 0.5)$ | SIS | 745 (275) | 0.00 | 492 (424) | 0.09 |
| | | SIRS | 738 (180) | 0.00 | 374 (434) | 0.12 |
| | | DC-SIS | 801 (219) | 0.00 | 256 (391) | 0.19 |
| | | Kendall's $\tau$-SIS | 239 (438) | 0.32 | 168 (465) | 0.39 |
| | | CC-SIS(0.5,0.5) | 226 (322) | 0.20 | 216 (465) | 0.30 |
| | | QC-SIS(0.5) | 752 (223) | 0.01 | 204 (471) | 0.31 |
| | | RC-SIS | 81 (277) | 0.40 | 19 (152) | 0.58 |

Table 2: Screening results for Example 2, where MMS represents the minimum number of the selected predictors that contain all the active predictors, its robust standard deviations (RSD) are given in the parenthesis, and $\mathcal{P}$ stands for the proportion of all the active predictors being selected with the screening threshold $d_n = \lfloor n/\log n \rfloor$.

|  |  | \multicolumn{6}{c}{$n=200$} | \multicolumn{6}{c}{$n=400$} |
| $\varepsilon$ | Method | $\mathcal{R}_1$ | $\mathcal{R}_2$ | $\mathcal{R}_3$ | $\mathcal{R}_4$ | MMS (RSD) | $\mathcal{P}$ | $\mathcal{R}_1$ | $\mathcal{R}_2$ | $\mathcal{R}_3$ | $\mathcal{R}_4$ | MMS (RSD) | $\mathcal{P}$ |
|---|---|---|---|---|---|---|---|---|---|---|---|---|---|
| Normal | SIS | 9 | 1 | 19 | 140 | 208(460) | 0.22 | 4 | 1 | 3 | 32 | 51(187) | 0.54 |
|  | SIRS | 824 | 997 | 45 | 55 | 997(11) | 0.00 | 893 | 1000 | 8 | 12 | 1000(0) | 0.00 |
|  | DC-SIS | 35 | 1 | 3 | 2 | 55(117) | 0.45 | 8 | 1 | 3 | 2 | 9(25) | 0.81 |
|  | Kendall's $\tau$-SIS | 24 | 1 | 3 | 5 | 56(124) | 0.42 | 5 | 1 | 2 | 3 | 6(17) | 0.85 |
|  | CC-SIS(0.5,0.5) | 126 | 1 | 3 | 4 | 187(285) | 0.18 | 62 | 1 | 2 | 3 | 69(211) | 0.50 |
|  | QC-SIS(0.5) | 93 | 1 | 4 | 4 | 192(298) | 0.22 | 39 | 1 | 3 | 2 | 53(157) | 0.55 |
|  | RC-SIS | 22 | 1 | 3 | 2 | 34(85) | 0.51 | 5 | 1 | 3 | 2 | 5(12) | 0.89 |
| Student | SIS | 16 | 1 | 14 | 120 | 256(521) | 0.23 | 4 | 1 | 4 | 81 | 92(370) | 0.47 |
|  | SIRS | 802 | 997 | 48 | 67 | 997(9) | 0.00 | 889 | 1000 | 6 | 12 | 1000(0) | 0.00 |
|  | DC-SIS | 46 | 1 | 3 | 2 | 55(149) | 0.42 | 8 | 1 | 3 | 2 | 8(15) | 0.87 |
|  | Kendall's $\tau$-SIS | 27 | 1 | 2 | 4 | 70(176) | 0.36 | 5 | 1 | 2 | 3 | 7(16) | 0.87 |
|  | CC-SIS(0.5,0.5) | 162 | 1 | 2 | 5 | 206(305) | 0.15 | 56 | 1 | 2 | 3 | 61(156) | 0.52 |
|  | QC-SIS(0.5) | 158 | 1 | 3 | 3 | 265(384) | 0.17 | 31 | 1 | 3 | 3 | 37(100) | 0.61 |
|  | RC-SIS | 26 | 1 | 3 | 2 | 38(82) | 0.50 | 6 | 1 | 3 | 2 | 7(11) | 0.91 |
| Cauchy | SIS | 78 | 1 | 210 | 427 | 713(442) | 0.05 | 30 | 1 | 71 | 294 | 532(515) | 0.18 |
|  | SIRS | 744 | 997 | 44 | 62 | 997(10) | 0.00 | 886 | 1000 | 9 | 19 | 1000(1) | 0.00 |
|  | DC-SIS | 44 | 1 | 3 | 2 | 73(153) | 0.41 | 9 | 1 | 3 | 2 | 9(34) | 0.81 |
|  | Kendall's $\tau$-SIS | 27 | 1 | 3 | 6 | 87(220) | 0.32 | 5 | 1 | 2 | 3 | 7(24) | 0.83 |
|  | CC-SIS(0.5,0.5) | 119 | 1 | 3 | 6 | 210(293) | 0.15 | 70 | 1 | 2 | 3 | 79(146) | 0.45 |
|  | QC-SIS(0.5) | 98 | 1 | 4 | 4 | 184(340) | 0.18 | 30 | 1 | 3 | 3 | 44(137) | 0.57 |
|  | RC-SIS | 26 | 1 | 3 | 2 | 40(85) | 0.48 | 6 | 1 | 3 | 2 | 6(10) | 0.90 |
| Mixed Normal | SIS | 18 | 1 | 28 | 294 | 392(485) | 0.15 | 3 | 1 | 5 | 130 | 166(415) | 0.38 |
|  | SIRS | 798 | 997 | 38 | 95 | 998(11) | 0.00 | 890 | 1000 | 7 | 27 | 1000(1) | 0.00 |
|  | DC-SIS | 58 | 1 | 3 | 2 | 72(153) | 0.39 | 6 | 1 | 3 | 2 | 6(13) | 0.87 |
|  | Kendall's $\tau$-SIS | 30 | 1 | 3 | 8 | 84(223) | 0.35 | 5 | 1 | 2 | 3 | 7(19) | 0.83 |
|  | CC-SIS(0.5,0.5) | 157 | 1 | 2 | 9 | 186(292) | 0.13 | 53 | 1 | 2 | 3 | 86(143) | 0.47 |
|  | QC-SIS(0.5) | 108 | 1 | 5 | 6 | 189(321) | 0.16 | 26 | 1 | 2 | 3 | 35(170) | 0.55 |
|  | RC-SIS | 37 | 1 | 4 | 2 | 67(125) | 0.41 | 6 | 1 | 3 | 2 | 6(10) | 0.89 |

Table 3: Screening results for Example 3, where $\mathcal{R}_j$ represents the rank for the $j$th active covariate, MMS represents the minimum number of the selected predictors that contain all the active predictors, its robust standard deviations (RSD) are given in the parenthesis, and $\mathcal{P}$ stands for the proportion of all the active predictors being selected with the screening threshold $d_n = \lfloor n/\log n \rfloor$.

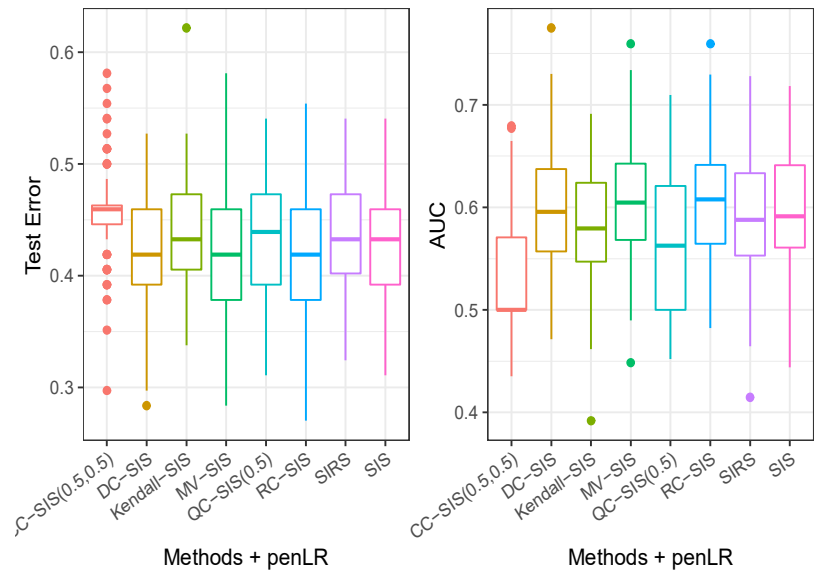

Figure 1: Results of test errors and AUCs for various two-stage methods for the lung cancer data.

| $\varepsilon$ | $R^2$ | Method | $\mathcal{R}_1$ | $\mathcal{R}_2$ | $\mathcal{R}_3$ | MMS (RSD) | $\mathcal{P}$ |
|---|---|---|---|---|---|---|---|
| $\frac{1}{3}Cauchy(0,1)$ | 0.05 | QPC-SIS(0.5) | 1 | 2 | 3 | 3(1) | 0.96 |
| | | CPC-SIS(0.5,0.5) | 1 | 2 | 3 | 3(1) | 0.93 |
| | | CQC-SIS(0.5) | 2 | 3 | 8 | 8(10) | 0.85 |
| | | NIS | 150 | 127 | 213 | 269(233) | 0.09 |
| | | RPC-SIS(L2) | 1 | 2 | 3 | 3(0) | 0.99 |
| | | RPC-SIS(L1) | 1 | 2 | 3 | 3(0) | 0.99 |
| | 0.1 | QPC-SIS(0.5) | 1 | 2 | 3 | 3(0) | 0.99 |
| | | CPC-SIS(0.5,0.5) | 1 | 2 | 3 | 3(1) | 0.96 |
| | | CQC-SIS(0.5) | 2 | 3 | 7 | 7(6) | 0.89 |
| | | NIS | 46 | 51 | 103 | 121(157) | 0.20 |
| | | RPC-SIS(L2) | 1 | 2 | 3 | 3(0) | 0.99 |
| | | RPC-SIS(L1) | 1 | 2 | 3 | 3(0) | 1.00 |
| | 0.3 | QPC-SIS(0.5) | 1 | 2 | 3 | 3(0) | 1.00 |
| | | CPC-SIS(0.5,0.5) | 1 | 2 | 3 | 3(0) | 1.00 |
| | | CQC-SIS(0.5) | 2 | 3 | 6 | 6(4) | 0.97 |
| | | NIS | 2 | 2 | 4 | 4(7) | 0.91 |
| | | RPC-SIS(L2) | 1 | 2 | 3 | 3(0) | 1.00 |
| | | RPC-SIS(L1) | 1 | 2 | 3 | 3(0) | 1.00 |
| student's $t(3)$ | 0.05 | QPC-SIS(0.5) | 22 | 31 | 122 | 136(265) | 0.28 |
| | | CPC-SIS(0.5,0.5) | 67 | 92 | 131 | 289(368) | 0.14 |
| | | CQC-SIS(0.5) | 110 | 121 | 250 | 337(361) | 0.08 |
| | | NIS | 118 | 163 | 254 | 340(308) | 0.07 |
| | | RPC-SIS(L2) | 9 | 17 | 58 | 92(192) | 0.35 |
| | | RPC-SIS(L1) | 8 | 16 | 58 | 96(196) | 0.35 |
| | 0.1 | QPC-SIS(0.5) | 3 | 6 | 21 | 25(77) | 0.60 |
| | | CPC-SIS(0.5,0.5) | 10 | 15 | 49 | 89(219) | 0.32 |
| | | CQC-SIS(0.5) | 18 | 30 | 77 | 106(187) | 0.28 |
| | | NIS | 16 | 19 | 72 | 91(144) | 0.31 |
| | | RPC-SIS(L2) | 2 | 2 | 7 | 8(20) | 0.78 |
| | | RPC-SIS(L1) | 2 | 2 | 7 | 8(21) | 0.77 |
| | 0.3 | QPC-SIS(0.5) | 1 | 2 | 3 | 3(6) | 0.91 |
| | | CPC-SIS(0.5,0.5) | 1 | 2 | 5 | 7(17) | 0.82 |
| | | CQC-SIS(0.5) | 5 | 6 | 20 | 20(49) | 0.61 |
| | | NIS | 2 | 2 | 3 | 3(2) | 0.93 |
| | | RPC-SIS(L2) | 1 | 2 | 3 | 3(0) | 0.99 |
| | | RPC-SIS(L1) | 1 | 2 | 3 | 3(0) | 1.00 |

Table 4: Screening results for Example 4, where $\mathcal{R}_j$ represents the rank for the $j$th active covariate, MMS represents the minimum number of the selected predictors that contain all the active predictors, its robust standard deviations (RSD) are given in the parenthesis, and $\mathcal{P}$ stands for the proportion of all the active predictors being selected with the screening threshold $d_n = \lfloor n/\log n \rfloor$.

|  |  |  | $w_0 = 0.5$ | | $w_0 = 0.8$ | |
| --- | --- | --- | --- | --- | --- | --- |
| $\varepsilon$ | Model | Method | MMS (RSD) | $\mathcal{P}$ | MMS (RSD) | $\mathcal{P}$ |
| $\frac{1}{3}Cauchy(0,1)$ | (d1) | QPC-SIS(0.5) | 191(333) | 0.19 | 6(10) | 0.85 |
|  |  | CPC-SIS(0.5,0.5) | 39(79) | 0.50 | 3(1) | 0.97 |
|  |  | CQC-SIS(0.5) | 216(254) | 0.05 | 15(9) | 0.86 |
|  |  | NIS | 296(365) | 0.17 | 23(155) | 0.54 |
|  |  | RPC-SIS(L2) | 6(15) | 0.81 | 3(0) | 0.98 |
|  |  | RPC-SIS(L1) | 5(10) | 0.86 | 3(0) | 1.00 |
|  | (d2) | QPC-SIS(0.5) | 189(327) | 0.22 | 5(11) | 0.81 |
|  |  | CPC-SIS(0.5,0.5) | 32(97) | 0.55 | 3(2) | 0.95 |
|  |  | CQC-SIS(0.5) | 199(256) | 0.06 | 15(11) | 0.85 |
|  |  | NIS | 290(347) | 0.14 | 10(120) | 0.60 |
|  |  | RPC-SIS(L2) | 10(24) | 0.75 | 3(0) | 1.00 |
|  |  | RPC-SIS(L1) | 7(16) | 0.80 | 3(0) | 1.00 |
|  | (d3) | QPC-SIS(0.5) | 476(396) | 0.01 | 67(126) | 0.37 |
|  |  | CPC-SIS(0.5,0.5) | 334(302) | 0.02 | 48(81) | 0.43 |
|  |  | CQC-SIS(0.5) | 561(304) | 0.00 | 96(151) | 0.16 |
|  |  | NIS | 451(313) | 0.02 | 76(257) | 0.35 |
|  |  | RPC-SIS(L2) | 142(175) | 0.22 | 15(14) | 0.83 |
|  |  | RPC-SIS(L1) | 119(168) | 0.24 | 15(11) | 0.85 |
| Student's $t(3)$ | (d1) | QPC-SIS(0.5) | 206(295) | 0.17 | 5(10) | 0.85 |
|  |  | CPC-SIS(0.5,0.5) | 40(77) | 0.48 | 3(2) | 0.95 |
|  |  | CQC-SIS(0.5) | 197(249) | 0.06 | 16(7) | 0.87 |
|  |  | NIS | 45(124) | 0.48 | 3(1) | 0.99 |
|  |  | RPC-SIS(L2) | 5(8) | 0.88 | 3(0) | 1.00 |
|  |  | RPC-SIS(L1) | 4(9) | 0.89 | 3(0) | 1.00 |
|  | (d2) | QPC-SIS(0.5) | 258(307) | 0.17 | 4(8) | 0.86 |
|  |  | CPC-SIS(0.5,0.5) | 44(93) | 0.47 | 3(2) | 0.97 |
|  |  | CQC-SIS(0.5) | 209(248) | 0.07 | 16(11) | 0.83 |
|  |  | NIS | 58(169) | 0.40 | 3(1) | 0.98 |
|  |  | RPC-SIS(L2) | 7(19) | 0.80 | 3(0) | 1.00 |
|  |  | RPC-SIS(L1) | 8(16) | 0.80 | 3(0) | 1.00 |
|  | (d3) | QPC-SIS(0.5) | 505(372) | 0.03 | 50(88) | 0.41 |
|  |  | CPC-SIS(0.5,0.5) | 319(319) | 0.04 | 49(64) | 0.40 |
|  |  | CQC-SIS(0.5) | 591(320) | 0.01 | 90(115) | 0.17 |
|  |  | NIS | 309(329) | 0.06 | 17(28) | 0.71 |
|  |  | RPC-SIS(L2) | 99(157) | 0.23 | 15(12) | 0.86 |
|  |  | RPC-SIS(L1) | 99(157) | 0.23 | 15(12) | 0.86 |

Table 5: Screening results for Example 5, where MMS represents the minimum number of the selected predictors that contain all the active predictors, its robust standard deviations (RSD) are given in the parenthesis, and $\mathcal{P}$ stands for the proportion of all the active predictors being selected with the screening threshold $d_n = \lfloor n/\log n \rfloor$.

|  | $w_0 = 0.8$ | | | | | $w_0 = 1.0$ | | | | |
|---|---|---|---|---|---|---|---|---|---|---|
| Method | $\mathcal{R}_1$ | $\mathcal{R}_2$ | $\mathcal{R}_3$ | $\mathcal{R}_4$ | MMS(RSD) | $\mathcal{R}_1$ | $\mathcal{R}_2$ | $\mathcal{R}_3$ | $\mathcal{R}_4$ | MMS(RSD) |
| | | | | | Setting 1 $(n, \rho_0) = (200, 0.4)$ | | | | | |
| QPC-SIS(0.5) | 182 | 122 | 164 | 454 | 665(297) | 3 | 1 | 2 | 221 | 232(276) |
| CPC-SIS(0.5,0.5) | 20 | 3 | 11 | 222 | 266(261) | 6 | 2 | 3 | 198 | 211(278) |
| CQC-SIS(0.5) | 317 | 243 | 246 | 373 | 741(262) | 40 | 31 | 33 | 143 | 172(214) |
| NIS | 268 | 271 | 312 | 416 | 667(276) | 72 | 32 | 56 | 199 | 358(401) |
| RPC-SIS(L2) | 10 | 2 | 5 | 248 | 289(303) | 3 | 1 | 2 | 131 | 135(200) |
| RPC-SIS(L1) | 4 | 1 | 2 | 156 | 165(247) | 3 | 1 | 2 | 121 | 121(182) |
| | | | | | Setting 2 $(n, \rho_0) = (200, 0.8)$ | | | | | |
| QPC-SIS(0.5) | 429 | 382 | 456 | 474 | 803(180) | 55 | 10 | 29 | 398 | 452(294) |
| CPC-SIS(0.5,0.5) | 135 | 94 | 157 | 344 | 520(254) | 74 | 17 | 48 | 355 | 442(290) |
| CQC-SIS(0.5) | 430 | 400 | 472 | 517 | 825(178) | 216 | 168 | 209 | 436 | 668(357) |
| NIS | 383 | 416 | 437 | 484 | 803(168) | 132 | 72 | 129 | 296 | 526(344) |
| RPC-SIS(L2) | 154 | 114 | 137 | 377 | 645(330) | 5 | 1 | 2 | 203 | 230(284) |
| RPC-SIS(L1) | 60 | 17 | 48 | 318 | 426(302) | 5 | 1 | 2 | 181 | 181(271) |
| | | | | | Setting 3 $(n, \rho_0) = (400, 0.8)$ | | | | | |
| QPC-SIS(0.5) | 396 | 444 | 414 | 445 | 820(160) | 10 | 2 | 4 | 322 | 336(330) |
| CPC-SIS(0.5,0.5) | 87 | 40 | 58 | 269 | 441(291) | 24 | 5 | 14 | 289 | 325(297) |
| CQC-SIS(0.5) | 453 | 419 | 448 | 471 | 808(184) | 278 | 208 | 235 | 399 | 669(338) |
| NIS | 402 | 429 | 459 | 445 | 805(169) | 103 | 101 | 123 | 315 | 497(381) |
| RPC-SIS(L2) | 82 | 20 | 70 | 240 | 529(376) | 3 | 1 | 2 | 100 | 105(217) |
| RPC-SIS(L1) | 21 | 2 | 7 | 266 | 306(294) | 3 | 1 | 2 | 86 | 86(158) |

Table 6: Screening results for Example 6, where $\mathcal{R}_j$ represents the rank for the $j$th active covariate, MMS represents the minimum number of the selected predictors that contain all the active predictors and its robust standard deviations (RSD) are given in the parenthesis.

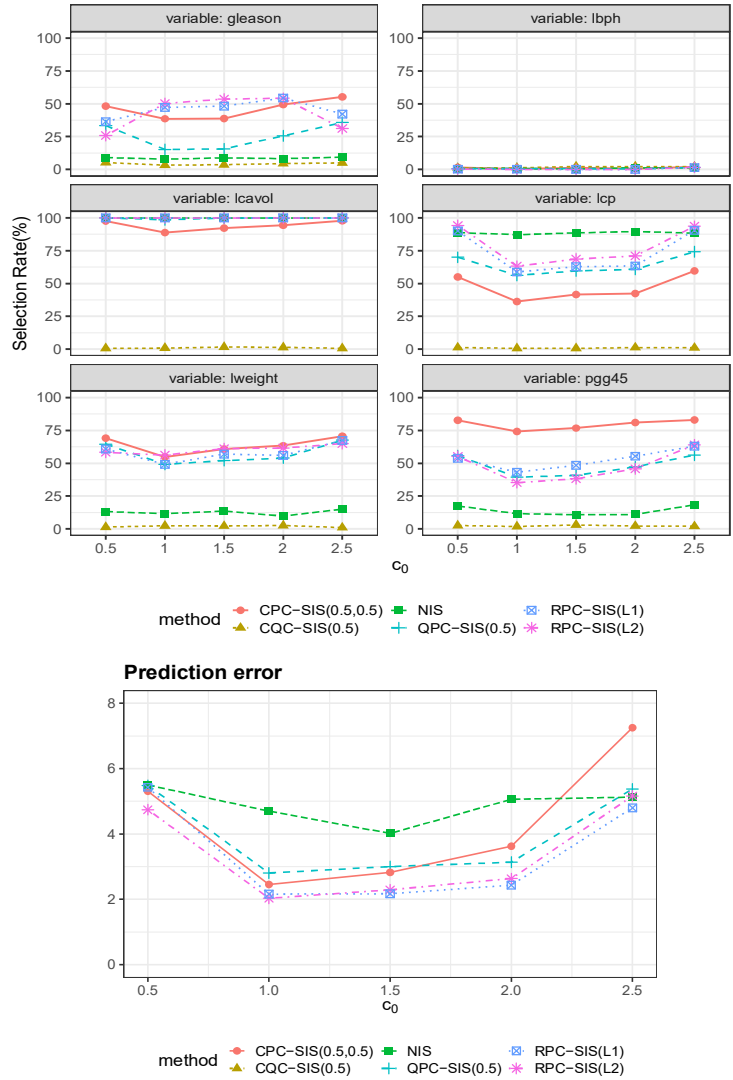

Figure 2: Selection rates and prediction errors by various methods across different values of $c_0$ over 500 replicates for specific antigen data, where the above figure displays the selection rates for each predictors with threshold parameter $d_n = 30$, and the below figure presents the prediction errors.

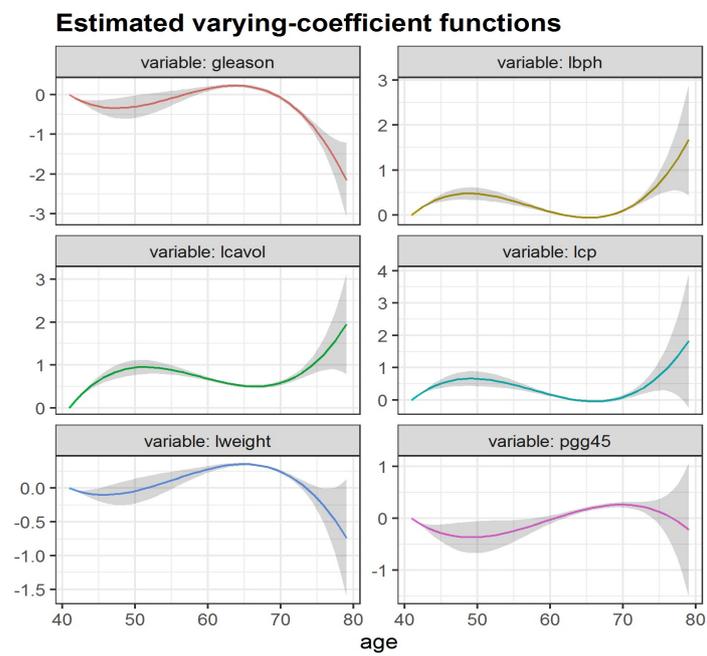

Figure 3: Estimated varying-coefficient functions with 95% confidence region for the original specific antigen data.

# Supplementary Materials to "A Robust Partial Correlation-based Screening Approach" *

Xiaochao Xia[1]

[1]*College of Mathematics and Statistics, Chongqing University, Chongqing, China*

In this supplementary materials, we provide some additional numerical results in the Appendix A as well as the detailed proofs of the main theorems that are stated in the main manuscript in the Appendix B.

## Appendix A: Additional Numerical Results and Discussion

### A1 Additional Numerical Results

**1) Additional simulations**

In order to investigate whether the proposed RC-SIS works for the discrete response data, we consider four additional examples as follows.

*Example S1* (Linear Logistic model) In this example, we consider DGP from a Logistic model with a *linear* structure. Let $\mathbf{X}_0$ be a latent covariate vector and $\mathbf{X}_0 \sim N(\mathbf{0}_{p_n}, \Sigma)$ with $\Sigma = (\sigma_{ij})_{1 \leq i,j \leq p_n}$ and $\sigma_{ij} = \rho_0^{|i-j|}$. We generate the response $Y$ from Bernoulli distribution with conditional probability $P(Y=1|\mathbf{X}) = \exp(\eta\{\mathbf{X}\})/(1+\exp(\eta\{\mathbf{X}\}))$, where $\eta\{\mathbf{X}\} = \eta\{\mathbf{X}\} = 2X_{01} + 1.5X_{02} + 2X_{0100} + 2X_{0400}$. We consider four cases for generating the observable covariates $\mathbf{X}$:

- (Case 1) $\mathbf{X} = \mathbf{X}_0$,

- (Case 2) $\mathbf{X} = 0.95\mathbf{X}_0 + 0.05\boldsymbol{\epsilon}$, where each element of $\boldsymbol{\epsilon}$ is independent and follows $t(3)$ distribution,

- (Case 3) same as that in Case 2 except that each element of $\boldsymbol{\epsilon}$ is distributed as $\frac{1}{3}Cauchy(0,1)$,

- (Case 4) same as that in Case 2 except that each element of $\boldsymbol{\epsilon}$ is distributed as $N(5,1)$.

---

For each case, the simulation results for various methods are reported in Table 1, from which, we can see that in Case 3, our RC-SIS performs best, and in other cases, the RC-SIS together with the DC-SIS works very well.

*Example S2* (Nonlinear Logistic model) In this example, we consider DGP from a Logistic model but with a *nonlinear* structure. Similar to Example S1, $\mathbf{X}_0 \sim N(\mathbf{0}_{p_n}, \Sigma)$ with $\Sigma = (\sigma_{ij})_{1 \leq i,j \leq p_n}$ and $\sigma_{ij} = \rho_0^{|i-j|}$. The response $Y$ follows from Bernoulli distribution with conditional probability $P(Y = 1|\mathbf{X}) = \exp(\eta\{\mathbf{X}\})/(1 + \exp(\eta\{\mathbf{X}\}))$, where $\eta\{\mathbf{X}\} = \eta\{\mathbf{X}\} = 2X_{01} + 2(X_{02}+0.5)^2 + 3\exp(-X_{0100}) + 6\sin(\pi X_{0400})$. For generating the observable covariates $\mathbf{X}$, we consider the same cases as in Example S1. For each case, the simulation results for various methods are reported in Table 2. We can observe from Table 2 that in most cases (Case 1, Case 2 and Case 4), the DC-SIS and the proposed RC-SIS perform very closely to each other but uniformly better than other five methods although it seems that the DC-SIS slightly outperforms the RC-SIS. Meanwhile, the RC-SIS performs obviously better than the DC-SIS in Case 3.

|  |  | $(n,\rho_0) = (200, 0.4)$ | | | | | | $(n,\rho_0) = (200, 0.8)$ | | | | | |
|---|---|---|---|---|---|---|---|---|---|---|---|---|---|
| **X** | Method | $\mathcal{R}_1$ | $\mathcal{R}_2$ | $\mathcal{R}_{100}$ | $\mathcal{R}_{400}$ | MMS (RSD) | $\mathcal{P}$ | $\mathcal{R}_1$ | $\mathcal{R}_2$ | $\mathcal{R}_{100}$ | $\mathcal{R}_{400}$ | MMS (RSD) | $\mathcal{P}$ |
| Case 1 | SIS | 1 | 2 | 3 | 3 | 4(0) | 1.00 | 1 | 2 | 6 | 5 | 8(3) | 0.99 |
|  | SIRS | 2 | 3 | 3 | 3 | 8(11) | 0.86 | 2 | 2 | 8 | 7 | 15(17) | 0.80 |
|  | DC-SIS | 1 | 2 | 3 | 3 | 4(0) | 1.00 | 1 | 2 | 6 | 5 | 8(3) | 0.99 |
|  | Kendall's $\tau$-SIS | 1000 | 999 | 998 | 998 | 1000(0) | 0.00 | 1000 | 999 | 995 | 996 | 1000(0) | 0.00 |
|  | CC-SIS(0.5,0.5) | 1 | 2 | 7 | 7 | 13(296) | 0.55 | 1 | 2 | 69 | 88 | 180(291) | 0.47 |
|  | QC-SIS(0.5) | 2 | 4 | 4 | 4 | 5(602) | 0.56 | 3 | 3 | 14 | 22 | 27(582) | 0.51 |
|  | RC-SIS | 1 | 2 | 3 | 3 | 4(0) | 1.00 | 1 | 2 | 6 | 5 | 8(4) | 0.99 |
| Case 2 | SIS | 1 | 2 | 3 | 3 | 4(1) | 1.00 | 1 | 2 | 6 | 6 | 9(4) | 0.97 |
|  | SIRS | 2 | 2 | 6 | 5 | 16(24) | 0.74 | 2 | 2 | 12 | 9 | 26(45) | 0.60 |
|  | DC-SIS | 1 | 2 | 3 | 3 | 4(0) | 1.00 | 1 | 2 | 6 | 6 | 9(4) | 0.98 |
|  | Kendall's $\tau$-SIS | 1000 | 999 | 998 | 998 | 1000(0) | 0.00 | 1000 | 999 | 995 | 995 | 1000(0) | 0.00 |
|  | CC-SIS(0.5,0.5) | 1 | 2 | 7 | 13 | 22(296) | 0.53 | 1 | 2 | 100 | 400 | 400(290) | 0.45 |
|  | QC-SIS(0.5) | 3 | 4 | 5 | 5 | 9(591) | 0.56 | 26 | 64 | 39 | 51 | 477(615) | 0.48 |
|  | RC-SIS | 1 | 2 | 3 | 3 | 4(0) | 1.00 | 1 | 2 | 6 | 6 | 9(4) | 0.98 |
| Case 3 | SIS | 1 | 2 | 3 | 3 | 5(34) | 0.74 | 1 | 2 | 6 | 6 | 11(25) | 0.73 |
|  | SIRS | 27 | 29 | 43 | 36 | 64(56) | 0.23 | 23 | 23 | 45 | 47 | 75(65) | 0.16 |
|  | DC-SIS | 1 | 2 | 3 | 3 | 4(9) | 0.81 | 1 | 2 | 6 | 5 | 10(22) | 0.77 |
|  | Kendall's $\tau$-SIS | 1000 | 999 | 998 | 998 | 1000(0) | 0.00 | 1000 | 999 | 996 | 995 | 1000(0) | 0.00 |
|  | CC-SIS(0.5,0.5) | 1 | 2 | 26 | 53 | 81(295) | 0.49 | 1 | 2 | 45 | 42 | 75(289) | 0.46 |
|  | QC-SIS(0.5) | 31 | 20 | 65 | 45 | 431(630) | 0.38 | 26 | 9 | 41 | 18 | 446(650) | 0.40 |
|  | RC-SIS | 1 | 2 | 3 | 3 | 4(0) | 1.00 | 1 | 2 | 5 | 6 | 8(3) | 0.97 |
| Case 4 | SIS | 2 | 2 | 3 | 3 | 4(3) | 0.86 | 1 | 2 | 6 | 7 | 10(8) | 0.82 |
|  | SIRS | 656 | 847 | 929 | 901 | 983(172) | 0.03 | 380 | 456 | 915 | 914 | 970(303) | 0.03 |
|  | DC-SIS | 1 | 2 | 3 | 3 | 4(1) | 1.00 | 1 | 2 | 6 | 6 | 8(4) | 0.97 |
|  | Kendall's $\tau$-SIS | 1000 | 999 | 998 | 998 | 1000(0) | 0.00 | 1000 | 999 | 996 | 995 | 1000(0) | 0.00 |
|  | CC-SIS(0.5,0.5) | 1 | 2 | 13 | 13 | 32(295) | 0.51 | 1 | 2 | 53 | 51 | 68(292) | 0.46 |
|  | QC-SIS(0.5) | 6 | 26 | 53 | 49 | 259(625) | 0.46 | 3 | 3 | 64 | 44 | 358(621) | 0.44 |
|  | RC-SIS | 1 | 2 | 3 | 3 | 4(0) | 1.00 | 1 | 2 | 5 | 6 | 8(4) | 0.99 |

Table 1: Screening results for Example S1, where $\mathcal{R}_j$ represents the rank for the $j$th active covariate, MMS represents the minimum number of the selected predictors that contain all the active predictors, its robust standard deviations (RSD) are given in the parenthesis, and $\mathcal{P}$ stands for the proportion of all the active predictors being selected with the screening threshold $d_n = \lfloor n/\log n \rfloor$.

| | | \multicolumn{6}{c|}{$(n, \rho_0) = (200, 0.4)$} | \multicolumn{6}{c}{$(n, \rho_0) = (200, 0.8)$} |
|---|---|---|---|---|---|---|---|---|---|---|---|---|---|
| **X** | Method | $\mathcal{R}_1$ | $\mathcal{R}_2$ | $\mathcal{R}_{100}$ | $\mathcal{R}_{400}$ | MMS (RSD) | $\mathcal{P}$ | $\mathcal{R}_1$ | $\mathcal{R}_2$ | $\mathcal{R}_{100}$ | $\mathcal{R}_{400}$ | MMS (RSD) | $\mathcal{P}$ |
| Case 1 | SIS | 17 | 16 | 1 | 519 | 547(361) | 0.03 | 8 | 8 | 1 | 426 | 439(387) | 0.03 |
| | SIRS | 30 | 32 | 1 | 524 | 547(298) | 0.02 | 15 | 12 | 2 | 392 | 410(361) | 0.03 |
| | DC-SIS | 19 | 6 | 1 | 13 | 34(67) | 0.53 | 7 | 5 | 1 | 16 | 24(23) | 0.71 |
| | Kendall's $\tau$-SIS | 992 | 997 | 1 | 743 | 999(4) | 0.00 | 998 | 999 | 1 | 757 | 1000(1) | 0.00 |
| | CC-SIS(0.5,0.5) | 28 | 12 | 1 | 6 | 84(148) | 0.35 | 8 | 7 | 2 | 10 | 37(58) | 0.51 |
| | QC-SIS(0.5) | 17 | 16 | 1 | 519 | 547(361) | 0.03 | 8 | 8 | 1 | 426 | 439(387) | 0.03 |
| | RC-SIS | 12 | 8 | 1 | 16 | 38(49) | 0.50 | 6 | 5 | 2 | 18 | 27(19) | 0.71 |
| Case 2 | SIS | 31 | 25 | 1 | 496 | 623(317) | 0.01 | 11 | 12 | 1 | 522 | 528(395) | 0.02 |
| | SIRS | 65 | 60 | 2 | 537 | 629(301) | 0.00 | 24 | 20 | 3 | 513 | 517(375) | 0.01 |
| | DC-SIS | 37 | 6 | 1 | 18 | 74(148) | 0.32 | 9 | 5 | 1 | 23 | 33(30) | 0.54 |
| | Kendall's $\tau$-SIS | 985 | 995 | 1 | 722 | 999(7) | 0.00 | 997 | 998 | 1 | 741 | 1000(2) | 0.00 |
| | CC-SIS(0.5,0.5) | 55 | 9 | 1 | 7 | 122(186) | 0.24 | 13 | 6 | 2 | 11 | 45(79) | 0.44 |
| | QC-SIS(0.5) | 31 | 25 | 1 | 496 | 623(317) | 0.01 | 11 | 12 | 1 | 522 | 528(395) | 0.02 |
| | RC-SIS | 25 | 9 | 1 | 22 | 59(90) | 0.34 | 8 | 5 | 2 | 26 | 36(33) | 0.54 |
| Case 3 | SIS | 31 | 22 | 1 | 483 | 552(368) | 0.03 | 12 | 13 | 2 | 484 | 576(370) | 0.01 |
| | SIRS | 69 | 75 | 21 | 468 | 555(340) | 0.00 | 42 | 44 | 21 | 478 | 485(336) | 0.02 |
| | DC-SIS | 34 | 7 | 1 | 19 | 137(373) | 0.27 | 10 | 6 | 1 | 23 | 55(255) | 0.37 |
| | Kendall's $\tau$-SIS | 988 | 996 | 1 | 738 | 999(6) | 0.00 | 998 | 998 | 1 | 750 | 1000(1) | 0.00 |
| | CC-SIS(0.5,0.5) | 49 | 12 | 1 | 9 | 109(231) | 0.25 | 9 | 7 | 2 | 10 | 48(76) | 0.46 |
| | QC-SIS(0.5) | 31 | 22 | 1 | 483 | 552(368) | 0.03 | 12 | 13 | 2 | 484 | 576(370) | 0.01 |
| | RC-SIS | 16 | 8 | 1 | 21 | 54(63) | 0.39 | 7 | 5 | 2 | 22 | 33(29) | 0.55 |
| Case 4 | SIS | 45 | 41 | 1 | 474 | 569(394) | 0.03 | 15 | 14 | 2 | 482 | 500(316) | 0.02 |
| | SIRS | 702 | 739 | 1 | 621 | 893(189) | 0.00 | 508 | 493 | 1 | 582 | 827(284) | 0.01 |
| | DC-SIS | 35 | 6 | 1 | 18 | 61(135) | 0.36 | 9 | 6 | 1 | 23 | 38(42) | 0.49 |
| | Kendall's $\tau$-SIS | 988 | 995 | 1 | 767 | 998(6) | 0.00 | 998 | 999 | 1 | 752 | 1000(1) | 0.00 |
| | CC-SIS(0.5,0.5) | 49 | 8 | 1 | 11 | 119(217) | 0.27 | 9 | 6 | 2 | 15 | 47(100) | 0.47 |
| | QC-SIS(0.5) | 45 | 41 | 1 | 474 | 569(394) | 0.03 | 15 | 14 | 2 | 482 | 500(316) | 0.02 |
| | RC-SIS | 17 | 10 | 1 | 23 | 58(89) | 0.36 | 8 | 6 | 2 | 26 | 42(34) | 0.45 |

Table 2: Screening results for Example S2, where $\mathcal{R}_j$ represents the rank for the $j$th active covariate, MMS represents the minimum number of the selected predictors that contain all the active predictors, its robust standard deviations (RSD) are given in the parenthesis, and $\mathcal{P}$ stands for the proportion of all the active predictors being selected with the screening threshold $d_n = \lfloor n/\log n \rfloor$.

*Example S3* (Linear Poisson model) In this example, we consider DGP from a Poisson model with a *linear* structure. Let $\mathbf{X}_0$ be a latent covariate vector and $\mathbf{X}_0 \sim N(\mathbf{0}_{p_n}, \Sigma)$ with $\Sigma = (\sigma_{ij})_{1 \leq i,j \leq p_n}$ and $\sigma_{ij} = \rho_0^{|i-j|}$. We generate the response $Y$ from Poisson distribution with conditional mean $E(Y|X) = \exp(\eta\{\mathbf{X}\})$, where $\eta\{\mathbf{X}\} = \eta\{\mathbf{X}\} = 2X_{01} + 1.5X_{02} + 2X_{0100} + 2X_{0400}$. We consider those four cases to generate the observable covariates $\mathbf{X}$ as in Example S1. Table 3 presents the corresponding simulation results for various methods and for each case. Clearly, it can be seen that by observing Table 3, the RC-SIS behaves most satisfactorily.

*Example S4* (Nonlinear Poisson model) In this example, we consider DGP from a Poisson model with a *nonlinear* structure. Different from Example S1, we let $\mathbf{X}_0 = (X_{0j})_{j=1}^{p_n}$ be generated as follows: $X_{0j} = (T_j + tU)/(1+t)$, where $T_j \sim_{i.i.d.} U(0,1)$ and $U \sim U(0,1)$ are independent. The value of $t$ is chosen such that $\text{corr}(X_{0j}, X_{0k}) = \rho_0$ for $j \neq k$. We generate the response $Y$ from Poisson distribution with conditional mean $E(Y|X) = \exp(\eta\{\mathbf{X}\})$, where $\eta\{\mathbf{X}\} = \eta\{\mathbf{X}\} = 1.5X_{01} + 0.5(X_{02} + 0.5)^2 + 1.5\exp(X_{0100}^2) + 1.5\sin(\pi X_{0400})$. We consider those four cases to generate the observable covariates $\mathbf{X}$ as in Example S1. For each case, the relevant simulation results for various methods are reported in Table 4. We can see that the RC-SIS performs best.

|  |  | $(n, \rho_0) = (200, 0.4)$ | | | | | | $(n, \rho_0) = (200, 0.8)$ | | | | | |
|---|---|---|---|---|---|---|---|---|---|---|---|---|---|
| **X** | Method | $\mathcal{R}_1$ | $\mathcal{R}_2$ | $\mathcal{R}_{100}$ | $\mathcal{R}_{400}$ | MMS (RSD) | $\mathcal{P}$ | $\mathcal{R}_1$ | $\mathcal{R}_2$ | $\mathcal{R}_{100}$ | $\mathcal{R}_{400}$ | MMS (RSD) | $\mathcal{P}$ |
| Case 1 | SIS | 5 | 11 | 28 | 30 | 169(220) | 0.15 | 3 | 3 | 43 | 67 | 157(249) | 0.18 |
|  | SIRS | 2 | 4 | 11 | 15 | 50(53) | 0.40 | 2 | 2 | 21 | 18 | 48(71) | 0.41 |
|  | DC-SIS | 2 | 3 | 8 | 7 | 30(79) | 0.53 | 2 | 2 | 14 | 24 | 52(135) | 0.45 |
|  | Kendall's $\tau$-SIS | 833 | 970 | 997 | 998 | 999(5) | 0.00 | 334 | 512 | 992 | 992 | 995(8) | 0.00 |
|  | CC-SIS(0.5,0.5) | 1 | 2 | 3 | 4 | 4(2) | 0.93 | 1 | 2 | 6 | 7 | 9(4) | 0.93 |
|  | QC-SIS(0.5) | 1 | 2 | 3 | 3 | 4(0) | 0.95 | 1 | 2 | 5 | 6 | 7(2) | 0.97 |
|  | RC-SIS | 1 | 2 | 3 | 3 | 4(0) | 1.00 | 1 | 2 | 5 | 5 | 6(1) | 1.00 |
| Case 2 | SIS | 8 | 15 | 37 | 42 | 179(245) | 0.14 | 5 | 5 | 83 | 76 | 220(286) | 0.10 |
|  | SIRS | 4 | 7 | 21 | 17 | 76(82) | 0.29 | 2 | 3 | 31 | 30 | 82(101) | 0.27 |
|  | DC-SIS | 2 | 4 | 11 | 9 | 60(113) | 0.44 | 2 | 2 | 35 | 25 | 101(169) | 0.28 |
|  | Kendall's $\tau$-SIS | 945 | 990 | 998 | 998 | 1000(1) | 0.00 | 521 | 714 | 995 | 993 | 997(6) | 0.00 |
|  | CC-SIS(0.5,0.5) | 1 | 2 | 4 | 3 | 4(2) | 0.93 | 1 | 2 | 6 | 7 | 11(13) | 0.82 |
|  | QC-SIS(0.5) | 1 | 2 | 3 | 3 | 4(0) | 0.95 | 1 | 2 | 6 | 6 | 8(5) | 0.89 |
|  | RC-SIS | 1 | 2 | 4 | 3 | 4(0) | 1.00 | 1 | 2 | 5 | 5 | 7(2) | 1.00 |
| Case 3 | SIS | 16 | 24 | 52 | 49 | 260(335) | 0.09 | 7 | 9 | 60 | 98 | 294(345) | 0.07 |
|  | SIRS | 50 | 57 | 83 | 78 | 139(94) | 0.01 | 43 | 46 | 95 | 91 | 150(111) | 0.01 |
|  | DC-SIS | 2 | 6 | 16 | 12 | 135(289) | 0.26 | 2 | 2 | 22 | 35 | 161(307) | 0.22 |
|  | Kendall's $\tau$-SIS | 925 | 989 | 998 | 998 | 1000(1) | 0.00 | 530 | 650 | 994 | 994 | 997(5) | 0.00 |
|  | CC-SIS(0.5,0.5) | 1 | 2 | 3 | 3 | 5(4) | 0.89 | 1 | 2 | 7 | 7 | 11(13) | 0.86 |
|  | QC-SIS(0.5) | 1 | 2 | 3 | 3 | 5(85) | 0.70 | 1 | 2 | 6 | 6 | 10(26) | 0.75 |
|  | RC-SIS | 1 | 2 | 3 | 3 | 4(0) | 1.00 | 1 | 2 | 6 | 5 | 7(2) | 1.00 |
| Case 4 | SIS | 18 | 50 | 100 | 71 | 321(287) | 0.03 | 12 | 12 | 138 | 112 | 312(338) | 0.02 |
|  | SIRS | 849 | 867 | 882 | 851 | 952(129) | 0.00 | 811 | 818 | 856 | 867 | 945(105) | 0.00 |
|  | DC-SIS | 2 | 4 | 16 | 8 | 45(96) | 0.45 | 2 | 2 | 23 | 23 | 83(156) | 0.35 |
|  | Kendall's $\tau$-SIS | 981 | 995 | 999 | 999 | 1000(0) | 0.00 | 709 | 825 | 996 | 995 | 998(4) | 0.00 |
|  | CC-SIS(0.5,0.5) | 1 | 2 | 3 | 4 | 5(4) | 0.90 | 1 | 2 | 7 | 7 | 12(8) | 0.86 |
|  | QC-SIS(0.5) | 1 | 2 | 3 | 4 | 4(8) | 0.81 | 2 | 2 | 8 | 6 | 11(21) | 0.76 |
|  | RC-SIS | 1 | 2 | 3 | 3 | 4(0) | 1.00 | 1 | 2 | 6 | 5 | 7(2) | 1.00 |

Table 3: Screening results for Example S3, where $\mathcal{R}_j$ represents the rank for the $j$th active covariate, MMS represents the minimum number of the selected predictors that contain all the active predictors, its robust standard deviations (RSD) are given in the parenthesis, and $\mathcal{P}$ stands for the proportion of all the active predictors being selected with the screening threshold $d_n = \lfloor n/\log n \rfloor$.

| **X** | Method | \multicolumn{6}{c}{$(n, \rho_0) = (200, 0.4)$} | \multicolumn{6}{c}{$(n, \rho_0) = (200, 0.8)$} |
| | | $\mathcal{R}_1$ | $\mathcal{R}_2$ | $\mathcal{R}_{100}$ | $\mathcal{R}_{400}$ | MMS (RSD) | $\mathcal{P}$ | $\mathcal{R}_1$ | $\mathcal{R}_2$ | $\mathcal{R}_{100}$ | $\mathcal{R}_{400}$ | MMS (RSD) | $\mathcal{P}$ |
|---|---|---|---|---|---|---|---|---|---|---|---|---|---|
| Case 1 | SIS | 2 | 11 | 1 | 977 | 977(59) | 0.00 | 5 | 18 | 1 | 999 | 999(6) | 0.00 |
| | SIRS | 931 | 862 | 989 | 290 | 995(14) | 0.00 | 805 | 738 | 921 | 279 | 969(65) | 0.00 |
| | DC-SIS | 2 | 4 | 1 | 723 | 732(296) | 0.00 | 3 | 11 | 1 | 931 | 931(130) | 0.00 |
| | Kendall's $\tau$-SIS | 2 | 4 | 1 | 483 | 493(424) | 0.07 | 2 | 4 | 1 | 313 | 313(459) | 0.15 |
| | CC-SIS(0.5,0.5) | 2 | 18 | 1 | 556 | 562(340) | 0.01 | 15 | 45 | 4 | 444 | 469(319) | 0.01 |
| | QC-SIS(0.5) | 2 | 8 | 1 | 795 | 799(295) | 0.01 | 7 | 32 | 1 | 695 | 707(362) | 0.00 |
| | RC-SIS | 2 | 4 | 1 | 10 | 15(27) | 0.72 | 2 | 5 | 1 | 8 | 14(30) | 0.72 |
| Case 2 | SIS | 110 | 213 | 14 | 746 | 811(228) | 0.00 | 369 | 397 | 172 | 659 | 834(191) | 0.00 |
| | SIRS | 697 | 646 | 833 | 424 | 941(78) | 0.00 | 583 | 491 | 720 | 414 | 881(128) | 0.00 |
| | DC-SIS | 25 | 130 | 2 | 585 | 679(318) | 0.00 | 364 | 386 | 190 | 612 | 801(174) | 0.00 |
| | Kendall's $\tau$-SIS | 2 | 5 | 1 | 466 | 466(403) | 0.06 | 2 | 21 | 1 | 430 | 454(374) | 0.05 |
| | CC-SIS(0.5,0.5) | 3 | 21 | 1 | 567 | 573(368) | 0.01 | 17 | 89 | 5 | 445 | 502(301) | 0.01 |
| | QC-SIS(0.5) | 77 | 234 | 48 | 585 | 762(253) | 0.00 | 461 | 419 | 318 | 542 | 825(170) | 0.00 |
| | RC-SIS | 2 | 4 | 1 | 42 | 59(111) | 0.41 | 2 | 15 | 1 | 61 | 91(159) | 0.26 |
| Case 3 | SIS | 160 | 220 | 18 | 685 | 823(174) | 0.00 | 458 | 428 | 380 | 589 | 837(148) | 0.00 |
| | SIRS | 720 | 619 | 819 | 384 | 906(68) | 0.00 | 546 | 591 | 677 | 417 | 887(131) | 0.00 |
| | DC-SIS | 102 | 179 | 2 | 575 | 781(207) | 0.00 | 458 | 421 | 419 | 538 | 848(135) | 0.00 |
| | Kendall's $\tau$-SIS | 2 | 4 | 1 | 536 | 536(416) | 0.03 | 2 | 18 | 1 | 361 | 376(344) | 0.05 |
| | CC-SIS(0.5,0.5) | 3 | 18 | 1 | 560 | 566(351) | 0.01 | 25 | 87 | 6 | 436 | 474(342) | 0.01 |
| | QC-SIS(0.5) | 164 | 250 | 25 | 613 | 800(221) | 0.00 | 467 | 457 | 459 | 492 | 836(143) | 0.00 |
| | RC-SIS | 2 | 4 | 1 | 61 | 76(124) | 0.38 | 2 | 14 | 1 | 46 | 75(119) | 0.34 |
| Case 4 | SIS | 46 | 98 | 2 | 736 | 758(264) | 0.00 | 285 | 326 | 52 | 795 | 853(180) | 0.00 |
| | SIRS | 807 | 679 | 912 | 475 | 963(66) | 0.00 | 589 | 542 | 759 | 409 | 899(120) | 0.00 |
| | DC-SIS | 4 | 24 | 1 | 593 | 595(358) | 0.01 | 185 | 255 | 34 | 713 | 770(248) | 0.00 |
| | Kendall's $\tau$-SIS | 2 | 3 | 1 | 404 | 404(427) | 0.10 | 2 | 5 | 1 | 344 | 350(455) | 0.12 |
| | CC-SIS(0.5,0.5) | 2 | 14 | 1 | 586 | 586(378) | 0.01 | 17 | 59 | 2 | 569 | 581(307) | 0.01 |
| | QC-SIS(0.5) | 31 | 116 | 12 | 499 | 605(358) | 0.01 | 375 | 357 | 207 | 583 | 777(193) | 0.00 |
| | RC-SIS | 2 | 4 | 1 | 6 | 11(25) | 0.75 | 2 | 7 | 1 | 4 | 13(28) | 0.72 |

Table 4: Screening results for Example S4, where $\mathcal{R}_j$ represents the rank for the $j$th active covariate, MMS represents the minimum number of the selected predictors that contain all the active predictors, its robust standard deviations (RSD) are given in the parenthesis, and $\mathcal{P}$ stands for the proportion of all the active predictors being selected with the screening threshold $d_n = \lfloor n/\log n \rfloor$.

**2) Application to Rats Data**

The rat data contain expression measurements of 31,099 gene probes and can be downloaded from ftp://ftp.ncbi.nlm.nih.gov/geo/series/GSE5nnn/GSE5680/matrix. This data set has been analyzed by several researchers. For example, Scheetz et al. (2006) analyzed this data for investigation of the gene regulation in the mammalian. Chiang et al. (2006) further studied this data and revealed that the expression of gene TRIM32 (probe 1389163 at) is an important factor to the cause of Bardet-Biedl syndrome, which was identified to have an vital impact on the human hereditary disease of the retina. Recently, Ma, Li and Tsai (2017) have also investigated this data for gene identification using a quantile partial correlation based variable screening method. Here, the interest of this study is to detect which gene probes have most important influence on the gene TRIM32. Thus, the gene TRIM32 is treated as the response $Y$. Following the approach of Ma, Li and Tsai (2017), we subsequently select 3000 genes with largest variances in expression values from the remaining genes. We use the standardized expression values of these 3000 genes as the covariates **X**. Thus, the sample size is $n = 120$ and the dimensionality is $p_n = 3000$.

We first employ those methods mentioned in the previous simulation study as well as the proposed RC-SIS to select the top $d_n = \lfloor n/\log n \rfloor = 25$ genes as the most relevant ones. For each method, we then carry out a variable selection based on the SCAD penalization of Fan and Li (2001) to further refine the screened genes. We refer to such a procedure as a two-stage approach. For instance, the approach "RC-SIS+SCAD" means a two-stage approach that performs RC-SIS in the first stage and implements a SCAD-penalized regularization using *cv.ncvreg* function in R package "*ncvreg*" in the second stage, where the tuning parameter involved in SCAD penalty is chosen by the BIC criterion of Wang, Li and Leng (2009) and the 10-fold cross-validation (CV), respectively. Other two-stage approaches are similarly referred to. Afterwards, we use the data based on the genes selected by each two-stage approach to assess the finite-sample performance. To this end, we consider 500 random partitions. For each partition, we randomly divide 120 observations into a training set and a testing set with a sample size ratio equal to 4 : 1. In the training set, we fit a specific regression model with the selected predictors by any two-stage procedure. In the testing set, we use the fitted regression model to compute the prediction error PE = $\frac{1}{n.test} \sum_{i=1}^{n.test}(y_i - \hat{y}_i)^2$, where $n.test$ denotes the size of testing sample and $\hat{y}_i$ is computed through fitting a nonparametric additive model using *gam* function in R package "*mgcv*". The corresponding results including the numbers of selected gene probes and the average of PEs and its standard error (s.e.) over 500 random partitions are reported in Table 5.

Looking at Table 5, we can observe that our proposed RC-SIS+SCAD has the most satisfactory performance since it uniformly outperforms other competitors in terms of PE and its standard errors. It is also interesting to see that when the tuning parameter in SCAD penalization is chosen by a BIC criterion, our RC-SIS+SCAD selects only two relevant genes (gene No. 2051 2132) and thus results in the simplest prediction model among all these two-stage methods. Meanwhile, when the tuning parameter in SCAD penalty is chosen by the 10-fold CV, our RC-SIS+SCAD still has the smallest prediction error, and simultaneously produces a model with only three predictors (gene No. 2051 2132 2647). We can further see that both the criteria BIC and 10-fold CV can identify two common genes (gene No. 2051 2132) corresponding to probes 1383110_at and 1389584_at, which are newly identified genes different from those in existing literature (Wang, Wu and Li (2012)). Last, we use these two identified genes to fit an additive model using R function *gam* and two fitted nonparametric functions are plotted in Figure 1, both of which display a significantly nonlinear relationship.

| Two-stage method | PE (s.e.) | Selected gene number |
|---|---|---|
| Tuning parameter in SCAD is chosen by BIC | | |
| SIS+SCAD | 0.0099 (0.0041) | 2132 2685 2061 |
| SIRS+SCAD | 0.0083 (0.0033) | 2647 2132 2051 1864 |
| DC-SIS+SCAD | 0.0083 (0.0033) | 2051 2647 2132 1864 |
| Kendall's $\tau$-SIS+SCAD | 0.0083 (0.0033) | 2051 2132 2647 1864 |
| CC-SIS(0.5,0.5)+SCAD | 0.0098 (0.0105) | 2051 2685 2416 2647 2717 2979 2417 |
| QC-SIS(0.5)+SCAD | 0.0087 (0.0034) | 2647 2051 |
| RC-SIS+SCAD | 0.0080 (0.0025) | 2051 2132 |
| Tuning parameter in SCAD is chosen by 10-fold CV | | |
| SIS+SCAD | 0.0085 (0.0046) | 2051 2132 2685 2952 2647 2061 2737 2665 2198 2654 1665 2210 |
| SIRS+SCAD | 0.0078 (0.0027) | 2647 2132 2051 1864 |
| DC-SIS+SCAD | 0.0083 (0.0029) | 2051 2647 2132 2685 1864 1752 |
| Kendall's $\tau$-SIS+SCAD | 0.0078 (0.0027) | 2051 2132 2647 1864 |
| CC-SIS(0.5,0.5)+SCAD | 0.0083 (0.0075) | 2051 2685 2379 2416 2647 2717 2979 2417 |
| QC-SIS(0.5)+SCAD | 0.0084 (0.0034) | 2647 2051 |
| RC-SIS+SCAD | 0.0077 (0.0027) | 2051 2132 2647 |

Table 5: Results of PE and the numbers of selected genes for rats dataset, where PE stands for prediction error computed through fitting a nonparametric additive model with smoothing spline and the values in the parenthesis represent standard errors (s.e.) of PE.

## A2 Discussion about comparison of RC-SIS and Kendall's $\tau$-SIS

We here provide some discussion about the comparison of our RC-SIS with Kendall's $\tau$-SIS. From our simulation results, one can observe that the finite-sample performance of our RC-SIS is comparable in some cases to that of Kendall's $\tau$-SIS. This is probably because the definitions of RC and Kendall's $\tau$ merely involve the distribution functions of random variable due to the use of indicator function. From the definitions, we know that RC is $\rho(y,x) = \frac{F_{Y,X}(y,x) - F_Y(y)F_X(x)}{\sqrt{F_Y(y)[1-F_Y(y)]F_X(x)[1-F_X(x)]}}$ and Kendall's $\tau$ correlation is $\tau = 12[P(X_1 \leq X_2, Y_1 \leq Y_2) - \frac{1}{4}]$, where $(X_1, Y_1), (X_2, Y_2)$ are independent copies of $(X, Y)$. On one hand, note that $E[F_{Y,X}(Y,X)] = P(X_1 \leq X_2, Y_1 \leq Y_2) = \frac{1}{12}\tau + \frac{1}{4}$. Then, by condition (D2),

$$\begin{aligned}
E\{\rho^2(Y,X)\} &\leq \frac{1}{K_1 K_2} E\{F_{Y,X}^2(Y,X) - 2F_Y(Y)F_X(X)F_{Y,X}(Y,X) + F_Y^2(Y)F_X^2(X)\} \\
&\leq \frac{1}{K_1 K_2} E\{F_{Y,X}(Y,X) - 2K_1 K_2 F_{Y,X}(Y,X) + F_Y^2(Y)F_X^2(X)\} \\
&= \frac{1}{K_1 K_2}(1 - 2K_1 K_2) E\{F_{Y,X}(Y,X)\} + \frac{1}{K_1 K_2} E\{F_Y^2(Y)F_X^2(X)\} \\
&= \frac{(1 - 2K_1 K_2)}{12 K_1 K_2}\tau + \frac{1 - 2K_1 K_2}{4 K_1 K_2} + \frac{1}{K_1 K_2} E\{F_Y^2(Y)F_X^2(X)\},
\end{aligned}$$

which means our proposed screener utility can be bounded by a constant times Kendall's $\tau$ correlation plus a constant. This provides a theoretical connection between RC and Kendall's $\tau$ to some extent. On the other hand, according to the definitions of RC and Kendall's $\tau$, the numerator in RC could play the same role as in Kendall's $\tau$. However, the terms in the denominator in RC may also have an impact on the magnitude of the screener utility, especially when the marginal distributions of $X$ and $Y$ are more heavy-tailed or outlying, because the denominator in RC involves the marginal probability distributions of random variables. Hence, our RC-SIS takes into consideration the information about both joint and marginal distributions of the response and each predictor, while Kendall's $\tau$-SIS only involves the joint distribution of the response and each predictor. For these reasons, our proposed RC-SIS may have better performance than Kendall's $\tau$-SIS.

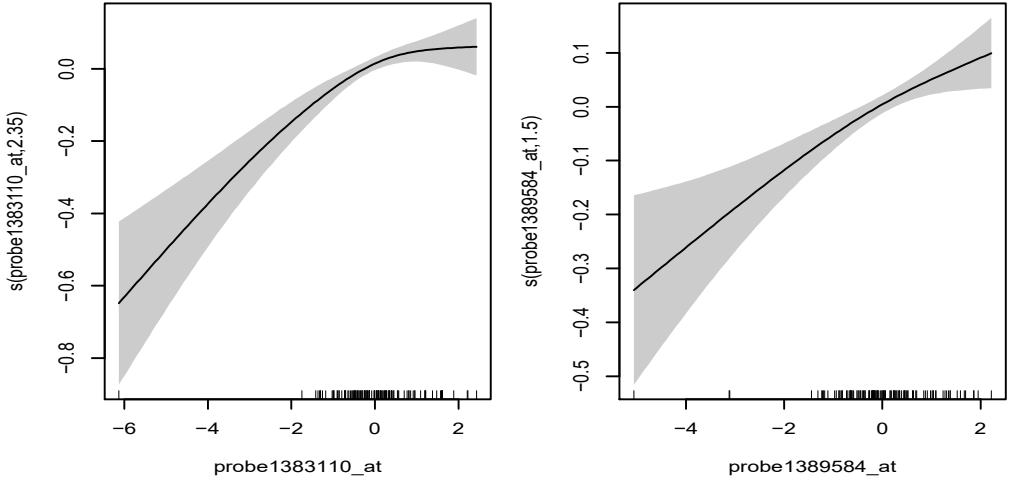

Figure 1: Estimated nonparametric functions by fitting an additive model on two selected gene probes for rats data.

# Appendix B: Proofs

## B1 Asymptotic Variance of $\widehat{\rho}(y, x)$

Denote $\xi_{1i}(y,x) = [I(X_i \leq x) - F_X(x)][I(Y_i \leq y) - F_Y(y)] - [F_{X,Y}(x,y) - F_X(x)F_Y(y)]$, $\xi_{2i}(x) = [I(X_i \leq x) - F_X(x)]^2 - [F_X(x) - F_X^2(x)]$ and $\xi_{3i}(y) = [I(Y_i \leq y) - F_Y(y)]^2 - [F_Y(y) - F_Y^2(y)]$. The asymptotic variance of $\widehat{\rho}(y,x)$ is $\Xi(x,y)/n$, where

$$
\begin{aligned}
\Xi(x,y) &= \{\theta_2^2(x)\theta_3^2(y)\}^{-1/2} \times \Big[ E\{\xi_{1i}(y,x)\xi_{1i}(y,x)\} + \frac{\theta_1^2(x,y)E\{\xi_{2i}^2(x)\}}{4\theta_2^2(x)} \\
&\quad + \frac{\theta_1^2(x,y)E\{\xi_{3i}^2(y)\}}{4\theta_3^2(y)} - \frac{\theta_1(x,y)E\{\xi_{1i}(y,x)\xi_{2i}(x)\}}{2\theta_2(x)} - \frac{\theta_1(x,y)E\{\xi_{1i}(y,x)\xi_{3i}(y)\}}{2\theta_3(y)} \\
&\quad - \frac{\theta_1(x,y)E\{\xi_{2i}(x)\xi_{1i}(y,x)\}}{2\theta_2(x)\theta_3(y)} + \frac{\theta_1^2(x,y)E\{\xi_{2i}(x)\xi_{3i}(y)\}}{4\theta_2(x)\theta_3(y)} \\
&\quad - \frac{\theta_1(x,y)E\{\xi_{3i}(y)\xi_{1i}(y,x)\}}{2\theta_3(y)} + \frac{\theta_1^2(x,y)E\{\xi_{3i}(y)\xi_{2i}(x)\}}{4\theta_3(y)\theta_2(x)} \Big],
\end{aligned}
$$

where $\theta_1(x,y) = F_{Y,X}(y,x) - F_Y(y)F_X(x)$, $\theta_2(x) = F_X(x) - F_X^2(x)$ and $\theta_3(y) = F_Y(y) - F_Y^2(y)$. A consistent estimate $\widehat{\Xi}(x,y)$ for $\Xi(x,y)$ can be constructed directly through replacing the marginal CDFs by their empirical CDFs and the involved expectations by the corresponding sample averages. Thus, a confidence interval with $1-\alpha$ confidence for $\rho(y_0, x_0)$ is constructed as $\widehat{\rho}(y_0, x_0) \pm z_{1-\alpha/2}\sqrt{\widehat{\Xi}(x_0, y_0)}/\sqrt{n}$, where $z_{1-\alpha/2}$ represents the $1-\alpha/2$ quantile of standard normal distribution.

## B2 Proofs of Main Results

**Proof of Proposition 3.1** We can first derive that, for any $(y,x) \in \mathcal{Y} \times \mathcal{X}$,

$$\sqrt{n}\{[\widehat{F}_{Y,X}(y,x) - \widehat{F}_Y(y)\widehat{F}_X(x)] - [F_{Y,X}(y,x) - F_Y(y)F_X(x)]\} = \frac{1}{\sqrt{n}}\sum_{i=1}^{n}\xi_{1i}(y,x) + o_p(1),$$

$$\sqrt{n}\{[\widehat{F}_X(x) - \widehat{F}_X^2(x)] - [F_X(x) - F_X^2(x)]\} = \frac{1}{\sqrt{n}}\sum_{i=1}^{n}\xi_{2i}(x) + o_p(1),$$

$$\sqrt{n}\{[\widehat{F}_Y(y) - \widehat{F}_Y^2(y)] - [F_Y(y) - F_Y^2(y)]\} = \frac{1}{\sqrt{n}}\sum_{i=1}^{n}\xi_{3i}(y) + o_p(1),$$

where $\xi_{1i}(y,x)$, $\xi_{2i}(x)$ and $\xi_{3i}(y)$ are given in Section 3. Since $\{\xi_{1i}(y,x), y \in \mathcal{Y}, x \in \mathcal{X}\}$, $\{\xi_{2i}(x), x \in \mathcal{X}\}$ and $\{\xi_{3i}(y), y \in \mathcal{Y}\}$ are collections of functions of indicator functions and bounded continuous functions, so they are Donsker-class (page 277, van der Vaart (1998)). By Donsker's Theorem, we have

$$\sqrt{n}\{[\widehat{F}_{Y,X}(y,x) - \widehat{F}_Y(y)\widehat{F}_X(x)] - [F_{Y,X}(y,x) - F_Y(y)F_X(x)]\} \overset{w}{\leadsto} \mathbb{G}_{Y,X}(y,x)$$

in $\ell^\infty(\mathcal{Y} \times \mathcal{X})$, where $\mathbb{G}_{Y,X}(y,x)$ is a Gaussian process with mean zero and covariance function $\text{cov}(\mathbb{G}_{Y,X}(y_1,x_1), \mathbb{G}_{Y,X}(y_2,x_2)) = E\{\xi_{1i}(y_1,x_1)\xi_{1i}(y_2,x_2)\}$, and, similarly,

$$\sqrt{n}\{[\widehat{F}_X(x) - \widehat{F}_X^2(x)] - [F_X(x) - F_X^2(x)]\} \overset{w}{\leadsto} \mathbb{G}_X(x),$$

in $\ell^\infty(\mathcal{X})$, where $\mathbb{G}_X(x)$ is a Gaussian process with mean zero and covariance function $\text{cov}(\mathbb{G}_X(x_1), \mathbb{G}_X(x_2)) = E\{\xi_{2i}(x_1)\xi_{2i}(x_2)\}$, and

$$\sqrt{n}\{[\widehat{F}_Y(y) - \widehat{F}_Y^2(y)] - [F_Y(y) - F_Y^2(y)]\} \overset{w}{\leadsto} \mathbb{G}_Y(y),$$

in $\ell^\infty(\mathcal{Y})$, where $\mathbb{G}_Y(y)$ is a Gaussian process with mean zero and covariance function $\text{cov}(\mathbb{G}_Y(y_1), \mathbb{G}_Y(y_2)) = E\{\xi_{3i}(y_1)\xi_{3i}(y_2)\}$. Denote $\mathbb{G} = (\mathbb{G}_{Y,X}(y,x), \mathbb{G}_X(x), \mathbb{G}_Y(y))^T$, $\mathbb{G}_n = (\widehat{F}_{Y,X}(y,x) - \widehat{F}_Y(y)\widehat{F}_X(x), \widehat{F}_X(x) - \widehat{F}_X^2(x), \widehat{F}_Y(y) - \widehat{F}_Y^2(y))^T$ and $\boldsymbol{\theta} \hat{=} (\theta_1, \theta_2, \theta_3)^T = (F_{Y,X}(y,x) - F_Y(y)F_X(x), F_X(x) - F_X^2(x), F_Y(y) - F_Y^2(y))^T$. Clearly, we have $\sqrt{n}(\mathbb{G}_n - \boldsymbol{\theta}) \overset{w}{\leadsto} \mathbb{G}$ in $\ell^\infty(\mathcal{Y} \times \mathcal{X})$.

On the other hand, define a three-variate function $\phi(x,y,z) = \frac{x}{\sqrt{yz}}$. Then, $\widehat{\rho}(y,x) = \phi(\mathbb{G}_n)$ and $\rho(y,x) = \phi(\boldsymbol{\theta})$. By the functional Delta method (Theorem 20.8 on page 297, van der Vaart (1998)), we can obtain

$$\sqrt{n}[\widehat{\rho}(y,x) - \rho(y,x)] \overset{w}{\leadsto} \phi'_{\boldsymbol{\theta}}(\mathbb{G})$$

in $\ell^\infty(\mathcal{Y} \times \mathcal{X})$, where $\phi'_{\boldsymbol{\theta}}(\cdot)$ is the Hadamard derivative (page 296, van der Vaart (1998)), namely, for any $\mathbf{h} \in \mathbb{R}^3$, $\phi'_{\boldsymbol{\theta}}(\mathbf{h}) = \lim_{t \to 0} \frac{\phi(\boldsymbol{\theta} + t\mathbf{h}) - \phi(\boldsymbol{\theta})}{t} = (\frac{1}{\sqrt{\theta_2 \theta_3}}, -\frac{\theta_1}{2\sqrt{\theta_2 \theta_3} \theta_2}, -\frac{\theta_1}{2\sqrt{\theta_2 \theta_3} \theta_3})\mathbf{h}$, and $\phi'_{\boldsymbol{\theta}}(\mathbb{G})$ is the Gaussian process with mean zero and covariance function

$$\begin{aligned}
&\Xi(x_1, y_1, x_2, y_2) \\
&\hat{=} \text{cov}(\phi'_{\boldsymbol{\theta}}(\mathbb{G})(x_1, y_1), \phi'_{\boldsymbol{\theta}}(\mathbb{G})(x_2, y_2)) \\
&= \{\theta_2(x_1)\theta_3(y_1)\theta_2(x_2)\theta_3(y_2)\}^{-1/2} \times \Big[ E\{\xi_{1i}(y_1,x_1)\xi_{1i}(y_2,x_2)\} \\
&\quad + \frac{\theta_1(x_1,y_1)\theta_1(x_2,y_2)E\{\xi_{2i}(x_1)\xi_{2i}(x_2)\}}{4\theta_2(x_1)\theta_2(x_2)} + \frac{\theta_1(x_1,y_1)\theta_1(x_2,y_2)E\{\xi_{3i}(y_1)\xi_{3i}(y_2)\}}{4\theta_3(y_1)\theta_3(y_2)}
\end{aligned}$$

$$
\begin{aligned}
&-\frac{\theta_1(x_2,y_2)E\{\xi_{1i}(y_1,x_1)\xi_{2i}(x_2)\}}{2\theta_2(x_2)} - \frac{E\{\theta_1(x_2,y_2)\xi_{1i}(y_1,x_1)\xi_{3i}(y_2)\}}{2\theta_3(y_2)}\\
&-\frac{\theta_1(x_1,y_1)E\{\xi_{2i}(x_1)\xi_{1i}(y_2,x_2)\}}{2\theta_2(x_1)\theta_3(y_2)} + \frac{\theta_1(x_1,y_1)\theta_1(x_2,y_2)E\{\xi_{2i}(x_1)\xi_{3i}(y_2)\}}{4\theta_2(x_1)\theta_3(y_2)}\\
&-\frac{E\{\theta_1(x_1,y_1)\xi_{3i}(y_1)\xi_{1i}(y_2,x_2)\}}{2\theta_3(y_1)} + \frac{\theta_1(x_1,y_1)\theta_1(x_2,y_2)E\{\xi_{3i}(y_1)\xi_{2i}(x_2)\}}{4\theta_3(y_1)\theta_2(x_2)}\Bigg].
\end{aligned}
$$

for any $(x_1,y_1),(x_2,y_2)\in \mathcal{Y}\times\mathcal{X}$. This proves the result. $\square$

**Proof of Theorem 3.1** To prove the first assertion, under the condition of Theorem 3.1, we know that $\mu_j^{RC}=0$. Recalling the definition of $\widehat{\mu}_j^{RC}$ and by Proposition 3.1, we have

$$
\begin{aligned}
n\widehat{\mu}_j^{RC} &= \int_{(x_j,y)\in \mathcal{X}_j\times\mathcal{Y}} \{\sqrt{n}\widehat{\rho}_j(x_j,y)\}^2 d\widehat{F}_{X_j,Y}(x_j,y)\\
&\xrightarrow{d} \int_{(x_j,y)\in \mathcal{X}_j\times\mathcal{Y}} [\phi'_{\boldsymbol{\theta}_j}(\mathbb{G})(x_j,y)]^2 dF_{X_j,Y}(x_j,y),
\end{aligned}
$$

where $\phi'_{\boldsymbol{\theta}_j}(\mathbb{G})(x_j,y)$ are zero-mean Gaussian process, similarly defined in Proposition 3.1. According to Kuo (1975, Ch. 1, § 2),

$$
n\omega_j^{-1}\widehat{\mu}_j^{RC} \xrightarrow{d} \sum_{k=1}^{\infty} \lambda_{k,j}^* \chi_k^2(1),
$$

where $\chi_k^2(1)$s are independent $\chi^2(1)$ random variables, $\lambda_{k,j}^*$ are non-negative constants that depend on the joint distribution of $(X_j,Y)$, with sum equal to one, and $\omega_j = E\Big[\int_{(x_j,y)\in\mathcal{X}_j\times\mathcal{Y}}[\phi'_{\boldsymbol{\theta}_j}(\mathbb{G})(x_j,y)]^2 dF_{X_j,Y}(x_j,y\Big] = E\{[\phi'_{\boldsymbol{\theta}_j}(\mathbb{G})(\widetilde{Y},\widetilde{X}_j)]^2\} = E\{\Xi(\widetilde{Y},\widetilde{X}_j,\widetilde{Y},\widetilde{X}_j)\}$, in which $(\widetilde{Y},\widetilde{X}_j)$ are independent copy of $(Y,X_j)$.

Next we prove the second assertion. To this end, note that

$$
\begin{aligned}
\widehat{\mu}_j^{RC} &= \int_{(x_j,y)\in\mathcal{X}_j\times\mathcal{Y}} [\widehat{\rho}_j(x_j,y) - \rho_j(x_j,y)]^2 d\widehat{F}_{X_j,Y}(x_j,y)\\
&\quad + 2\int_{(x_j,y)\in\mathcal{X}_j\times\mathcal{Y}} [\widehat{\rho}_j(x_j,y) - \rho_j(x_j,y)]\rho_j(x_j,y) d\widehat{F}_{X_j,Y}(x_j,y)\\
&\quad + \frac{1}{n}\sum_{i=1}^n \rho_j^2(X_{ij},Y_i)\\
&=: I_1 + I_2 + I_3,
\end{aligned}
$$

where the definitions of $I_1, I_2, I_3$ are clear from the context. If $X_j$ and $Y$ are not independent and by

Proposition 3.1, we know that

$$nI_1 = \int_{(x_j,y)\in\mathcal{X}_j\times\mathcal{Y}} [\sqrt{n}(\widehat{\rho}_j(x_j,y) - \rho_j(x_j,y))]^2 \mathrm{d}\widehat{F}_{X_j,Y}(x_j,y)$$
$$\xrightarrow{d} \int_{(x_j,y)\in\mathcal{X}_j\times\mathcal{Y}} [\phi'_{\boldsymbol{\theta}_j}(\mathbb{G})(x_j,y)]^2 \mathrm{d}F_{X_j,Y}(x_j,y) = O_p(1),$$

which implies $I_1 = o_p(n^{-1/2})$, and

$$\sqrt{n}I_2 = 2\int_{(x_j,y)\in\mathcal{X}_j\times\mathcal{Y}} \sqrt{n}(\widehat{\rho}_j(x_j,y) - \rho_j(x_j,y))\rho_j(x_j,y)\mathrm{d}\widehat{F}_{X_j,Y}(x_j,y)$$
$$\xrightarrow{d} 2\int_{(x_j,y)\in\mathcal{X}_j\times\mathcal{Y}} [\phi'_{\boldsymbol{\theta}_j}(\mathbb{G})(x_j,y) + o_p(1)]\rho_j(x_j,y)\mathrm{d}F_{X_j,Y}(x_j,y) + o_p(1)$$
$$= 2\frac{1}{\sqrt{n}}\sum_{i=1}^n \int_{(x_j,y)\in\mathcal{X}_j\times\mathcal{Y}} \rho_j(x_j,y)V_{ij}(y,x_j)\mathrm{d}F_{X_j,Y}(x_j,y) + o_p(1),$$

where $V_{ij}(y,x_j) = \frac{\xi_{1ij}(y,x_j)}{\sqrt{\theta_{2j}\theta_3}} - \frac{\theta_{1j}\xi_{2ij}(x_j)}{2\sqrt{\theta_{2j}\theta_3}\theta_{2j}} - \frac{\theta_{1j}\xi_{3i}(y)}{2\sqrt{\theta_{2j}\theta_3}\theta_3}$, and $\xi_{1ij}(y,x_j) = [I(X_{ij} \leq x_j) - F_{X_j}(x_j)][I(Y_i \leq y) - F_Y(y)] - [F_{X_j,Y}(x_j,y) - F_{X_j}(x_j)F_Y(y)]$, $\xi_{2ij}(x_j) = [I(X_{ij} \leq x_j) - F_{X_j}(x_j)]^2 - [F_{X_j}(x_j) - F_{X_j}^2(x_j)]$ and $\xi_{3i}(y) = [I(Y_i \leq y) - F_Y(y)]^2 - [F_Y(y) - F_Y^2(y)]$, and $(\theta_{1j}, \theta_{2j}, \theta_3) = (F_{Y,X_j}(y,x_j) - F_Y(y)F_{X_j}(x_j), F_{X_j}(x_j) - F_{X_j}^2(x_j), F_Y(y) - F_Y^2(y))$. Hence, it follows that

$$\sqrt{n}[\widehat{\mu}_j^{RC} - u_j^{RC}] = \frac{1}{\sqrt{n}}\sum_{i=1}^n \Big(2\int_{(x_j,y)\in\mathcal{X}_j\times\mathcal{Y}} \rho_j(x_j,y)V_{ij}(y,x_j)\mathrm{d}F_{X_j,Y}(x_j,y)$$
$$+ [\rho_j^2(X_{ij},Y_i) - E\{\rho_j^2(X_{ij},Y_i))\}]\Big) + o_p(1)$$
$$=: \frac{1}{\sqrt{n}}\sum_{i=1}^n Z_{ij} + o_p(1).$$

By Slutsky's theorem and the central limit theorem, $\sqrt{n}[\widehat{\mu}_j^{RC} - u_j^{RC}]$ converges in distribution to $N(0, \Delta_j)$, where $\Delta_j = \mathrm{var}(Z_{ij})$. This completes the proof of Theorem 3.1. $\square$

**Lemma A.1.** *(Hoeffding's inequality, Lemma 14.11, Bühlmann and van de Geer (2011)) Let $Y_1, \ldots, Y_n$ be independent random variables satisfying $P(a_i \leq Y_i \leq b_i) = 1$ for some $a_i$ and $b_i$ for all $i = 1, \ldots, n$, where $a_i < b_i$. Then, for all $\epsilon > 0$,*

$$P\Big(\Big|\sum_{i=1}^n [Y_i - E(Y_i)]\Big| \geq \epsilon\Big) \leq 2\exp\Big(-\frac{2\epsilon^2}{\sum_{i=1}^n (b_i - a_i)^2}\Big).$$

**Lemma A.2.** *(Bernstein's inequality, Lemma 2.2.11, van der Vaart and Wellner (1996)) For independent*

random variables $Y_1, \ldots, Y_n$ with mean zero and $E\{|Y_i|^r\} \leq r! K^{r-2} v_i/2$ for every $r \geq 2$, $i = 1, \ldots, n$ and some constants $K, v_i$. Then, for $x > 0$, we have

$$P(|Y_1 + \cdots + Y_n| > x) \leq 2 \exp\Big(-\frac{x^2}{2(v + Kx)}\Big),$$

for $v \geq \sum_{i=1}^n v_i$.

**Lemma A.3.** *(DKW inequality, Massart (1990))* Let $F_n$ be the empirical CDF function and $F$ be the CDF. Then, for every $\epsilon$,

$$P\big(\sqrt{n}\|F_n - F\|_\infty > \epsilon\big) \leq 2 \exp\big(-2\epsilon^2\big),$$

where $\|F_n - F\|_\infty = \sup_y |F_n(y) - F(y)|$ is the Kolmogorov-Smirnov statistic.

**Lemma A.4.** Suppose that $\{(Y_i, X_{ij})_{i=1}^n\}$ are iid copies of $(Y, X_j)$ and let $\widehat{F}(y, x_j) = \frac{1}{n} \sum_{i=1}^n I(Y_i \leq y, X_{ij} \leq x_j)$ be empirical CDF and $F(y, x_j)$ be the joint CDF of $(Y, X_j)$, and assume that condition (D1) holds, then there exists a uniform constant $\tilde{c}_1$ such that, for any $\epsilon$ such that $n\epsilon \to \infty$ and $n\epsilon^2/\log n \to \infty$ as $n \to \infty$,

$$\max_{1 \leq j \leq p_n} P\Big(\sup_{(y, x_j) \in \mathcal{Y} \times \mathcal{X}_j} |\widehat{F}(y, x_j) - F(y, x_j)| > \epsilon\Big) \leq 6 \exp(-\tilde{c}_1 n \epsilon^2).$$

**Proof of Lemma A.4** By the definitions of $U_{ij}$ and $U_{i0}$, we note that $U_{ij} \sim U(0,1)$ and $U_{i0} \sim U(0,1)$. Then

$$P\Big(\sup_{(y, x_j) \in \mathcal{Y} \times \mathcal{X}_j} |\widehat{F}(y, x_j) - F(y, x_j)| > \epsilon\Big) = P\Big(\sup_{(u_0, u_j) \in \Omega} |\widehat{F}_{0j}(u_0, u_j) - F_{0j}(u_0, u_j)| > \epsilon\Big),$$

where $\Omega = \{(u, v) : 0 < u < 1, 0 < v < 1\}$.

Next, we partition $\Omega$ into $K_n = n^2$ small rectangles $\{\Omega_{k,l}\}_{k,l=1}^n$, where $\Omega_{k,l} = \{(u, v) : (k-1)R_n < u \leq kR_n, (l-1)R_n < v \leq lR_n\}$ and $R_n = 1/n$. By this, we can see that $\Omega$ can be covered by $\cup_{k=1}^n \cup_{l=1}^n \Omega_{k,l}$, and the rectangles $\{\Omega_{k,l}\}_{k,l=1}^n$ are disjoint and $\Omega_{k,l}$ has the center $(a_k, b_l)$ with $a_k = (k-1)R_n + \frac{R_n}{2}$ and $b_l = (l-1)R_n + \frac{R_n}{2}$. It follows that $\sup_{(u_0, u_j) \in \Omega} |\widehat{F}_{0j}(u_0, u_j) - F_{0j}(u_0, u_j)| \leq \max_{1 \leq k \leq n, 1 \leq l \leq n} \sup_{(u_0, u_j) \in \Omega_{k,l}} |\widehat{F}_{0j}(u_0, u_j) - F_{0j}(u_0, u_j)|$. This implies

$$P\Big(\sup_{(u_0, u_j) \in \Omega} |\widehat{F}_{0j}(u_0, u_j) - F_{0j}(u_0, u_j)| > \epsilon\Big)$$

$$\leq \sum_{k=1}^{n}\sum_{l=1}^{n} P\Big(\sup_{(u_0,u_j)\in\Omega_{k,l}} |\widehat{F}_{0j}(u_0,u_j) - F_{0j}(u_0,u_j)| > \epsilon\Big)$$

$$\leq \sum_{k=1}^{n}\sum_{l=1}^{n} P\Big(|\widehat{F}_{0j}(a_k,b_l) - F_{0j}(a_k,b_l)| > \frac{\epsilon}{2}\Big)$$

$$+ \sum_{k=1}^{n}\sum_{l=1}^{n} P\Big(\sup_{(u_0,u_j)\in\Omega_{k,l}} |\widehat{F}_{0j}(u_0,u_j) - \widehat{F}_{0j}(a_k,b_l) - F_{0j}(u_0,u_j) + F_{0j}(a_k,b_l)| > \frac{\epsilon}{2}\Big)$$

$$\triangleq \Pi_{nj}^{(1)} + \Pi_{nj}^{(2)} \text{ (say)}. \tag{B.1}$$

On one hand, for the term $\Pi_{nj}^{(1)}$, by Lemma A.1, we have

$$\max_{1\leq k\leq n, 1\leq l\leq n} P\Big(|\widehat{F}_{0j}(a_k,b_l) - F_{0j}(a_k,b_l)| > \frac{\epsilon}{2}\Big) \leq 2\exp(-\frac{n\epsilon^2}{32}),$$

which yields

$$\max_{1\leq j\leq p_n} \Pi_{nj}^{(1)} \leq 2n^2 \exp(-\frac{n\epsilon^2}{32}) = 2\exp(-(\frac{n\epsilon^2}{32} - 2\log n)) \leq 2\exp(-c_1 n\epsilon^2), \tag{B.2}$$

for some uniform constant $c_1 > 0$. On the other hand, for the term $\Pi_{nj}^{(2)}$, we note that

$$\sup_{(u_0,u_j)\in\Omega_{k,l}} |\widehat{F}_{0j}(u_0,u_j) - \widehat{F}_{0j}(a_k,b_l) - F_{0j}(u_0,u_j) + F_{0j}(a_k,b_l)|$$

$$\leq \sup_{(u_0,u_j)\in\Omega_{k,l}} |\widehat{F}_{0j}(u_0,u_j) - \widehat{F}_{0j}(a_k,b_l)| + \sup_{(u_0,u_j)\in\Omega_{k,l}} |F_{0j}(u_0,u_j) - F_{0j}(a_k,b_l)|$$

$$\triangleq A_{nkl,1} + A_{kl,2} \text{ (say)},$$

where the first term $A_{nkl,1}$ can be further bounded by

$$\begin{aligned}
A_{nkl,1} &\leq \sup_{(u_0,u_j)\in\Omega_{k,l}} \frac{1}{n}\sum_{i=1}^{n} |I(U_{i0}\leq u_0, U_{ij}\leq u_j) - I(U_{i0}\leq a_k, U_{ij}\leq b_l)| \\
&\leq \sup_{(k-1)R_n < u_0 \leq kR_n} \frac{1}{n}\sum_{i=1}^{n} |I(U_{i0}\leq u_0) - I(U_{i0}\leq a_k)| \\
&\quad + \sup_{(l-1)R_n < u_j \leq lR_n} \frac{1}{n}\sum_{i=1}^{n} |I(U_{ij}\leq u_j) - I(U_{ij}\leq b_l)| \\
&\leq \frac{1}{n}\sum_{i=1}^{n} I(a_k - \frac{R_n}{2} \leq U_{i0} \leq a_k + \frac{R_n}{2}) + \frac{1}{n}\sum_{i=1}^{n} I(b_l - \frac{R_n}{2} \leq U_{ij} \leq b_l + \frac{R_n}{2}),
\end{aligned}$$

in which the last line uses the fact that for any $y$ and any $\delta > 0$, $\sup_{|y_1-y|<\delta} |I(Y<y_1) - I(Y<y)| \leq I(y-\delta < Y < y+\delta)$, and for the second term $A_{kl,2}$, by condition (D1), there exists a point $(u_0^*, u_j^*)$ between

$(u_0, u_j)$ and $(a_k, b_l)$ such that

$$\begin{aligned}A_{kl,2} &= \sup_{(u_0,u_j)\in\Omega_{k,l}} \Big|\frac{\partial F_{0j}(u_0^*, u_j^*)}{\partial u_0}(u_0 - a_k) + \frac{\partial F_{0j}(u_0^*, u_j^*)}{\partial u_j}(u_j - b_l)\Big| \\ &\leq M_1 \frac{R_n}{2} + M_2 \frac{R_n}{2} \leq \frac{\max(M_1, M_2)}{n}.\end{aligned}$$

Therefore, letting $\frac{\epsilon}{2} - \frac{\max(M_1, M_2)}{n} \geq \frac{\epsilon}{3}$ for sufficiently large $n$, we have

$$\begin{aligned}\Pi_{nj}^{(2)} &\leq \sum_{k=1}^{n}\sum_{l=1}^{n} P\Big(\frac{1}{n}\sum_{i=1}^{n} I(a_k - \frac{R_n}{2} \leq U_{i0} \leq a_k + \frac{R_n}{2}) \\ &\quad + \frac{1}{n}\sum_{i=1}^{n} I(b_l - \frac{R_n}{2} \leq U_{ij} \leq b_l + \frac{R_n}{2}) > \frac{\epsilon}{3}\Big) \\ &\leq \sum_{k=1}^{n}\sum_{l=1}^{n} P\Big(\frac{1}{n}\sum_{i=1}^{n} I(a_k - \frac{R_n}{2} \leq U_{i0} \leq a_k + \frac{R_n}{2}) > \frac{\epsilon}{6}\Big) \\ &\quad + \sum_{k=1}^{n}\sum_{l=1}^{n} P\Big(\frac{1}{n}\sum_{i=1}^{n} I(b_l - \frac{R_n}{2} \leq U_{ij} \leq b_l + \frac{R_n}{2}) > \frac{\epsilon}{6}\Big).\end{aligned}$$

It suffices to bound those two terms on the right-hand side of the above inequality. In what follows, we only derive an exponential upper bound of the second term since the first term can be dealt with in the same manner. Let $\omega_{ijl} = I(b_l - \frac{R_n}{2} \leq U_{ij} \leq b_l + \frac{R_n}{2})$. It follows that $E\omega_{ijl} = F_{U_{ij}}(b_l + \frac{R_n}{2}) - F_{U_{ij}}(b_l - \frac{R_n}{2}) = R_n = \frac{1}{n}$. Then, there exists a uniform constant $c_2$

$$\begin{aligned}&\max_{1\leq j\leq p_n} \max_{1\leq l\leq n} P\Big(\frac{1}{n}\sum_{i=1}^{n} I(b_l - \frac{R_n}{2} \leq U_{ij} \leq b_l + \frac{R_n}{2}) > \frac{\epsilon}{6}\Big) \\ &= \max_{1\leq j\leq p_n} \max_{1\leq l\leq n} P\Big(\frac{1}{n}\sum_{i=1}^{n} \omega_{ijl} - E\omega_{ijl} > \frac{\epsilon}{6} - \frac{1}{n}\Big) \\ &\leq \max_{1\leq j\leq p_n} \max_{1\leq l\leq n} P\Big(\Big|\frac{1}{n}\sum_{i=1}^{n} \omega_{ijl} - E\omega_{ijl}\Big| > \frac{\epsilon}{6} - \frac{1}{n}\Big) \\ &\leq \max_{1\leq j\leq p_n} \max_{1\leq l\leq n} P\Big(\Big|\frac{1}{n}\sum_{i=1}^{n} \omega_{ijl} - E\omega_{ijl}\Big| > \frac{\epsilon}{8}\Big) \\ &\leq 2\exp(-c_2 n\epsilon^2)\end{aligned}$$

provided $\frac{\epsilon}{6} - \frac{1}{n} \geq \frac{\epsilon}{8}$, where the last line holds by Lemma A.1. Thus,

$$\begin{aligned}&\max_{1\leq j\leq p_n} \sum_{l=1}^{n}\sum_{l=1}^{n} P\Big(\frac{1}{n}\sum_{i=1}^{n} I(b_l - \frac{R_n}{2} \leq U_{ij} \leq b_l + \frac{R_n}{2}) > \frac{\epsilon}{6}\Big) \\ &\leq 2n^2 \exp(-c_2 n\epsilon^2) = 2\exp(-(c_2 n\epsilon^2 - 2\log n)) \leq 2\exp(-c_3 n\epsilon^2)\end{aligned}$$

for some uniform constant $c_3 > 0$ and sufficiently large $n$. Similarly, there exists a uniform constant $c_4$ such that

$$\sum_{k=1}^{n}\sum_{l=1}^{n} P\Big(\frac{1}{n}\sum_{i=1}^{n} I(a_k - \frac{R_n}{2} \leq U_{i0} \leq a_k + \frac{R_n}{2}) > \frac{\epsilon}{6}\Big) \leq 2\exp(-c_4 n\epsilon^2)$$

for sufficiently large $n$. Stacking these two results yields

$$\max_{1\leq j\leq p_n} \Pi_{nj}^{(2)} \leq 2\exp(-c_3 n\epsilon^2) + 2\exp(-c_4 n\epsilon^2) \leq 4\exp(-c_5 n\epsilon^2), \tag{B.3}$$

where $c_5 = \min(c_3, c_4)$. Putting (B.2) and (B.3) together gives the desired result. $\square$

**Proof of Theorem 3.2** Firstly, we note that Lemma A.3 implies that

$$\max_{1\leq j\leq p_n} \sup_{x_j \in \mathcal{X}_j} P\Big(|\widehat{F}_{X_j}(x_j) - F_{X_j}(x_j)| > \epsilon\Big)$$
$$\leq \max_{1\leq j\leq p_n} P\Big(\sup_{x_j \in \mathcal{X}_j} |\widehat{F}_{X_j}(x_j) - F_{X_j}(x_j)| > \epsilon\Big) \leq 2\exp(-2n\epsilon^2) \tag{B.4}$$

and

$$\sup_{y\in\mathcal{Y}} P\Big(|\widehat{F}_Y(y) - F_Y(y)| > \epsilon\Big)$$
$$\leq P\Big(\sup_{y\in\mathcal{Y}} |\widehat{F}_Y(y) - F_Y(y)| > \epsilon\Big) \leq 2\exp(-2n\epsilon^2). \tag{B.5}$$

Since $|\widehat{F}_{X_j}(x_j)\widehat{F}_Y(y) - F_{X_j}(x_j)F_Y(y)| \leq |[\widehat{F}_{X_j}(x_j) - F_{X_j}(x_j)]\widehat{F}_Y(y)| + |F_{X_j}(x_j)[\widehat{F}_Y(y) - F_Y(y)]| \leq |\widehat{F}_{X_j}(x_j) - F_{X_j}(x_j)| + |\widehat{F}_Y(y) - F_Y(y)|$, so for any $\epsilon > 0$ and using (B.4) and (B.5), we have,

$$\max_{1\leq j\leq p_n} \sup_{(x_j,y)\in\mathcal{X}_j\times\mathcal{Y}} P\Big(|\widehat{F}_{X_j}(x_j)\widehat{F}_Y(y) - F_{X_j}(x_j)F_Y(y)| > \epsilon\Big)$$
$$\leq \max_{1\leq j\leq p_n} \sup_{x_j\in\mathcal{Y}} P\Big(|\widehat{F}_{X_j}(x_j) - F_{X_j}(x_j)| > \epsilon/2\Big) + \sup_{y\in\mathcal{Y}} P\Big(|\widehat{F}_Y(y) - F_Y(y)| > \epsilon/2\Big)$$
$$\leq 2\exp(-n\epsilon^2/2) + 2\exp(-n\epsilon^2/2) = 4\exp(-n\epsilon^2/2) \leq 4\exp(-c_6 n\epsilon^2) \tag{B.6}$$

for some uniform constant $c_6 \leq 1/2$. On the other hand, because $|\widehat{F}_{X_j}(x_j) + F_{X_j}(x_j)| \leq |\widehat{F}_{X_j}(x_j)| + |F_{X_j}(x_j)| \leq 2$, we obtain

$$\max_{1\leq j\leq p_n} \sup_{x_j\in\mathcal{X}_j} P\Big(|\widehat{F}_{X_j}^2(x_j) - F_{X_j}^2(x_j)| > \epsilon\Big)$$

$$\leq \max_{1\leq j\leq p_n} \sup_{x_j\in\mathcal{X}_j} P\Big(|\widehat{F}_{X_j}(x_j) - F_{X_j}(x_j)| > \epsilon/2\Big)$$
$$\leq 2\exp(-n\epsilon^2/2) \leq 2\exp(-c_7 n\epsilon^2) \tag{B.7}$$

for some uniform constant $c_7 \leq 1/2$. Similarly, we have

$$\sup_{y\in\mathcal{Y}} P\Big(|\widehat{F}_Y^2(y) - F_Y^2(y)| > \epsilon\Big) \leq 2\exp(-c_8 n\epsilon^2) \tag{B.8}$$

for some uniform constant $c_8 \leq 1/2$.

Using (B.4) and (B.7), then there exists a uniform constant $c_9$ such that

$$\max_{1\leq j\leq p_n} \sup_{x_j\in\mathcal{X}_j} P\Big(|\{\widehat{F}_{X_j}(x_j) - \widehat{F}_{X_j}^2(x_j)\} - \{F_{X_j}(x_j) - F_{X_j}^2(x_j)\}| > \epsilon\Big)$$
$$\leq \max_{1\leq j\leq p_n} \sup_{x_j\in\mathcal{X}_j} P\Big(|\widehat{F}_{X_j}(x_j) - F_{X_j}(x_j)| > \epsilon/2\Big)$$
$$+ \max_{1\leq j\leq p_n} \sup_{x_j\in\mathcal{X}_j} P\Big(|\widehat{F}_{X_j}^2(x_j) - F_{X_j}^2(x_j)| > \epsilon/2\Big)$$
$$\leq 2\exp(-n\epsilon^2/2) + 2\exp(-c_7 n\epsilon^2/4) \leq 4\exp(-c_9 n\epsilon^2), \tag{B.9}$$

where $c_9 = \min(1/2, c_7/4)$. Similarly, using (B.5) and (B.8), we have

$$\sup_{y\in\mathcal{Y}} P\Big(|\{\widehat{F}_Y(y) - \widehat{F}_Y^2(y)\} - \{F_Y(y) - F_Y^2(y)\}| > \epsilon\Big)$$
$$\leq \sup_{y\in\mathcal{Y}} P\Big(|\widehat{F}_Y(y) - F_Y(y)| > \epsilon/2\Big)$$
$$+ \sup_{y\in\mathcal{Y}} P\Big(|\widehat{F}_Y^2(y) - F_Y^2(y)| > \epsilon/2\Big)$$
$$\leq 2\exp(-n\epsilon^2/2) + 2\exp(-c_8 n\epsilon^2/4) \leq 4\exp(-c_{10} n\epsilon^2), \tag{B.10}$$

where $c_{10} = \min(1/2, c_8/4)$. Invoking (B.9) and (B.10) and using the fact that $x - x^2 \leq 1/4$ for any $x \in (0,1)$, we have

$$\max_{1\leq j\leq p_n} \sup_{(x_j,y)\in\mathcal{X}_j\times\mathcal{Y}} P\Big(|\{\widehat{F}_{X_j}(x_j) - \widehat{F}_{X_j}^2(x_j)\}\{\widehat{F}_Y(y) - \widehat{F}_Y^2(y)\}$$
$$-\{F_{X_j}(x_j) - F_{X_j}^2(x_j)\}\{F_Y(y) - F_Y^2(y)\}| > \epsilon\Big)$$
$$\leq \max_{1\leq j\leq p_n} \sup_{x_j\in\mathcal{X}_j} P\Big(|[\{\widehat{F}_{X_j}(x_j) - \widehat{F}_{X_j}^2(x_j)\} - \{F_{X_j}(x_j) - F_{X_j}^2(x_j)\}][\widehat{F}_Y(y) - \widehat{F}_Y^2(y)]| > \epsilon/2\Big)$$
$$+ \max_{1\leq j\leq p_n} \sup_{y\in\mathcal{Y}} P\Big(|[\{\widehat{F}_Y(y) - \widehat{F}_Y^2(y)\} - \{F_Y(y) - F_y^2(y)\}][\widehat{F}_{X_j}(x_j) - \widehat{F}_{X_j}^2(x_j)]| > \epsilon/2\Big)$$

$$
\begin{aligned}
&\leq \max_{1\leq j\leq p_n} \sup_{x_j\in\mathcal{X}_j} P\Big(\big|\{\widehat{F}_{X_j}(x_j) - \widehat{F}^2_{X_j}(x_j)\} - \{F_{X_j}(x_j) - F^2_{X_j}(x_j)\}\big| > 2\epsilon\Big) \\
&\quad + \max_{1\leq j\leq p_n} \sup_{y\in\mathcal{Y}} P\Big(\big|\{\widehat{F}_Y(y) - \widehat{F}^2_Y(y)\} - \{F_Y(y) - F^2_y(y)\}\big| > 2\epsilon\Big) \\
&\leq 4\exp(-4c_9 n\epsilon^2) + 4\exp(-4c_{10} n\epsilon^2) \leq 8\exp(-c_{11} n\epsilon^2),
\end{aligned} \tag{B.11}
$$

where $c_{11} = 4\min(c_9, c_{10})$. Furthermore, since $|\widehat{F}_{Y,X_j}(y,x_j) - \widehat{F}_Y(y)\widehat{F}_{X_j}(x_j) + F_{Y,X_j}(y,x_j) - F_Y(y)F_{X_j}(x_j)| \leq 4$, so

$$
\begin{aligned}
&\max_{1\leq j\leq p_n} \sup_{(y,x_j)\in\mathcal{Y}\times\mathcal{X}_j} P\Big(\big|\{\widehat{F}_{Y,X_j}(y,x_j) - \widehat{F}_Y(y)\widehat{F}_{X_j}(x_j)\}^2 - \{F_{Y,X_j}(y,x_j) - F_Y(y)F_{X_j}(x_j)\}^2\big| > \epsilon\Big) \\
&\leq \max_{1\leq j\leq p_n} \sup_{(y,x_j)\in\mathcal{Y}\times\mathcal{X}_j} P\Big(\big|\{\widehat{F}_{Y,X_j}(y,x_j) - \widehat{F}_Y(y)\widehat{F}_{X_j}(x_j)\} - \{F_{Y,X_j}(y,x_j) - F_Y(y)F_{X_j}(x_j)\}\big| > \epsilon/4\Big) \\
&\leq \max_{1\leq j\leq p_n} \sup_{(y,x_j)\in\mathcal{Y}\times\mathcal{X}_j} P\Big(\big|\widehat{F}_{Y,X_j}(y,x_j) - F_{Y,X_j}(y,x_j)\big| > \epsilon/8\Big) \\
&\quad + \max_{1\leq j\leq p_n} \sup_{(y,x_j)\in\mathcal{Y}\times\mathcal{X}_j} P\Big(\big|\widehat{F}_Y(y)\widehat{F}_{X_j}(x_j) - F_Y(y)F_{X_j}(x_j)\big| > \epsilon/8\Big) \\
&\leq 6\exp(-\tilde{c}_1 n\epsilon^2/64) + 4\exp(-c_6 n\epsilon^2/64) \leq 10\exp(-c_{12} n\epsilon^2),
\end{aligned} \tag{B.12}
$$

where $c_{12} = \min(\tilde{c}_1, c_6)/64$, and the last inequality holds due to Lemma A.4 and (B.6).

Secondly, we are going to derive a uniform upper bound for the tail probability

$$\max_{1\leq j\leq p_n} \sup_{(y,x_j)\in\mathcal{Y}\times\mathcal{X}_j} P(|\widehat{\rho}^2_j(y,x_j) - \rho^2_j(y,x_j)| > \epsilon).$$

To this end, define $N_{nj1}(y,x_j) = \{\widehat{F}_{Y,X_j}(y,x_j) - \widehat{F}_Y(y)\widehat{F}_{X_j}(x_j)\}^2$, $N_{nj2}(y,x_j) = \{\widehat{F}_Y(y) - \widehat{F}^2_Y(y)\}\{\widehat{F}_{X_j}(x_j) - \widehat{F}^2_{X_j}(x_j)\}$, $N_{j1}(y,x_j) = \{F_{Y,X_j}(y,x_j) - F_Y(y)F_{X_j}(x_j)\}^2$ and $N_{j2}(y,x_j) = \{F_Y(y) - F^2_Y(y)\}\{F_{X_j}(x_j) - F^2_{X_j}(x_j)\}$. Then, we have $\widehat{\rho}^2_j(y,x_j) = \frac{N_{nj1}(y,x_j)}{N_{nj2}(y,x_j)}$ and $\rho^2_j(y,x_j) = \frac{N_{j1}(y,x_j)}{N_{j2}(y,x_j)}$. Since

$$|\widehat{\rho}^2_j(y,x_j) - \rho^2_j(y,x_j)| \leq \left|\frac{N_{nj1}(y,x_j) - N_{j1}(y,x_j)}{N_{nj2}(y,x_j)}\right| + \left|\frac{\{N_{nj2}(y,x_j) - N_{j2}(y,x_j)\}N_{j1}(y,x_j)}{N_{nj2}(y,x_j)N_{j2}(y,x_j)}\right|,$$

it follows that

$$
\begin{aligned}
&\max_{1\leq j\leq p_n} \sup_{(y,x_j)\in\mathcal{Y}\times\mathcal{X}_j} P(|\widehat{\rho}^2_j(y,x_j) - \rho^2_j(y,x_j)| > \epsilon) \\
&\leq \max_{1\leq j\leq p_n} \sup_{(y,x_j)\in\mathcal{Y}\times\mathcal{X}_j} P\Big(\Big|\frac{N_{nj1}(y,x_j) - N_{j1}(y,x_j)}{N_{nj2}(y,x_j)}\Big| > \frac{\epsilon}{2}\Big) \\
&\quad + \max_{1\leq j\leq p_n} \sup_{(y,x_j)\in\mathcal{Y}\times\mathcal{X}_j} P\Big(\Big|\frac{\{N_{nj2}(y,x_j) - N_{j2}(y,x_j)\}N_{j1}(y,x_j)}{N_{nj2}(y,x_j)N_{j2}(y,x_j)}\Big| > \frac{\epsilon}{2}\Big)
\end{aligned}
$$

$$\triangleq II_{n1} + II_{n2}, \text{ (say)}. \tag{B.13}$$

Before dealing with the first term $II_{n1}$, we observe that $\min_{1\leq j\leq p_n} \inf_{(y,x_j)\in \mathcal{Y}\times\mathcal{X}_j} N_{j2}(y,x_j) \geq K_1 K_2 > 0$ by condition (D2). Then, it follows that

$$\begin{aligned}
II_{n1} &\leq \max_{1\leq j\leq p_n} \sup_{(y,x_j)\in\mathcal{Y}\times\mathcal{X}_j} P\Big(\Big|\frac{N_{nj1}(y,x_j) - N_{j1}(y,x_j)}{N_{nj2}(y,x_j)}\Big| > \frac{\epsilon}{2}, |N_{nj2}(y,x_j)| \geq \frac{K_1 K_2}{2}\Big) \\
&\quad + \max_{1\leq j\leq p_n} \sup_{(y,x_j)\in\mathcal{Y}\times\mathcal{X}_j} P\Big(|N_{nj2}(y,x_j)| < \frac{K_1 K_2}{2}\Big) \\
&\leq \max_{1\leq j\leq p_n} \sup_{(y,x_j)\in\mathcal{Y}\times\mathcal{X}_j} P\Big(|N_{nj1}(y,x_j) - N_{j1}(y,x_j)| > \frac{K_1 K_2 \epsilon}{4}\Big) \\
&\quad + \max_{1\leq j\leq p_n} \sup_{(y,x_j)\in\mathcal{Y}\times\mathcal{X}_j} P\Big(|N_{nj2}(y,x_j)| < \frac{K_1 K_2}{2}\Big) \\
&\triangleq II_{n1,1} + II_{n1,2}, \text{ (say)},
\end{aligned}$$

where the first term on the right-hand side can be bounded further by

$$II_{n1,1} \leq 10 \exp(-c_{12} K_1^2 K_2^2 n\epsilon^2/16) \tag{B.14}$$

using (B.12), and since $N_{nj2}(y, x_j) > 0$ for any $\epsilon \leq K_1 K_2 / 2$, the second term is bounded by

$$\begin{aligned}
II_{n1,2} &\leq \max_{1\leq j\leq p_n} \sup_{(y,x_j)\in\mathcal{Y}\times\mathcal{X}_j} P\Big(N_{nj2}(y,x_j) < K_1 K_2 - \epsilon\Big) \\
&\leq \max_{1\leq j\leq p_n} \sup_{(y,x_j)\in\mathcal{Y}\times\mathcal{X}_j} P\Big(N_{nj2}(y,x_j) < N_{j2}(y,x_j) - \epsilon\Big) \\
&\leq \max_{1\leq j\leq p_n} \sup_{(y,x_j)\in\mathcal{Y}\times\mathcal{X}_j} P\Big(N_{j2}(y,x_j) - N_{nj2}(y,x_j) > \epsilon\Big) \\
&\leq \max_{1\leq j\leq p_n} \sup_{(y,x_j)\in\mathcal{Y}\times\mathcal{X}_j} P\Big(|N_{nj2}(y,x_j) - N_{j2}(y,x_j)| > \epsilon\Big) \\
&\leq 8 \exp(-c_{11} n\epsilon^2), \tag{B.15}
\end{aligned}$$

using (B.11). Thus, combining (B.14) and (B.15), we have

$$II_{n1} \leq 10 \exp(-c_{12} K_1^2 K_2^2 n\epsilon^2/16) + 8\exp(-c_{11} n\epsilon^2) \leq 18 \exp(-c_{13} n\epsilon^2), \tag{B.16}$$

where $c_{13} = \min(c_{12} K_1^2 K_2^2/16, c_{11})$. Next, we consider the second term $II_{n2}$ in (B.13). Because $\min_{1\leq j\leq p_n} \inf_{(y,x_j)\in\mathcal{Y}\times\mathcal{X}_j} N_{j2}(y,x_j) \geq K_1 K_2 > 0$ and $\max_{1\leq j\leq p_n} \sup_{(y,x_j)\in\mathcal{Y}\times\mathcal{X}_j} N_{j1}(y,x_j) \leq 1/16$. Similar-

ly, we have

$$
\begin{aligned}
II_{n2} &\leq \max_{1\leq j\leq p_n} \sup_{(y,x_j)\in \mathcal{Y}\times\mathcal{X}_j} P\Big(\big|N_{nj2}(y,x_j) - N_{j2}(y,x_j)\big| > N_{nj2}(y,x_j) \times 8K_1K_2\epsilon\Big) \\
&\leq \max_{1\leq j\leq p_n} \sup_{(y,x_j)\in \mathcal{Y}\times\mathcal{X}_j} P\Big(\big|N_{nj2}(y,x_j) - N_{j2}(y,x_j)\big| > 4K_1^2K_2^2\epsilon\Big) \\
&\quad + \max_{1\leq j\leq p_n} \sup_{(y,x_j)\in \mathcal{Y}\times\mathcal{X}_j} P\Big(\big|N_{nj2}(y,x_j)\big| < \frac{K_1K_2}{2}\Big) \\
&\leq 8\exp(-16K_1^4K_2^4 c_{11} n\epsilon^2) + 8\exp(-c_{11} n\epsilon^2) \\
&\leq 16\exp(-c_{14} n\epsilon^2), \tag{B.17}
\end{aligned}
$$

where $c_{14} = \min(16K_1^4K_2^4 c_{11}, c_{11})$, and the last inequality is satisfied thanks to (B.11) and (B.15). Therefore, substituting (B.16) and (B.17) into (B.13), we immediately conclude that there exists a uniform constant $c_{15} \leq \min(c_{13}, c_{14})$ such that

$$
\max_{1\leq j\leq p_n} \sup_{(y,x_j)\in \mathcal{Y}\times\mathcal{X}_j} P(|\widehat{\rho}_j^2(y,x_j) - \rho_j^2(y,x_j)| > \epsilon) \leq 34\exp(-c_{15} n\epsilon^2). \tag{B.18}
$$

Finally, we are ready to prove the theorem. Recall that $\widehat{u}_j^{RC} = \frac{1}{n}\sum_{i=1}^n \widehat{\rho}_j^2(Y_i, X_{ij})$ and $u_j^{RC} = E\{\rho_j^2(Y_i, X_{ij})\}$. Then, it follows that

$$
\max_{1\leq j\leq p_n} P(|\widehat{u}_j^{RC} - u_j^{RC}| > \epsilon) \leq \max_{1\leq j\leq p_n} P(|Q_{nj1}| > \epsilon/2) + \max_{1\leq j\leq p_n} P(|Q_{nj2}| > \epsilon/2), \tag{B.19}
$$

where $Q_{nj1} = \frac{1}{n}\sum_{i=1}^n [\widehat{\rho}_j^2(Y_i, X_{ij}) - \rho_j^2(Y_i, X_{ij})]$ and $Q_{nj2} = \frac{1}{n}\sum_{i=1}^n [\rho_j^2(Y_i, X_{ij}) - E\{\rho_j^2(Y_i, X_{ij})\}]$. On one hand, using (B.18), we have

$$
\begin{aligned}
\max_{1\leq j\leq p_n} P(|Q_{nj1}| > \epsilon/2) &\leq \max_{1\leq j\leq p_n} P\Big(\frac{1}{n}\sum_{i=1}^n \big|\widehat{\rho}_j^2(Y_i, X_{ij}) - \rho_j^2(Y_i, X_{ij})\big| > \epsilon/2\Big) \\
&\leq \max_{1\leq j\leq p_n} P\Big(\max_{1\leq i\leq n} |\widehat{\rho}_j^2(Y_i, X_{ij}) - \rho_j^2(Y_i, X_{ij})| > \epsilon/2\Big) \\
&\leq \max_{1\leq j\leq p_n} \sum_{i=1}^n P\Big(\big|\widehat{\rho}_j^2(Y_i, X_{ij}) - \rho_j^2(Y_i, X_{ij})\big| > \epsilon/2\Big) \\
&\leq n \max_{1\leq j\leq p_n} \sup_{(y,x_j)\in \mathcal{Y}\times\mathcal{X}_j} P\Big(\big|\widehat{\rho}_j^2(y,x_j) - \rho_j^2(y,x_j)\big| > \epsilon/2\Big) \\
&\leq 34\exp(-c_{15} n\epsilon^2/4).
\end{aligned}
$$

For the second term on the right-hand side of (B.19), since $|\rho_j^2(Y_i, X_{ij}) - E\{\rho_j^2(Y_i, X_{ij})\}| \leq 2$ and by Lemma

A.1, we obtain

$$\max_{1\leq j\leq p_n} P(|Q_{nj2}| > \epsilon/2) \leq 2\exp(-n\epsilon^2/32).$$

Consequently, with the above results, we obtain

$$\max_{1\leq j\leq p_n} P(|\widehat{u}_j^{RC} - u_j^{RC}| > \epsilon) \leq 34\exp(-c_{15}n\epsilon^2/4) + 2\exp(-n\epsilon^2/32) \leq 36\exp(-c_{16}n\epsilon^2)$$

for some constant $c_{16} \leq \min(c_{15}/4, 1/32)$. This further implies

$$P(\max_{1\leq j\leq p_n} |\widehat{u}_j^{RC} - u_j^{RC}| > \epsilon) \leq p_n \max_{1\leq j\leq p_n} P(|\widehat{u}_j^{RC} - u_j^{RC}| > \epsilon) \leq 36 p_n \exp(-c_{16}n\epsilon^2),$$

which completes the proof of Theorem 3.2. $\square$

**Proof of Theorem 3.3** Prove the assertion (i). By Theorem 3.2 and the choice of $\varsigma_n$ and using condition (D3), we have

$$\begin{aligned}
P(\mathcal{M}_{1*} \subset \widehat{\mathcal{M}}_a) &\geq P\Big(\min_{j\in\mathcal{M}_{1*}} \widehat{u}_j^{RC} > \varsigma_n\Big) \geq P\Big(\min_{j\in\mathcal{M}_{1*}} (\widehat{u}_j^{RC} - u_j^{RC}) > \varsigma_n - \min_{j\in\mathcal{M}_{1*}} u_j^{RC}\Big) \\
&\geq P\Big(\min_{j\in\mathcal{M}_{1*}} u_j^{RC} - \max_{j\in\mathcal{M}_{1*}} |\widehat{u}_j^{RC} - u_j^{RC}| > \varsigma_n\Big) \geq 1 - P\Big(\max_{j\in\mathcal{M}_{1*}} |\widehat{u}_j^{RC} - u_j^{RC}| \geq \varsigma_n\Big) \\
&\geq 1 - 36 s_n \exp(-c_{16} n^{1-2\kappa}).
\end{aligned}$$

Next, prove the assertion (ii). By the definition of $u_j^{RC}$ and condition (D2), we have

$$\begin{aligned}
\sum_{j=1}^{p_n} I(u_j^{RC} \geq \delta n^{-\kappa}) &\leq \sum_{j=1}^{p_n} \delta^{-1} n^\kappa u_j^{RC} \leq (\delta K_1 K_2)^{-1} n^\kappa \sum_{j=1}^{p_n} E\{\pi_j(Y, X_j)^2\} \\
&= O(n^\kappa E\|\boldsymbol{\pi}(Y, \mathbf{X})\|^2) = O(n^{\kappa+\iota}),
\end{aligned}$$

which means the size of $\{j : u_j^{RC} \geq \delta n^{-\kappa}\}$ cannot exceed $O(n^{\kappa+\iota})$ for any constant $\delta > 0$. Thus, on the event $\mathcal{A}_n = \{\max_{1\leq j\leq p_n} |\widehat{u}_j^{RC} - u_j^{RC}| \leq \delta n^{-\kappa}\}$, the size of $\{j : \widehat{u}_j^{RC} \geq 2\delta n^{-\kappa}\}$ cannot exceed the size of $\{j : u_j^{RC} \geq \delta n^{-\kappa}\}$, which is bounded by $O(n^{\kappa+\iota})$. Then, by taking $\delta = C$, there exists a positive constant $c_{17}$ such that

$$P(|\widehat{\mathcal{M}}_a| \leq O(n^{\kappa+\iota})) \geq P(\mathcal{A}_n) = 1 - 36 p_n \exp(-c_{17} n^{1-2\kappa})$$

Last, prove the assertion (iii). Since

$$
\begin{aligned}
|\widehat{\mathcal{M}}_a| &= \sum_{j \in \mathcal{M}_{1*}} I(\widehat{u}_j^{RC} \geq \varsigma_n) + \sum_{j \notin \mathcal{M}_{1*}} I(\widehat{u}_j^{RC} \geq \varsigma_n) \\
&\leq s_n + \sum_{j \notin \mathcal{M}_{1*}} I(\widehat{u}_j^{RC} \geq \varsigma_n),
\end{aligned}
$$

it follows that

$$
P(|\widehat{\mathcal{M}}_a| > s_n) \leq \sum_{j \notin \mathcal{M}_{1*}} P(\widehat{u}_j^{RC} \geq \varsigma_n) \leq 36(p_n - s_n)\exp(-\tilde{c}_2 n^{1-2\kappa}),
$$

for some constant $\tilde{c}_2 > 0$. This together with the part (i) immediately implies the part (iii). Thus, the proof is complete. $\square$

**Lemma A.5.** *Assume that the random variable $Y$ satisfies $P(Y = y_k) = p_k$, $k = 1, \ldots, J$. Let $F_Y(y)$ and $\widehat{F}_Y(y)$ be the CDF and empirical CDF of $Y$, respectively. Then, for every $\epsilon$,*

$$
P\big(\sup_{y \in \mathbb{R}} |\widehat{F}_Y(y) - F_Y(y)| > \epsilon\big) \leq 2J \exp\big(-2n\epsilon^2/J^2\big),
$$

**Proof of Lemma A.5** Let $\widehat{p}_k = \frac{1}{n}\sum_{i=1}^n I(Y_i = y_k)$. We note that $\sup_{y \in \mathbb{R}} |\widehat{F}_Y(y) - F_Y(y)| = \sup_{y \in \mathbb{R}} \big|\sum_{k=1}^J \widehat{p}_k I\{y_k \leq y\} - \sum_{k=1}^J p_k I y_k \leq y\big| \leq J \max_{1 \leq k \leq J} |\widehat{p}_k - p_k|$. Thus, by Lemma A.1, we have

$$
\begin{aligned}
P\big(\sup_{y \in \mathbb{R}} |\widehat{F}_Y(y) - F_Y(y)| > \epsilon\big) &\leq P\big(J \max_{1 \leq k \leq J} |\widehat{p}_k - p_k| > \epsilon\big) \\
&\leq \sum_{k=1}^J P\big(\big|\frac{1}{n}\sum_{i=1}^n [I(Y_i = y_k) - P(Y = y_k)]\big| > \epsilon/J\big) \\
&\leq 2J\exp(-2n\epsilon^2/J^2),
\end{aligned}
$$

which proves Lemma A.5. $\square$

**Lemma A.6.** *Assume that the random variable $Y$ satisfies $P(Y = y_k) = p_k$, $k = 1, \ldots, J$. Suppose that $\{(Y_i, X_i)_{i=1}^n\}$ are i.i.d. copies of $(Y, X)$ and let $\widehat{F}(y, x) = \frac{1}{n}\sum_{i=1}^n I(Y_i \leq y, X_i \leq x_j)$ be empirical joint CDF and $F(y, x)$ be the joint CDF of $(Y, X)$. Then there exists a uniform constant $\tilde{c}_1^*$ such that, for any $\epsilon$ such that $n\epsilon/J \to \infty$ and $\frac{n\epsilon^2}{J^2 \log \max(n, J)} \to \infty$ as $n \to \infty$,*

$$
P\Big(\sup_{(y,x) \in \mathcal{Y} \times \mathcal{X}} |\widehat{F}(y, x) - F(y, x)| > \epsilon\Big) \leq \exp(-\tilde{c}_1^* n\epsilon^2 J^{-2}).
$$

**Proof of Lemma A.6** Let $g_{ik}(u) \triangleq I(Y_i = y_k, U_i \leq u)$ and $U_i = F_X(X_i)$ and $u = F_X(x)$, $i = 1, \ldots, n$, where

$U_i \sim Uniform(0,1)$. First note that

$$\sup_{(y,x)\in\mathcal{Y}\times\mathcal{X}} |\widehat{F}(y,x) - F(y,x)| \leq J \max_{1\leq k\leq J} \sup_x \left|\frac{1}{n}\sum_{i=1}^n [I(Y_i = y_k, X_i \leq x) - P(Y = y_k, X \leq x)]\right|$$

$$= J \max_{1\leq k\leq J} \sup_{u\in(0,1)} \left|\frac{1}{n}\sum_{i=1}^n [g_{ik}(u) - Eg_{ik}(u)]\right|.$$

Denote $\{u_l, l = 0, 1, 2, \ldots, n\}$ by equally spaced grid points on the interval $(0,1)$. That is, $u_l = l/n$. Then,

$$\sup_{u\in(0,1)} \left|\frac{1}{n}\sum_{i=1}^n [g_{ik}(u) - Eg_{ik}(u)]\right| = \max_{1\leq l\leq n} \sup_{u\in(u_{l-1},u_l]} \left|\frac{1}{n}\sum_{i=1}^n [g_{ik}(u) - Eg_{ik}(u)]\right|$$

$$\leq \max_{1\leq l\leq n} \left|\frac{1}{n}\sum_{i=1}^n [g_{ik}(u_{l-1}) - Eg_{ik}(u_{l-1})]\right|$$

$$+ \max_{1\leq l\leq n} \sup_{u\in(u_{l-1},u_l]} \left|\frac{1}{n}\sum_{i=1}^n I(Y_i = y_k, u_{l-1} < U_i \leq u)\right|$$

$$+ \max_{1\leq l\leq n} \sup_{u\in(u_{l-1},u_l]} \left|\frac{1}{n}\sum_{i=1}^n [P(Y = y_k, U \leq u) - P(Y = y_k, U \leq u_{l-1})]\right|$$

$$\leq \max_{1\leq l\leq n} \left|\frac{1}{n}\sum_{i=1}^n [g_{ik}(u_{l-1}) - Eg_{ik}(u_{l-1})]\right|$$

$$+ \max_{1\leq l\leq n} \frac{1}{n}\sum_{i=1}^n I(u_{l-1} < U_i \leq u_{l-1} + \frac{1}{n})$$

$$+ \max_{1\leq l\leq n} \sup_{u\in(u_{l-1},u_l]} P(u_{l-1} < U \leq u_{l-1} + \frac{1}{n})$$

$$= \max_{1\leq l\leq n} \left|\frac{1}{n}\sum_{i=1}^n [g_{ik}(u_{l-1}) - Eg_{ik}(u_{l-1})]\right|$$

$$+ \max_{1\leq l\leq n} \frac{1}{n}\sum_{i=1}^n I(u_{l-1} < U_i \leq u_{l-1} + \frac{1}{n}) + \frac{1}{n}$$

Note that the condition that $n\epsilon/J \to \infty$ as $n \to \infty$ implies that we can pick $\epsilon > 0$ and sufficiently large $n$ such that $\frac{\epsilon}{2J} - \frac{1}{n} \geq \frac{\epsilon}{3J}$. It follows that

$$P\left(\sup_{(y,x)\in\mathcal{Y}\times\mathcal{X}} |\widehat{F}(y,x) - F(y,x)| > \epsilon\right) \leq P\left(\max_{1\leq k\leq J} \sup_{u\in(0,1)} \left|\frac{1}{n}\sum_{i=1}^n [g_{ik}(u) - Eg_{ik}(u)]\right| > \frac{\epsilon}{J}\right)$$

$$\leq P\left(\max_{1\leq k\leq J} \max_{1\leq l\leq n} \left|\frac{1}{n}\sum_{i=1}^n [g_{ik}(u_{l-1}) - Eg_{ik}(u_{l-1})]\right| > \frac{\epsilon}{2J}\right)$$

$$+ P\left(\max_{1\leq k\leq J} \max_{1\leq l\leq n} \frac{1}{n}\sum_{i=1}^n I(u_{l-1} < U_i \leq u_{l-1} + \frac{1}{n}) > \frac{\epsilon}{3J}\right)$$

$$\leq \sum_{k=1}^{J}\sum_{l=1}^{n} P\Big(\Big|\frac{1}{n}\sum_{i=1}^{n}[g_{ik}(u_{l-1}) - Eg_{ik}(u_{l-1})]\Big| > \frac{\epsilon}{2J}\Big)$$
$$+ \sum_{l=1}^{n} P\Big(\frac{1}{n}\sum_{i=1}^{n} I(u_{l-1} < U_i \leq u_{l-1} + \frac{1}{n}) > \frac{\epsilon}{3J}\Big)$$
$$=: \Delta_{n1} + \Delta_{n2} \text{ (say)}.$$

For $\Delta_{n1}$, since $0 \leq g_{ik}(u_{l-1}) \leq 1$, by using Lemma A.1, we have

$$\Delta_{n1} \leq Jn\exp(-cn\epsilon^2 J^{-2})$$
$$= \exp(-(cn\epsilon^2 J^{-2} - \log J - \log n))$$
$$\leq \exp(-cn\epsilon^2 J^{-2}/2)$$

for some constant $c > 0$ and all sufficiently large $n$. For $\Delta_{n2}$, we note that $EI(u_{l-1} < U_i \leq u_{l-1} + \frac{1}{n}) = F_U(u_{l-1} + \frac{1}{n}) - F_U(u_{l-1}) = \frac{1}{n}$. It follows that

$$\Delta_{n2} = \sum_{l=1}^{n} P\Big(\frac{1}{n}\sum_{i=1}^{n}[I(u_{l-1} < U_i \leq u_{l-1} + \frac{1}{n}) - EI(u_{l-1} < U_i \leq u_{l-1} + \frac{1}{n})] > \frac{\epsilon}{3J} - \frac{1}{n}\Big)$$
$$\leq \sum_{l=1}^{n} P\Big(\Big|\frac{1}{n}\sum_{i=1}^{n}[I(u_{l-1} < U_i \leq u_{l-1} + \frac{1}{n}) - EI(u_{l-1} < U_i \leq u_{l-1} + \frac{1}{n})]\Big| > \frac{\epsilon}{6J}\Big)$$
$$\leq n\exp(-c'n\epsilon^2 J^{-2}) \leq \exp(-c'n\epsilon^2 J^{-2}/2)$$

for some constant $c' > 0$ and all sufficiently large $n$. Hence, combining above arguments gives the desired result immediately. $\square$

**Proof of Theorem 3.4** Following the arguments in Theorem 3.2, and by Lemmas A.5 and A.6, we can obtain that there exist two uniform constants $c_{18}$ and $c_{19}$ such that

$$\max_{1 \leq j \leq p_n} P\Big(|\widehat{u}_j^{RC} - u_j^{RC}| > Cn^{-\kappa}\Big) \leq c_{18}\exp(-c_{19}n^{1-2\kappa}/J^2).$$

Similar to Theorem 3.3, we have

$$P(\mathcal{M}_{1*} \subset \widehat{\mathcal{M}}_a) \geq 1 - P\Big(\max_{j \in \mathcal{M}_{1*}} |\widehat{u}_j^{RC} - u_j^{RC}| \geq \varsigma_n\Big)$$
$$\geq 1 - c_{18}s_n\exp(-c_{19}n^{1-2\kappa}/J^2),$$

which results in part (i) if $\log p_n = o(n^{1-2\kappa}/J^2)$. Parts (ii) and (iii) can follow the arguments of Theorem

3.3 directly. $\square$

Below, in order to prove the sure screening property of our proposed RPC-SIS, we need to introduce some properties of B-spline approximation. For each $j = 1, \cdots, p_n$ and $k = 1, \ldots, L_n$, we have that (i) $B_k(z) \geq 0$ and $\sum_{k=1}^{L_n} B_k(z) = 1$ for $z \in \mathcal{Z}$; (ii) there exist positive constants $C_1, C_2$ such that for any $\mathbf{a} \in \mathbb{R}^{L_n}$,

$$\frac{C_1}{L_n}\|\mathbf{a}\|^2 \leq \int \mathbf{a}^T \mathbf{B}(z)(\mathbf{B}(z))^T \mathbf{a} dz \leq \frac{C_2}{L_n}\|\mathbf{a}\|^2.$$

Under property (ii), we further have that (iii) there exist positive constants $C_3$ and $C_4$ such that for $k = 1, \ldots, L_n$,

$$C_3 L_n^{-1} \leq E\{(B_k(Z))^2\} \leq C_4 L_n^{-1}, \tag{B.20}$$

where $C_3 = C_1 \bar{M}_1$ and $C_4 = C_2 \bar{M}_2$, and $\bar{M}_1$ and $\bar{M}_2$ are the constants given in condition (D5); and (iv)

$$C_3 L_n^{-1} \leq \lambda_{\min}\big(E\{\mathbf{B}(Z)\mathbf{B}(Z)^T\}\big) \leq \lambda_{\max}\big(E\{\mathbf{B}(Z)\mathbf{B}(Z)^T\}\big) \leq C_4 L_n^{-1}, \tag{B.21}$$

where $\lambda_{\min}(\cdot)$ and $\lambda_{\max}(\cdot)$ denotes the smallest and largest eigenvalues of a symmetric matrix, respectively.

**Lemma A.7.** *Let $m_0(z) = E(Y|Z = z)$ and $m_j(z) = E(X_j|Z = z), j = 1, \ldots, p_n$, and assume that conditions (D4)-(D6) hold. For every $0 \leq j \leq p_n$, then there exist two uniform constants $\tilde{c}_2$ and $\tilde{c}_3$ such that for any $\epsilon > 0$ such that $n\epsilon^2 L_n^{-4}/\log L_n \to \infty$ and $L_n^{d-1/2}\epsilon \to \infty$ as $n \to \infty$,*

$$P\Big(\sup_{z \in \mathcal{Z}} |\widehat{m}_j(z) - m_j(z)| > \epsilon\Big) \leq \tilde{c}_2 \exp(-\tilde{c}_3 n L_n^{-4} \epsilon^2).$$

**Proof of Lemma A.7** We only present the proof details for $m_j(z), j > 0$. Let $\boldsymbol{\gamma}_j^M = \operatorname{argmin}_{\boldsymbol{\gamma}} E\{(X_j - \mathbf{B}(Z)^T \boldsymbol{\gamma}_j)^2\}$. Then $\boldsymbol{\gamma}_j^M = [E\{\mathbf{B}(Z)\mathbf{B}(Z)^T\}]^{-1} E\{\mathbf{B}(Z)X_j\}$. By the B-spline theory (de Boor (2001)), there exists an approximation $\mathbf{B}(z)^T \boldsymbol{\gamma}_j^*$ in $\mathcal{B}$ (see condition (D4)) to $m_j(z)$ such that $m_j(z) = \mathbf{B}(z)^T \boldsymbol{\gamma}_j^* + \eta_j(z)$, where $\eta_j(z)$ denotes the approximation error fulfilling $\max_j \sup_{z \in \mathcal{Z}} |\eta_j(z)| \leq C_1 L_n^{-d}$ for some positive constant $C_1$. It follows that

$$\begin{aligned}
\sup_{z \in \mathcal{Z}} |\widehat{m}_j(z) - m_j(z)| &= \sup_{z \in \mathcal{Z}} |\mathbf{B}(z)^T(\widehat{\boldsymbol{\gamma}}_j - \boldsymbol{\gamma}_j^*) - \eta_j(z)| \\
&\leq \|\widehat{\boldsymbol{\gamma}}_j - \boldsymbol{\gamma}_j^*\| + \sup_{z \in \mathcal{Z}} |\eta_j(z)| \\
&\leq \|\widehat{\boldsymbol{\gamma}}_j - \boldsymbol{\gamma}_j^M\| + \|\boldsymbol{\gamma}_j^M - \boldsymbol{\gamma}_j^*\| + C_1 L_n^{-d},
\end{aligned} \tag{B.22}$$

where the second line is due to the fact that $\|\mathbf{B}(z)\| \leq 1$ by the properties of the above B-spline. We first consider the second term in equality (B.22). Using (B.21), we have

$$
\begin{aligned}
\|\boldsymbol{\gamma}_j^M - \boldsymbol{\gamma}_j^*\| &= \|[E\{\mathbf{B}(Z)\mathbf{B}(Z)^T\}]^{-1}E\{\mathbf{B}(Z)X_j\} - \boldsymbol{\gamma}_j^*\| \\
&= \|[E\{\mathbf{B}(Z)\mathbf{B}(Z)^T\}]^{-1}E\{\mathbf{B}(Z)m_j(Z)\} - \boldsymbol{\gamma}_j^*\| \\
&\leq \|[E\{\mathbf{B}(Z)\mathbf{B}(Z)^T\}]^{-1}E\{\mathbf{B}(Z)\eta_j(Z)\}\| \\
&\leq C_3^{-1}L_n C_1 L_n^{-d}\|E\mathbf{B}(Z)\| \\
&\leq C_1 C_3^{-1} L_n^{\frac{1}{2}-d}.
\end{aligned}
\tag{B.23}
$$

Next, we consider the first term in equality (B.22). Denote $\mathbf{D}_n = \frac{1}{n}\sum_{i=1}^n \mathbf{B}(Z_i)\mathbf{B}(Z_i)^T$, $\mathbf{H}_{nj} = \frac{1}{n}\sum_{i=1}^n \mathbf{B}(Z_i)X_{ij}$, $\mathbf{D} = E\{\mathbf{B}(Z)\mathbf{B}(Z)^T\}$ and $\mathbf{H}_j = E\{\mathbf{B}(Z)X_j\}$. Then, it follows that

$$
\begin{aligned}
\widehat{\boldsymbol{\gamma}}_j - \boldsymbol{\gamma}_j^M &= \mathbf{D}_n^{-1}\mathbf{H}_{nj} - \mathbf{D}^{-1}\mathbf{H}_j \\
&= \mathbf{D}_n^{-1}(\mathbf{H}_{nj} - \mathbf{H}_j) + \mathbf{D}_n^{-1}(\mathbf{D} - \mathbf{D}_n)\mathbf{D}^{-1}\mathbf{H}_j \\
&\triangleq I_{nj1} + I_{nj2} \text{ (say)}.
\end{aligned}
$$

It suffices to bound two terms on the right-hand side of the above equation. On the other hand, according to Lemma 7 of Fan, Ma and Dai (2014), we have that for any positive constant $C_5$, there exists a positive constant $C_6$ such that

$$
P\Big(\|\mathbf{D}_n^{-1}\| \geq (1+C_5)\|\mathbf{D}^{-1}\|\Big) \leq 6L_n^2 \exp(-C_6 L_n^{-3}n),
\tag{B.24}
$$

where $\|\cdot\|$ denotes the operator norm, i.e., $\|\mathbf{A}\| = \sqrt{\lambda_{\max}(\mathbf{A}^T\mathbf{A})}$ for any matrix $\mathbf{A}$; and that for any $\delta > 0$, there exist positive constants $C_7$ and $C_8$ such that

$$
P\Big(\|\mathbf{D}_n - \mathbf{D}\| \geq \frac{L_n\delta}{n}\Big) \leq 6L_n^2 \exp\Big(-\frac{\delta^2}{C_7 n L_n^{-1} + C_8\delta}\Big).
\tag{B.25}
$$

Inequality (B.24) with inequality (B.21) implies

$$
P\Big(\|\mathbf{D}_n^{-1}\| \geq (1+C_5)C_3^{-1}L_n\Big) \leq 6L_n^2 \exp(-C_6 L_n^{-3}n).
\tag{B.26}
$$

Let $\xi_{ijk} = B_k(Z_i)X_{ij} - E\{B_k(Z_i)X_{ij}\}$. Then, we are going to establish the exponential bound for the tail probability of $\frac{1}{n}\sum_{i=1}^{n}\xi_{ijk}$, the $k$th component of $\mathbf{H}_{nj} - \mathbf{H}_j$. To this end, we first note that condition (D6) implies that for any integer $s \geq 2$, $E(|X_j|^s|Z) = \int_0^\infty P(|X_j| > x^{1/s}|Z)\mathrm{d}x \leq \bar{K}_1 \int_0^\infty \exp(-\bar{K}_1^{-1} x^{1/s}\mathrm{d}x) \leq \bar{K}_1^{s+1} s!$ by change of variables. This in conjunction with (B.20) gives that $E(\xi_{ijk}) = 0$ and for any integer $s \geq 2$,

$$
\begin{aligned}
E|\xi_{ijk}|^s &\leq 2^s E\{|B_k(Z_i)X_{ij}|^s\} = 2^s E\{|B_k(Z_i)|^s E(|X_{ij}|^s|Z)\} \\
&\leq 2^s \bar{K}_1^{s+1} s! E\{|B_k(Z_i)|^2\} \\
&\leq C_4 \bar{K}_1 (2\bar{K}_1)^s s!/L_n \\
&= \frac{8 C_4 \bar{K}_1^3}{2 L_n} (2\bar{K}_1)^{s-2} s!.
\end{aligned}
$$

An application of Lemma A.2 yields that for any $\delta > 0$,

$$P\Big(\Big|\frac{1}{n}\sum_{i=1}^{n}\xi_{ijk}\Big| > \frac{\delta}{n}\Big) \leq 2\exp\Big(-\frac{\delta^2}{16 C_4 \bar{K}_1^3 n L_n^{-1} + 4\bar{K}_1\delta}\Big) \tag{B.27}$$

for sufficiently large $n$. Thus, using (B.27), we have

$$
\begin{aligned}
P\Big(\|\mathbf{H}_{nj} - \mathbf{H}_j\| > \frac{L_n^{1/2}\delta}{n}\Big) &\leq \sum_{k=1}^{L_n} P\Big(\Big|\frac{1}{n}\sum_{i=1}^{n}\xi_{ijk}\Big| > \frac{\delta}{n}\Big) \\
&\leq 2 L_n \exp\Big(-\frac{\delta^2}{16 C_4 \bar{K}_1^3 n L_n^{-1} + 4\bar{K}_1\delta}\Big).
\end{aligned}
$$

This together with (B.26) gives

$$
\begin{aligned}
&P\Big(\|I_{nj1}\| > \frac{(1+C_5)C_3^{-1} L_n^{3/2}\delta}{n}\Big) \\
&\leq P\Big(\|\mathbf{D}_n^{-1}(\mathbf{H}_{nj} - \mathbf{H}_j)\| > \frac{(1+C_5)C_3^{-1} L_n^{3/2}\delta}{n}, \|\mathbf{D}_n^{-1}\| < (1+C_5)C_3^{-1} L_n\Big) \\
&\quad + P\Big(\|\mathbf{D}_n^{-1}\| \geq (1+C_5)C_3^{-1} L_n\Big) \\
&\leq P\Big(\|\mathbf{H}_{nj} - \mathbf{H}_j\| > \frac{L_n^{1/2}\delta}{n}\Big) + P\Big(\|\mathbf{D}_n^{-1}\| \geq (1+C_5)C_3^{-1} L_n\Big) \\
&\leq 2 L_n \exp\Big(-\frac{\delta^2}{16 C_4 \bar{K}_1^3 n L_n^{-1} + 4\bar{K}_1\delta}\Big) + 6 L_n^2 \exp(-C_6 L_n^{-3} n).
\end{aligned}
\tag{B.28}
$$

It remains to consider $I_{nj2}$. Since $E\{B_k(Z)\} \leq C_0 L_n^{-1}$ for some positive constant $C_0$, we have $|E\{B_k(Z)X_j\}| \leq$

$E\{B_k(Z)[E(|X_j|^2|Z)]^{1/2}\} \leq (2\bar{K}_1^3)^{1/2} E\{B_k(Z)\} \leq C_7 L_n^{-1}$, where $C_7 = C_0(2\bar{K}_1^3)^{1/2}$. Thus,

$$\|\mathbf{H}_j\| = \Big(\sum_{k=1}^{L_n} \big[E\{B_k(Z)X_j\}\big]^2\Big)^{1/2} \leq C_7 L_n^{-1/2}. \tag{B.29}$$

Combining the results in (B.21), (B.25), (B.26) and (B.29), we can obtain, for any $\delta > 0$,

$$\begin{aligned}
P\Big(\|I_{nj2}\| &> \frac{(1+C_5)C_3^{-2}C_7 L_n^{5/2}\delta}{n}\Big) \\
&\leq P\Big(\|\mathbf{D}_n^{-1}\|\|\mathbf{D}-\mathbf{D}_n)\| > \frac{(1+C_5)C_3^{-1}L_n^2\delta}{n}\Big) \\
&\leq P\Big(\|\mathbf{D}_n^{-1}\| \geq (1+C_5)C_3^{-1}L_n\Big) + P\Big(\|\mathbf{D}_n - \mathbf{D}\| \geq \frac{L_n \delta}{n}\Big) \\
&\leq 6L_n^2 \exp(-C_6 L_n^{-3} n) + 6L_n^2 \exp\Big(-\frac{\delta^2}{C_7 n L_n^{-1} + C_8 \delta}\Big).
\end{aligned} \tag{B.30}$$

Therefore, putting (B.28) and (B.30) together gives

$$\begin{aligned}
P\Big(\|\widehat{\boldsymbol{\gamma}}_j - \boldsymbol{\gamma}_j^M\| &\geq \frac{(1+C_5)C_3^{-1}L_n^{3/2}\delta + (1+C_5)C_3^{-2}C_7 L_n^{5/2}\delta}{n}\Big) \\
&\leq P\Big(\|I_{nj1}\| > \frac{(1+C_5)C_3^{-1}L_n^{3/2}\delta}{n}\Big) + P\Big(\|I_{nj2}\| > \frac{(1+C_5)C_3^{-2}C_7 L_n^{5/2}\delta}{n}\Big) \\
&\leq 12 L_n^2 \exp(-C_6 L_n^{-3} n) + 6 L_n^2 \exp\Big(-\frac{\delta^2}{C_7 n L_n^{-1} + C_8 \delta}\Big) \\
&\quad + 2 L_n \exp\Big(-\frac{\delta^2}{16 C_4 \bar{K}_1^3 n L_n^{-1} + 4\bar{K}_1 \delta}\Big).
\end{aligned} \tag{B.31}$$

Pick an $\epsilon$ such that $\frac{(1+C_5)C_3^{-1}L_n^{3/2}\delta + (1+C_5)C_3^{-2}C_7 L_n^{5/2}\delta}{n} = \frac{\epsilon}{2}$ and $C_1 L_n^{-d} + C_1 C_3^{-1} L_n^{\frac{1}{2}-d} \leq \frac{\epsilon}{2}$. Then, by invoking (B.22), (B.23) and (B.31), we can conclude the desired result. $\square$

**Proof of Theorem 3.5** We follow the proofs of Theorems 3.2 and 3.3 to show this theorem. Recall that $\varepsilon_{i0} = Y_i - E(Y|Z_i)$, $\varepsilon_{ij} = X_{ij} - E(X_{ij}|Z_i)$, $\widehat{\varepsilon}_{i0} = Y_i - \widehat{E}(Y|Z_i)$, and $\widehat{\varepsilon}_{ij} = X_{ij} - \widehat{E}(X_{ij}|Z_i)$. Thus, $\varepsilon_0 = \varepsilon_{Y|Z} = Y - E(Y|Z)$ and $\varepsilon_j = \varepsilon_{X_j|Z} = X_j - E(X_j|Z)$. Also, note that $\widehat{u}_j^{RPC} = \frac{1}{n}\sum_{i=1}^n \widehat{\varrho}_j^2(\widehat{\varepsilon}_{i0}, \widehat{\varepsilon}_{ij})$ and $u_j^{RPC} = E\{\varrho_j^2(\varepsilon_0, \varepsilon_j)\}$. Then, it follows that

$$\max_{1 \leq j \leq p_n} P(|\widehat{u}_j^{RPC} - u_j^{RPC}| > \epsilon) \leq \sum_{\ell=1}^{3} \max_{1 \leq j \leq p_n} P(|\bar{Q}_{nj\ell}| > \epsilon/3), \tag{B.32}$$

where

$$\bar{Q}_{nj1} = \frac{1}{n}\sum_{i=1}^{n}[\widehat{\varrho}_j^2(\widehat{\varepsilon}_{i0}, \widehat{\varepsilon}_{ij}) - \varrho_j^2(\widehat{\varepsilon}_{i0}, \widehat{\varepsilon}_{ij})],$$

$$\bar{Q}_{nj2} = \frac{1}{n}\sum_{i=1}^{n}[\varrho_j^2(\widehat{\varepsilon}_{i0},\widehat{\varepsilon}_{ij}) - \varrho_j^2(\varepsilon_{i0},\varepsilon_{ij})]$$

and

$$\bar{Q}_{nj3} = \frac{1}{n}\sum_{i=1}^{n}[\varrho_j^2(\varepsilon_{i0},\varepsilon_{ij}) - E\{\varrho_j^2(\varepsilon_0,\varepsilon_j)\}].$$

It is sufficient to find an exponential bound for each of three probabilities on the right-hand side of inequality (B.32). To this end, the rest of the proof consists of three steps.

**Step 1**. Derive an exponential bound for $\max_{1\leq j\leq p_n} P(|\bar{Q}_{nj1}| > \epsilon/3)$. For this, we need to bound the following three tail probabilities: for any $\epsilon > 0$,

$$\max_{1\leq j\leq p_n}\sup_{(u,v)\in\mathcal{E}_0\times\mathcal{E}_j} P\Big(\big|\widehat{F}_{\widehat{\varepsilon}_0,\widehat{\varepsilon}_j}(u,v) - F_{\varepsilon_0,\varepsilon_j}(u,v)\big| > \epsilon\Big), \tag{B.33}$$

$$\sup_{u\in\mathcal{E}_0} P\Big(\big|\widehat{F}_{\widehat{\varepsilon}_0}(u) - F_{\varepsilon_0}(u)\big| > \epsilon\Big) \tag{B.34}$$

and

$$\max_{1\leq j\leq p_n}\sup_{v\in\mathcal{E}_j} P\Big(\big|\widehat{F}_{\widehat{\varepsilon}_j}(v) - F_{\varepsilon_j}(v)\big| > \epsilon\Big), \tag{B.35}$$

which are the key to the entire proof. We only provide the details in deriving the exponential bound for the first tail probability since the other two terms can be dealt with similarly. To this end, we note that

$$\max_{1\leq j\leq p_n}\sup_{(u,v)\in\mathcal{E}_0\times\mathcal{E}_j} P\Big(\big|\widehat{F}_{\widehat{\varepsilon}_0,\widehat{\varepsilon}_j}(u,v) - F_{\varepsilon_0,\varepsilon_j}(u,v)\big| > \epsilon\Big)$$

$$\leq \max_{1\leq j\leq p_n} P\Big(\sup_{(u,v)\in\mathcal{E}_0\times\mathcal{E}_j}\big|\widehat{F}_{\widehat{\varepsilon}_0,\widehat{\varepsilon}_j}(u,v) - F_{\varepsilon_0,\varepsilon_j}(u,v)\big| > \epsilon\Big)$$

$$\leq \max_{1\leq j\leq p_n} P\Big(\sup_{(u,v)\in\mathcal{E}_0\times\mathcal{E}_j}\big|\widehat{F}_{\widehat{\varepsilon}_0,\widehat{\varepsilon}_j}(u,v) - \widehat{F}_{\varepsilon_0,\varepsilon_j}(u,v)\big| > \frac{\epsilon}{2}\Big)$$

$$+ \max_{1\leq j\leq p_n} P\Big(\sup_{(u,v)\in\mathcal{E}_0\times\mathcal{E}_j}\big|\widehat{F}_{\varepsilon_0,\varepsilon_j}(u,v) - F_{\varepsilon_0,\varepsilon_j}(u,v)\big| > \frac{\epsilon}{2}\Big)$$

$$\triangleq \Delta_{n1} + \Delta_{n2} \text{ (say)}.$$

For the term $\Delta_{n1}$, since

$$\sup_{(u,v)\in\mathcal{E}_0\times\mathcal{E}_j}\big|\widehat{F}_{\widehat{\varepsilon}_0,\widehat{\varepsilon}_j}(u,v) - \widehat{F}_{\varepsilon_0,\varepsilon_j}(u,v)\big|$$

$$
\begin{aligned}
&\leq \sup_{(u,v)\in\mathcal{E}_0\times\mathcal{E}_j} \Big|\frac{1}{n}\sum_{i=1}^n \{I(\widehat{\varepsilon}_{i0}\leq u) - I(\varepsilon_{i0}\leq u)\}I(\widehat{\varepsilon}_{ij}\leq v)\Big| \\
&\quad + \sup_{(u,v)\in\mathcal{E}_0\times\mathcal{E}_j} \Big|\frac{1}{n}\sum_{i=1}^n \{I(\widehat{\varepsilon}_{ij}\leq v) - I(\varepsilon_{ij}\leq v)\}I(\varepsilon_{i0}\leq u)\Big| \\
&\leq \sup_{u\in\mathcal{E}_0}\frac{1}{n}\sum_{i=1}^n \big|I(\widehat{\varepsilon}_{i0}\leq u) - I(\varepsilon_{i0}\leq u)\big| + \sup_{v\in\mathcal{E}_j}\frac{1}{n}\sum_{i=1}^n \big|I(\widehat{\varepsilon}_{ij}\leq v) - I(\varepsilon_{ij}\leq v)\big| \\
&\leq \frac{1}{n}\sum_{i=1}^n \big|I(\widehat{\varepsilon}_{i0}\leq u^*) - I(\varepsilon_{i0}\leq u^*)\big| + \frac{1}{n}\sum_{i=1}^n \big|I(\widehat{\varepsilon}_{ij}\leq v^*) - I(\varepsilon_{ij}\leq v^*)\big|
\end{aligned}
$$

for some $u^*\in\mathcal{E}_0$ and $v^*\in\mathcal{E}_j$, so we can obtain

$$
\begin{aligned}
\Delta_{n1} &\leq P\Big(\frac{1}{n}\sum_{i=1}^n \big|I(\widehat{\varepsilon}_{i0}\leq u^*) - I(\varepsilon_{i0}\leq u^*)\big| > \frac{\epsilon}{4}\Big) \\
&\quad + \max_{1\leq j\leq p_n} P\Big(\frac{1}{n}\sum_{i=1}^n \big|I(\widehat{\varepsilon}_{ij}\leq v^*) - I(\varepsilon_{ij}\leq v^*)\big| > \frac{\epsilon}{4}\Big) \\
&\triangleq \Delta_{n1}^{(1)} + \Delta_{n1}^{(2)}.
\end{aligned}
\tag{B.36}
$$

For the first term on the right-hand side of (B.36), it can be easily seen that $\widehat{\varepsilon}_{i0} - \varepsilon_{i0} = E(Y|Z_i) - \widehat{E}(Y|Z_i) = m_0(Z_i) - \widehat{m}_0(Z_i)$. Define an event $\mathcal{H}_{n0} = \{\sup_{z\in\mathcal{Z}}|\widehat{m}_0(z) - m_0(z)|\leq \delta\}$ for some small $\delta > 0$. It follows that

$$
\begin{aligned}
\Delta_{n1}^{(1)} &= P\Big(\frac{1}{n}\sum_{i=1}^n \big|I(\widehat{\varepsilon}_{i0}\leq u^*) - I(\varepsilon_{i0}\leq u^*)\big| > \frac{\epsilon}{4}, \mathcal{H}_{n0}\Big) \\
&\quad + P\Big(\frac{1}{n}\sum_{i=1}^n \big|I(\widehat{\varepsilon}_{i0}\leq u^*) - I(\varepsilon_{i0}\leq u^*)\big| > \frac{\epsilon}{4}, \mathcal{H}_{n0}^c\Big) \\
&\leq P\Big(\frac{1}{n}\sum_{i=1}^n I(u^* - \delta \leq \varepsilon_{i0}\leq u^* + \delta) > \frac{\epsilon}{4}\Big) \\
&\quad + P\Big(\sup_{z\in\mathcal{Z}}|\widehat{m}_0(z) - m_0(z)| > \delta\Big).
\end{aligned}
\tag{B.37}
$$

Let $\varphi_{ij} = I(u^* - \delta \leq \varepsilon_{i0}\leq u^* + \delta)$. Then, by mean value theorem and condition (D7), we have $E(\varphi_{ij}) = F_{\varepsilon_0}(u^* + \delta) - F_{\varepsilon_0}(u^* - \delta) = 2f_{\varepsilon_0}(u^{**})\delta \leq 2\delta\sup_u f_{\varepsilon_0}(u) \leq 2\delta M_{f_\varepsilon}$ for some $u^{**}$ between $u^* - \delta$ and $u^* + \delta$. By Lemma A.1, there exists a positive constant $C_9$ such that

$$
\begin{aligned}
&P\Big(\frac{1}{n}\sum_{i=1}^n I(u^* - \delta \leq \varepsilon_{i0}\leq u^* + \delta) > \frac{\epsilon}{4}\Big) \\
&\leq P\Big(\frac{1}{n}\sum_{i=1}^n [\varphi_{ij} - E(\varphi_{ij})] > \frac{\epsilon}{4} - 2\delta M_{f_\varepsilon}\Big)
\end{aligned}
$$

$$\leq P\Big(\Big|\frac{1}{n}\sum_{i=1}^{n}[\varphi_{ij} - E(\varphi_{ij})]\Big| > \frac{\epsilon}{8}\Big)$$

$$\leq 2\exp(-C_9 n\epsilon^2)$$

for sufficiently large $n$, provided that $\frac{\epsilon}{8} \geq 2\delta M_{f_\varepsilon}$. This in conjunction with inequality (B.37) and Lemma A.7 leads to that

$$\Delta_{n1}^{(1)} \leq 2\exp(-C_9 n\epsilon^2) + \tilde{c}_2 \exp(-\tilde{c}_3 n L_n^{-4}\delta^2) \tag{B.38}$$

for sufficiently large $n$, provided that $\frac{\epsilon}{8} \geq 2\delta M_{f_\varepsilon}$ and $n\delta^2 L_n^{-4}/\log L_n \to \infty$ and $L_n^{d-1/2}\delta \to \infty$ as $n \to \infty$. Taking $\delta = \frac{\epsilon}{16 M_{f_\varepsilon}}$, it follows from the inequality (B.38) that there exist two positive constants $\tilde{c}_4$ and $\tilde{c}_5$ such that

$$\begin{aligned}\Delta_{n1}^{(1)} &\leq 2\exp(-C_9 n\epsilon^2) + \tilde{c}_2 \exp(-\tilde{c}_3 n L_n^{-4}\epsilon^2/(256 M_{f_\varepsilon}^2)) \\ &\leq \tilde{c}_4 \exp(-\tilde{c}_5 n L_n^{-4}\epsilon^2),\end{aligned} \tag{B.39}$$

provided $n\epsilon^2 L_n^{-4}/\log L_n \to \infty$ and $L_n^{d-1/2}\epsilon \to \infty$ as $n \to \infty$, where the last line is because $nL_n^{-4}\epsilon^2 = o(n\epsilon^2)$. For $\Delta_{n1}^{(2)}$, using similar arguments as above, we can obtain

$$\Delta_{n1}^{(2)} \leq \tilde{c}_6 \exp(-\tilde{c}_7 n L_n^{-4}\epsilon^2), \tag{B.40}$$

for some positive constants $\tilde{c}_6$ and $\tilde{c}_7$ and sufficiently large $n$. As a result, combining (B.39) and (B.40) gives

$$\Delta_{n1} \leq (\tilde{c}_4 + \tilde{c}_6)\exp(-\min(\tilde{c}_5, \tilde{c}_7) n L_n^{-4}\epsilon^2), \tag{B.41}$$

On the other hand, following the proof of Lemma A.4 and by condition (D7), we can obtain that there exists a positive constant $\tilde{c}_8$ such that

$$\Delta_{n2} \leq 6\exp(-\tilde{c}_8 n\epsilon^2), \tag{B.42}$$

provided that $n\epsilon \to \infty$ and $n\epsilon^2/\log n \to \infty$ as $n \to \infty$. Hence, it follows from (B.41) and (B.42) that

$$\max_{1\leq j\leq p_n} \sup_{(u,v)\in \mathcal{E}_0 \times \mathcal{E}_j} P\Big(\big|\widehat{F}_{\widehat{\varepsilon}_0,\widehat{\varepsilon}_j}(u,v) - F_{\varepsilon_0,\varepsilon_j}(u,v)\big| > \epsilon\Big)$$

$$\leq (\tilde{c}_4 + \tilde{c}_6) \exp(-\min(\tilde{c}_5, \tilde{c}_7) n L_n^{-4} \epsilon^2) + 6 \exp(-\tilde{c}_8 n \epsilon^2)$$

$$\leq \tilde{c}_9 \exp(-\tilde{c}_{10} n L_n^{-4} \epsilon^2) \tag{B.43}$$

for some positive constants $\tilde{c}_9$ and $\tilde{c}_{10}$ and sufficiently large $n$.

Analogously, using the above arguments, we can derive that

$$\sup_{u \in \mathcal{E}_0} P\Big(\big|\widehat{F}_{\widehat{\varepsilon}_0}(u) - F_{\varepsilon_0}(u)\big| > \epsilon\Big) \leq \tilde{c}_{11} \exp(-\tilde{c}_{12} n L_n^{-4} \epsilon^2) \tag{B.44}$$

for some positive constants $\tilde{c}_{11}$ and $\tilde{c}_{12}$, and that

$$\max_{1 \leq j \leq p_n} \sup_{v \in \mathcal{E}_j} P\Big(\big|\widehat{F}_{\widehat{\varepsilon}_j}(v) - F_{\varepsilon_j}(v)\big| > \epsilon\Big) \leq \tilde{c}_{13} \exp(-\tilde{c}_{14} n L_n^{-4} \epsilon^2). \tag{B.45}$$

for some positive constants $\tilde{c}_{13}$ and $\tilde{c}_{14}$.

Then, following the same arguments in the proof of Theorem 3.2 and using (B.43)-(B.45), by a tedious derivation, we can prove that under the conditions of Theorem 3.4, there exist two positive constants $\tilde{c}_{15}$ and $\tilde{c}_{16}$ such that

$$\max_{1 \leq j \leq p_n} \sup_{(u,v) \in \mathcal{E}_0 \times \mathcal{E}_j} P(|\widehat{\varrho}_j^2(u,v) - \varrho_j^2(u,v)| > \epsilon) \leq \tilde{c}_{15} \exp(-\tilde{c}_{16} n L_n^{-4} \epsilon^2) \tag{B.46}$$

for sufficiently large $n$. Hence, it follows from (B.18) that

$$\begin{aligned}
\max_{1 \leq j \leq p_n} P(|\bar{Q}_{nj1}| > \epsilon/3) &\leq \max_{1 \leq j \leq p_n} P\Big(\frac{1}{n}\sum_{i=1}^n \big|\widehat{\varrho}_j^2(\widehat{\varepsilon}_{i0}, \widehat{\varepsilon}_{ij}) - \rho_j^2(\widehat{\varepsilon}_{i0}, \widehat{\varepsilon}_{ij})\big| > \epsilon/3\Big) \\
&\leq \max_{1 \leq j \leq p_n} \sum_{i=1}^n P\Big(\big|\widehat{\varrho}_j^2(\widehat{\varepsilon}_{i0}, \widehat{\varepsilon}_{ij}) - \rho_j^2(\widehat{\varepsilon}_{i0}, \widehat{\varepsilon}_{ij})\big| > \epsilon/3\Big) \\
&\leq n \max_{1 \leq j \leq p_n} \sup_{(u,v) \in \mathcal{E}_0 \times \mathcal{E}_j} P\Big(\big|\widehat{\varrho}_j^2(u,v) - \rho_j^2(u,v)\big| > \epsilon/3\Big) \\
&\leq \tilde{c}_{15} n \exp(-\tilde{c}_{16} n L_n^{-4} \epsilon^2/9) \\
&= \tilde{c}_{15} \exp(-(\tilde{c}_{16} n L_n^{-4} \epsilon^2/9 - \log n)) \\
&\leq \tilde{c}_{15} \exp(-\tilde{c}_{17} n L_n^{-4} \epsilon^2)
\end{aligned}$$

for some positive constant $\tilde{c}_{17}$ and sufficiently large $n$.

**Step 2**. Derive an exponential bound for $\max_{1 \leq j \leq p_n} P(|\bar{Q}_{nj2}| > \epsilon/3)$. Let $S_{j1}(u,v) = \{F_{\varepsilon_0, \varepsilon_j}(u,v) - F_{\varepsilon_0}(u) F_{\varepsilon_j}(v)\}^2$ and $S_{j2}(u,v) = \{F_{\varepsilon_0}(u) - F_{\varepsilon_0}^2(u)\}\{F_{\varepsilon_j}(v) - F_{\varepsilon_j}^2(v)\}$. Then, $\varrho_j^2(\widehat{\varepsilon}_{i0}, \widehat{\varepsilon}_{ij}) = S_{j1}(\widehat{\varepsilon}_{i0}, \widehat{\varepsilon}_{ij})/S_{j2}(\widehat{\varepsilon}_{i0}, \widehat{\varepsilon}_{ij})$ and $\varrho_j^2(\varepsilon_{i0}, \varepsilon_{ij}) = S_{j1}(\varepsilon_{i0}, \varepsilon_{ij})/S_{j2}(\varepsilon_{i0}, \varepsilon_{ij})$. Obviously, it holds that $K_3 K_4 \leq \min_j \inf_{u,v} S_{j2}(u,v) \leq \max_j \sup_{u,v} S_{j2}(u,v) \leq 1/16$ by condition (D8). Consequently, it follows that

$$
\begin{aligned}
P(|\bar{Q}_{nj2}| > \epsilon/3) &\leq P\Big(\frac{1}{n}\sum_{i=1}^{n} |\varrho_j^2(\widehat{\varepsilon}_{i0}, \widehat{\varepsilon}_{ij}) - \varrho_j^2(\varepsilon_{i0}, \varepsilon_{ij})| > \frac{\epsilon}{3}\Big) \\
&\leq \sum_{i=1}^{n} P\Big(|\varrho_j^2(\widehat{\varepsilon}_{i0}, \widehat{\varepsilon}_{ij}) - \varrho_j^2(\varepsilon_{i0}, \varepsilon_{ij})| > \frac{\epsilon}{3}\Big) \\
&\leq \sum_{i=1}^{n} P\Big(|S_{j1}(\widehat{\varepsilon}_{i0}, \widehat{\varepsilon}_{ij}) - S_{j1}(\varepsilon_{i0}, \varepsilon_{ij})| > \frac{8K_3^2 K_4^2 \epsilon}{3}\Big) \\
&\quad + \sum_{i=1}^{n} P\Big(|S_{j2}(\widehat{\varepsilon}_{i0}, \widehat{\varepsilon}_{ij}) - S_{j2}(\varepsilon_{i0}, \varepsilon_{ij})| > \frac{8K_3^2 K_4^2 \epsilon}{3}\Big) \\
&\triangleq \Lambda_{j1} + \Lambda_{j2} \text{ (say)}.
\end{aligned}
$$

For the first term $\Lambda_{j1}$, it can be shown that

$$
\begin{aligned}
|S_{j1}(\widehat{\varepsilon}_{i0}, \widehat{\varepsilon}_{ij}) - S_{j1}(\varepsilon_{i0}, \varepsilon_{ij})| &\leq 4|F_{\varepsilon_0, \varepsilon_j}(\widehat{\varepsilon}_{i0}, \widehat{\varepsilon}_{ij}) - F_{\varepsilon_0, \varepsilon_j}(\varepsilon_{i0}, \varepsilon_{ij})| \\
&\quad + 4|F_{\varepsilon_0}(\widehat{\varepsilon}_{i0}) - F_{\varepsilon_0}(\varepsilon_{i0})| + 4|F_{\varepsilon_j}(\widehat{\varepsilon}_{ij}) - F_{\varepsilon_j}(\varepsilon_{ij})| \\
&\leq 4M_{F_1}|\widehat{\varepsilon}_{i0} - \varepsilon_{i0}| + 4M_{F_2}|\widehat{\varepsilon}_{ij} - \varepsilon_{ij}| \\
&\quad + 4M_{f_\varepsilon}|\widehat{\varepsilon}_{i0} - \varepsilon_{i0}| + 4M_{f_\varepsilon}|\widehat{\varepsilon}_{ij} - \varepsilon_{ij}| \\
&= 4(M_{F_1} + M_{f_\varepsilon})|\widehat{m}_0(Z_i) - m_0(Z_i)| \\
&\quad + 4(M_{F_2} + M_{f_\varepsilon})|\widehat{m}_j(Z_i) - m_j(Z_i)|
\end{aligned}
$$

by condition (C7) and mean value theorem. Thus, it follows that

$$
\begin{aligned}
\Lambda_{j1} &\leq \sum_{i=1}^{n} P\Big(|\widehat{m}_0(Z_i) - m_0(Z_i)| > \frac{K_3^2 K_4^2 \epsilon}{3(M_{F_1} + M_{f_\varepsilon})}\Big) \\
&\quad + \sum_{i=1}^{n} P\Big(|\widehat{m}_j(Z_i) - m_j(Z_i)| > \frac{K_3^2 K_4^2 \epsilon}{3(M_{F_2} + M_{f_\varepsilon})}\Big).
\end{aligned}
$$

Then, by Lemma A.7, we get that there exist positive constants $\tilde{c}_{18}$ and $\tilde{c}_{19}$ such that

$$
\max_{1 \leq j \leq p_n} \Lambda_{j1} \leq 2\tilde{c}_{18} \exp(-\tilde{c}_{19} n L_n^{-4} \epsilon^2)
$$

for all sufficiently large $n$, where $\tilde{c}_{19}$ depends merely on $K_3, K_4, M_{F_1}, M_{F_2}$ and $M_{f_\varepsilon}$. Similarly, by condition

(C7), mean value theorem and Lemma A.7, we can derive that

$$\max_{1\leq j\leq p_n} \Lambda_{j2} \leq 2\tilde{c}_{20}\exp(-\tilde{c}_{21}nL_n^{-4}\epsilon^2)$$

for some positive constants $\tilde{c}_{20}$ and $\tilde{c}_{21}$ and sufficiently large $n$. A combination of above two results gives that

$$\max_{1\leq j\leq p_n} P(|\bar{Q}_{nj2}|>\epsilon/3) \leq 2(\tilde{c}_{18}+\tilde{c}_{20})\exp(-\min(\tilde{c}_{19},\tilde{c}_{21})nL_n^{-4}\epsilon^2)$$

for all sufficiently large $n$.

**Step 3**. Derive an exponential bound for $\max_{1\leq j\leq p_n} P(|\bar{Q}_{nj3}|>\epsilon/3)$. Since $|\varrho_j^2(\varepsilon_{i0},\varepsilon_{ij})-E\{\varrho_j^2(\varepsilon_0,\varepsilon_j)\}|\leq 2$, so we can obtain that

$$\max_{1\leq j\leq p_n} P(|\bar{Q}_{nj3}|>\epsilon/3) \leq 2\exp(-n\epsilon^2/72)$$

for all sufficiently large $n$ by employing Lemma A.1 directly.

Therefore, taking $\epsilon=\bar{C}n^{-\tau}$ in (B.32) and stacking **Steps 1-3**, we can conclude that

$$\begin{aligned}\max_{1\leq j\leq p_n} P(|\widehat{u}_j^{RPC}-u_j^{RPC}|>\bar{C}n^{-\tau}) &\leq \tilde{c}_{15}\exp(-\tilde{c}_{17}\bar{C}^2n^{1-2\tau}L_n^{-4})+2\exp(-\bar{C}^2n^{1-2\tau}/72)\\ &\quad +2(\tilde{c}_{18}+\tilde{c}_{20})\exp(-\min(\tilde{c}_{19},\tilde{c}_{21})\bar{C}^2n^{1-2\tau}L_n^{-4})\\ &\leq \tilde{c}_{22}\exp(-\tilde{c}_{23}n^{1-2\tau}L_n^{-4})\end{aligned}$$

for all sufficiently large $n$, where $\tilde{c}_{22}=\tilde{c}_{15}+2(\tilde{c}_{18}+\tilde{c}_{20}+1)$ and $\tilde{c}_{23}=\bar{C}^2\min(\tilde{c}_{17},1/72,\min(\tilde{c}_{19},\tilde{c}_{21}))$. This further implies the following consistency inequality for the screening utility:

$$P(\max_{1\leq j\leq p_n}|\widehat{u}_j^{RPC}-u_j^{RPC}|>\bar{C}n^{-\tau})\leq \tilde{c}_{22}p_n\exp(-\tilde{c}_{23}n^{1-2\tau}L_n^{-4}).$$

Based on such a result, we can show the results (i)-(iii) in Theorem 3.3 through using the same arguments in the proof of Theorem 3.2. The details are omitted. Thus, we complete the proof. □

**Lemma A.8.** *Let $m_0(z)=\mathrm{median}(Y|Z=z)$ and $m_j(z)=\mathrm{median}(X_j|Z=z), j=1,\ldots,p_n$. Under conditions (D4)-(D5) and (D10), for every $0\leq j\leq p_n$, then there exist two uniform constants $c_1^\dagger$ and $c_2^\dagger$* such that for any $\epsilon > 0$ fulfilling $n\epsilon^2 L_n^{-2} \to \infty$ and $L_n^{d/2}\epsilon \to \infty$ as $n \to \infty$,

$$P\Big(\sup_{z \in \mathcal{Z}} |\widehat{m}_j(z) - m_j(z)| > \epsilon\Big) \leq \exp(-c_1^\dagger n L_n^{-3} \epsilon^2) + 2\exp(-c_2^\dagger n L_n^{-2} \epsilon^4).$$

**Proof of Lemma A.8** We only consider the case for $j > 0$. Let $\boldsymbol{\gamma}_j^M = \mathrm{argmin}_{\boldsymbol{\gamma}_j} E\{|X_j - \mathbf{B}(Z)^T \boldsymbol{\gamma}_j|\}$ and $m_j^M(z) = \mathbf{B}(z)^T \boldsymbol{\gamma}_j^M$. Recall that $\widehat{m}_j(z) = \mathbf{B}(z)^T \widehat{\boldsymbol{\gamma}}_j$, where $\widehat{\boldsymbol{\gamma}}_j = \mathrm{argmin}_{\boldsymbol{\gamma}_j} \frac{1}{n} \sum_{i=1}^n |X_{ij} - \mathbf{B}(Z_i)^T \boldsymbol{\gamma}_j|$.

We first show that

$$\sup_{z \in \mathcal{Z}} \big|m_j(z) - m_j^M(z)\big| = O(L_n^{-d/2}). \tag{B.47}$$

To this end, by the B-spline theory (de Boor (2001)), there exists a $\boldsymbol{\gamma}_j^* \in \mathbb{R}^{L_n}$ such that $\sup_z |m_j(z) - \mathbf{B}(z)^T \boldsymbol{\gamma}_j^*| \leq \bar{c}_1 L_n^{-d}$ for some positive constant $\bar{c}_2$. Note that by the definition of $m_j$ and the triangle inequality, we have that for any $\boldsymbol{\gamma}_j \in \mathbb{R}^{L_n}$,

$$\begin{aligned}
E_{X_j|Z=z}\{|X_j - m_j(Z)|\} &\leq E_{X_j|Z=z}\{|X_j - \mathbf{B}(Z)^T \boldsymbol{\gamma}_j|\} \\
&\leq E_{X_j|Z=z}\{|X_j - m_j(Z)|\} + E_{X_j|Z=z}\{|m_j(Z) - \mathbf{B}(Z)^T \boldsymbol{\gamma}_j|\}.
\end{aligned}$$

Also, by the definition of $\boldsymbol{\gamma}_j^M$, we have

$$\begin{aligned}
E_{X_j|Z=z}\{|X_j - m_j(Z)|\} &\leq E_{X_j|Z=z}\{|X_j - \mathbf{B}(Z)^T \boldsymbol{\gamma}_j^M|\} \\
&\leq E_{X_j|Z=z}\{|X_j - \mathbf{B}(Z)^T \boldsymbol{\gamma}_j^*|\} \\
&\leq E_{X_j|Z=z}\{|X_j - m_j(Z)|\} + E_{X_j|Z=z}\{|m_j(Z) - \mathbf{B}(Z)^T \boldsymbol{\gamma}_j^*|\}.
\end{aligned}$$

Thus, it follows that

$$\begin{aligned}
\big|E_{X_j|Z=z}\{|X_j - \mathbf{B}(Z)^T \boldsymbol{\gamma}_j^M|\} &- E_{X_j|Z=z}\{|X_j - m_j(Z)|\}\big| \\
&\leq E_{X_j|Z=z}\{|m_j(Z) - \mathbf{B}(Z)^T \boldsymbol{\gamma}_j^*|\} \leq \bar{c}_1 L_n^{-d}.
\end{aligned} \tag{B.48}$$

Applying the identity that $|x-y| - |x| = -2y[\frac{1}{2} - I(x<0)] + 2\int_0^y \{I(x \leq s) - I(x \leq 0)\} \mathrm{d}s$ (Knight (1998), page 758), the left-hand side of (B.48) is equal to

$$2\Big|E_{X_j|Z=z}\Big\{\int_0^{\mathbf{B}(Z)^T \boldsymbol{\gamma}_j^M - m_j(Z)} \{I(X_j - m_j(Z) \leq s) - I(X_j - m_j(Z) \leq 0)\}\mathrm{d}s\Big\}\Big|$$

$$
\begin{aligned}
&= 2\Big|\Big\{\int_0^{\mathbf{B}(z)^T\boldsymbol{\gamma}_j^M - m_j(z)} \{F_{X_j|Z=z}(m_j(z)+s) - F_{X_j|Z=z}(m_j(z))\}\mathrm{d}s\Big\}\Big|\\
&= 2\Big|\Big\{\int_0^{\mathbf{B}(z)^T\boldsymbol{\gamma}_j^M - m_j(z)} f_{X_j|Z=z}(t)s\mathrm{d}s\Big\}\Big|\\
&\geq \bar{c}_2|\mathbf{B}(z)^T\boldsymbol{\gamma}_j^M - m_j(z)|^2
\end{aligned}
$$

for some positive constant $\bar{c}_2$, where $t$ is a point between $m_j(z)$ and $\mathbf{B}(z)^T\boldsymbol{\gamma}_j^M$, the third line uses mean value theorem, and the last line holds by condition (D10). Thus, the above implies (B.47).

Next, we are going to derive an upper bound for $P\big(\sup_{z\in\mathcal{Z}}|\widehat{m}_j(z) - m_j^M(z)| > \epsilon/2\big)$. Note that $\sup_{z\in\mathcal{Z}}|\widehat{m}_j(z) - m_j^M(z)| \leq \|\widehat{\boldsymbol{\gamma}}_j - \boldsymbol{\gamma}_j^M\|$. Thus, it is sufficient to bound $P\big(\|\widehat{\boldsymbol{\gamma}}_j - \boldsymbol{\gamma}_j^M\| > \epsilon/2\big)$. In what follows, we follow the proof of Lemma 8.3 of He, Wang and Hong (2013) to derive the bound. Denote $W_n(\boldsymbol{\gamma}_j) \hat{=} \frac{1}{n}\sum_{i=1}^n\{|X_{ij} - \mathbf{B}(Z_i)^T\boldsymbol{\gamma}_j| - |X_{ij}|\}$ and $W(\boldsymbol{\gamma}_j) \hat{=} E\{|X_j - \mathbf{B}(Z)^T\boldsymbol{\gamma}_j| - |X_j|\}$. Using Lemma 8.2 of He, Wang and Hong (2013), we have that for any $\epsilon > 0$,

$$
P\big(\|\widehat{\boldsymbol{\gamma}}_j - \boldsymbol{\gamma}_j^M\| > \epsilon\big) \leq P\Big(\sup_{\|\boldsymbol{\gamma}_j - \boldsymbol{\gamma}_j^M\|\leq\epsilon} |W_n(\boldsymbol{\gamma}_j) - W(\boldsymbol{\gamma}_j)| \geq \frac{1}{2}\inf_{\|\boldsymbol{\gamma}_j - \boldsymbol{\gamma}_j^M\|=\epsilon}\{W(\boldsymbol{\gamma}_j) - W(\boldsymbol{\gamma}_j^M)\}\Big).
$$

Let $\boldsymbol{\gamma}_j = \boldsymbol{\gamma}_j^M + \epsilon\mathbf{u}$. Then, using the identity of Knight (1998) (page 758) and condition (D10), we can obtain that

$$
\begin{aligned}
\inf_{\|\boldsymbol{\gamma}_j - \boldsymbol{\gamma}_j^M\|=\epsilon}\{W(\boldsymbol{\gamma}_j) - W(\boldsymbol{\gamma}_j^M)\} &= \inf_{\|\mathbf{u}\|=1}\{W(\boldsymbol{\gamma}_j^M + \epsilon\mathbf{u}) - W(\boldsymbol{\gamma}_j^M)\}\\
&= \inf_{\|\mathbf{u}\|=1} 2E\Big\{\int_0^{\epsilon\mathbf{B}(Z)^T\mathbf{u}} \{F_{X_j|Z}(\mathbf{B}(Z)^T\boldsymbol{\gamma}_j^M + s) - F_{X_j|Z}(\mathbf{B}(Z)^T\boldsymbol{\gamma}_j^M)\}\mathrm{d}s\Big\}\\
&= \inf_{\|\mathbf{u}\|=1} 2E\Big\{\int_0^{\epsilon\mathbf{B}(Z)^T\mathbf{u}} f_{X_j|Z=z}(t)s\mathrm{d}s\Big\}\\
&\geq \bar{c}_2\epsilon^2\lambda_{\min}(E\{\mathbf{B}(Z)^T\mathbf{B}(Z)\}) = \bar{c}_2\epsilon^2 L_n^{-1},
\end{aligned}
$$

where $t$ is a point between $\mathbf{B}(Z)^T\boldsymbol{\gamma}_j^M$ and $\mathbf{B}(Z)^T\boldsymbol{\gamma}_j^M + s$. Thus, it follows that

$$
\begin{aligned}
P\big(\|\widehat{\boldsymbol{\gamma}}_j - \boldsymbol{\gamma}_j^M\| > \epsilon\big) &\leq P\Big(\sup_{\|\boldsymbol{\gamma}_j - \boldsymbol{\gamma}_j^M\|\leq\epsilon} |W_n(\boldsymbol{\gamma}_j) - W(\boldsymbol{\gamma}_j)| \geq \frac{1}{2}\bar{c}_2\epsilon^2 L_n^{-1}\Big)\\
&\leq P\Big(\sup_{\|\boldsymbol{\gamma}_j - \boldsymbol{\gamma}_j^M\|\leq\epsilon} \big|[W_n(\boldsymbol{\gamma}_j) - W_n(\boldsymbol{\gamma}_j^M)] - [W(\boldsymbol{\gamma}_j) - W(\boldsymbol{\gamma}_j^M)]\big| \geq \frac{1}{4}\bar{c}_2\epsilon^2 L_n^{-1}\Big)\\
&\quad + P\Big(\big|W_n(\boldsymbol{\gamma}_j^M) - W(\boldsymbol{\gamma}_j^M)\big| \geq \frac{1}{4}\bar{c}_2\epsilon^2 L_n^{-1}\Big)\\
&=: I_{n1} + I_{n2} \text{ (say)}. \quad\quad\quad\quad\quad\text{(B.49)}
\end{aligned}
$$

For the term $I_{n1}$, note that $[W_n(\boldsymbol{\gamma}_j) - W_n(\boldsymbol{\gamma}_j^M)] - [W(\boldsymbol{\gamma}_j) - W(\boldsymbol{\gamma}_j^M)] = \frac{1}{n}\sum_{i=1}^n[T_{ij}(\boldsymbol{\gamma}_j) - E\{T_{ij}(\boldsymbol{\gamma}_j)\}]$, where $T_{ij}(\boldsymbol{\gamma}_j) = |X_{ij} - \mathbf{B}(Z_i)^T\boldsymbol{\gamma}_j| - |X_{ij} - \mathbf{B}(Z_i)^T\boldsymbol{\gamma}_j^M|$. Obviously, using the triangle inequality and the fact $\|B_k\|_\infty \leq 1$, we have $\sup_{\|\boldsymbol{\gamma}_j - \boldsymbol{\gamma}_j^M\| \leq \epsilon} |T_{ij}(\boldsymbol{\gamma}_j)| \leq \sup_{\|\boldsymbol{\gamma}_j - \boldsymbol{\gamma}_j^M\| \leq \epsilon} |\mathbf{B}(Z_i)^T(\boldsymbol{\gamma}_j - \boldsymbol{\gamma}_j^M)| \leq \sup_{\|\boldsymbol{\gamma}_j - \boldsymbol{\gamma}_j^M\| \leq \epsilon} \|\mathbf{B}(Z_i)\|\|\boldsymbol{\gamma}_j - \boldsymbol{\gamma}_j^M\| \leq L_n^{1/2}\epsilon$. Then, it follows that

$$
\begin{aligned}
E\Big\{&\sup_{\|\boldsymbol{\gamma}_j-\boldsymbol{\gamma}_j^M\|\leq\epsilon}\Big|[W_n(\boldsymbol{\gamma}_j) - W_n(\boldsymbol{\gamma}_j^M)] - [W(\boldsymbol{\gamma}_j) - W(\boldsymbol{\gamma}_j^M)]\Big|\Big\} \\
&\leq 2E\Big\{\sup_{\|\boldsymbol{\gamma}_j-\boldsymbol{\gamma}_j^M\|\leq\epsilon}\Big|\frac{1}{n}\sum_{i=1}^n e_i T_{ij}(\boldsymbol{\gamma}_j)\Big|\Big\} \\
&\leq 4E\Big\{\sup_{\|\boldsymbol{\gamma}_j-\boldsymbol{\gamma}_j^M\|\leq\epsilon}\Big|\frac{1}{n}\sum_{i=1}^n e_i \mathbf{B}(Z_i)^T(\boldsymbol{\gamma}_j - \boldsymbol{\gamma}_j^M)\Big|\Big\} \\
&\leq 4\epsilon E\Big\|\frac{1}{n}\sum_{i=1}^n e_i \mathbf{B}(Z_i)^T\Big\| \\
&\leq \bar{c}_3 n^{-1/2}\epsilon
\end{aligned}
$$

for some positive constant $\bar{c}_3$, where the first inequality applies the symmetrization theorem (Lemma 2.3.1, van der Vaart and Wellner (1996)) and $\{e_i\}_{i=1}^n$ denote a Rademacher sequence (i.i.d random variables with taking values $\pm 1$ with probability $\frac{1}{2}$), the second inequality applies the contraction theorem (Leddoux and Talagrand (1991)), and the third follows by Cauchy-Schwarz inequality, and the last line uses the fact in (B.20). Rewrite $U_n \hat{=} \sup_{\|\boldsymbol{\gamma}_j - \boldsymbol{\gamma}_j^M\| \leq \epsilon} \Big|[W_n(\boldsymbol{\gamma}_j) - W_n(\boldsymbol{\gamma}_j^M)] - [W(\boldsymbol{\gamma}_j) - W(\boldsymbol{\gamma}_j^M)]\Big|$. Therefore, an application of the concentration theorem (Massart (2000)) gives

$$
\begin{aligned}
I_{n1} &= P\Big(U_n \geq \frac{1}{4}\bar{c}_2\epsilon^2 L_n^{-1}\Big) \\
&\leq P\Big(U_n - E\{U_n\} \geq \frac{1}{4}\bar{c}_2\epsilon^2 L_n^{-1} - \bar{c}_3 n^{-1/2}\epsilon\Big) \\
&\leq P\Big(U_n - E\{U_n\} \geq \frac{1}{8}\bar{c}_2\epsilon^2 L_n^{-1}\Big) \\
&\leq \exp(-\bar{c}_4 n L_n^{-3}\epsilon^2) \qquad\qquad\qquad\qquad\qquad\qquad\qquad\qquad\qquad (B.50)
\end{aligned}
$$

for all sufficiently large $n$, where the last second line is due to the condition of this lemma, and $\bar{c}_4 = \frac{\bar{c}_2^2}{512}$. On the other hand, for the term $I_{n2}$, write $W_n(\boldsymbol{\gamma}_j^M) - W(\boldsymbol{\gamma}_j^M) = \frac{1}{n}\sum_{i=1}^n[\psi_{ij} - E\{\psi_{ij}\}]$ where $\psi_{ij} = |X_{ij} - \mathbf{B}(Z_i)^T\boldsymbol{\gamma}_j^M| - |X_{ij}|$. Note that conditions (D4) and (D5) imply that $m_j(Z)$ is bounded uniformly in $j$, that is, $\sup_z |m_j(z)| \leq \bar{c}_5$ for some uniform constant $\bar{c}_5$. So $|\psi_{ij}| \leq |\mathbf{B}(Z_i)^T\boldsymbol{\gamma}_j^M| \leq |m_j(Z_i)| + |m_j(Z_i) - \mathbf{B}(Z_i)^T\boldsymbol{\gamma}_j^M| \leq \bar{c}_5 + O(L_n^{-d/2}) \leq \bar{c}_6$ for some uniform constant $\bar{c}_6 > 0$ and sufficiently large $n$. Thus, it follows

that $|\psi_{ij} - E\psi_{ij}| \leq 2\bar{c}_6$. By Lemma A.1, we obtain

$$
\begin{aligned}
I_{n2} &= P\Big(\big|W_n(\boldsymbol{\gamma}_j^M) - W(\boldsymbol{\gamma}_j^M)\big| \geq \frac{1}{4}\bar{c}_2\epsilon^2 L_n^{-1}\Big) \\
&= P\Big(\big|\sum_{i=1}^n [\psi_{ij} - E\{\psi_{ij}\}]\big| \geq \frac{1}{4}\bar{c}_2 n\epsilon^2 L_n^{-1}\Big) \\
&\leq 2\exp(-\bar{c}_7 n\epsilon^4 L_n^{-2})
\end{aligned}
\tag{B.51}
$$

for some positive constant $\bar{c}_7$. Combining the above results in (B.47) and (B.49)-(B.51), we can obtain the desired result. $\square$

**Proof of Theorem 3.6** The proof of part (i) can be finished by directly following the steps in the proof of Theorem 3.4 and using Lemma A.8. We outline the details by pointing out some key modifications here. To this end, we just need to modify the tail probability bounds related to $\bar{Q}_{nj1}$ and $\bar{Q}_{nj2}$, which in turn depend on the tail probability bounds for the terms $\widehat{F}_{\widehat{\varepsilon}_0,\widehat{\varepsilon}_j}(u,v)$, $\widehat{F}_{\widehat{\varepsilon}_0}(u)$ and $\widehat{F}_{\widehat{\varepsilon}_j}(v)$, appeared in the proof of Theorem 3.3. Specifically, applying Lemma A.8, we can show that

$$
\begin{aligned}
&\max_{1\leq j\leq p_n} \sup_{(u,v)\in\mathcal{E}_0\times\mathcal{E}_j} P\Big(\big|\widehat{F}_{\widehat{\varepsilon}_0,\widehat{\varepsilon}_j}(u,v) - F_{\varepsilon_0,\varepsilon_j}(u,v)\big| > \epsilon\Big) \\
&\leq 12\exp(-\bar{c}_8 nL_n^{-3}\epsilon^2) + 4\exp(-\bar{c}_9 nL_n^{-2}\epsilon^4)
\end{aligned}
\tag{B.52}
$$

for some positive constants $\bar{c}_8$ and $\bar{c}_9$,

$$
\sup_{u\in\mathcal{E}_0} P\Big(\big|\widehat{F}_{\widehat{\varepsilon}_0}(u) - F_{\varepsilon_0}(u)\big| > \epsilon\Big) \leq 12\exp(-\bar{c}_{10} nL_n^{-3}\epsilon^2) + 4\exp(-\bar{c}_{11} nL_n^{-2}\epsilon^4)
\tag{B.53}
$$

for some positive constants $\bar{c}_{10}$ and $\bar{c}_{11}$, and

$$
\max_{1\leq j\leq p_n} \sup_{v\in\mathcal{E}_j} P\Big(\big|\widehat{F}_{\widehat{\varepsilon}_j}(v) - F_{\varepsilon_j}(v)\big| > \epsilon\Big) \leq 12\exp(-\bar{c}_{12} nL_n^{-3}\epsilon^2) + 4\exp(-\bar{c}_{13} nL_n^{-2}\epsilon^4)
\tag{B.54}
$$

for some positive constants $\bar{c}_{12}$ and $\bar{c}_{13}$, provided that $nL_n^{-4}\epsilon^2 \to \infty$, $nL_n^{-2}\epsilon^4 \to \infty$ and $L_n^{d/2}\epsilon \to \infty$ as $n\to\infty$. Using (B.52)-(B.54) and conditions (D7)-(D9), we can show that there exist some positive constants $\bar{c}_{14}$, $\bar{c}_{15}$, $\bar{c}_{16}$ and $\bar{c}_{17}$ such that

$$
\max_{1\leq j\leq p_n} \sup_{(u,v)\in\mathcal{E}_0\times\mathcal{E}_j} P(|\widehat{\varrho}_j^2(u,v) - \varrho_j^2(u,v)| > \epsilon) \leq \bar{c}_{14}\exp(-\bar{c}_{15} nL_n^{-3}\epsilon^2) + \bar{c}_{16}\exp(-\tilde{c}_{17} nL_n^{-2}\epsilon^4).
$$

Accordingly, we can obtain

$$\max_{1\leq j\leq p_n} P(|\bar{Q}_{nj1}| > \epsilon/3) \leq \bar{c}_{18}\exp(-\bar{c}_{19}nL_n^{-3}\epsilon^2) + \bar{c}_{20}\exp(-\bar{c}_{21}nL_n^{-2}\epsilon^4)$$

for some positive constants $\bar{c}_{18}, \bar{c}_{19}, \bar{c}_{20}$ and $\bar{c}_{21}$, and similarly, we have

$$\max_{1\leq j\leq p_n} P(|\bar{Q}_{nj2}| > \epsilon/3) \leq \bar{c}_{22}\exp(-\bar{c}_{23}nL_n^{-3}\epsilon^2) + \bar{c}_{24}\exp(-\bar{c}_{25}nL_n^{-2}\epsilon^4)$$

for some positive constants $\bar{c}_{22}, \bar{c}_{23}, \bar{c}_{24}$ and $\bar{c}_{25}$. Also, note that we have

$$\max_{1\leq j\leq p_n} P(|\bar{Q}_{nj3}| > \epsilon/3) \leq 2\exp(-n\epsilon^2/72)$$

for all sufficiently large $n$. Therefore, applying the above results and the arguments in the proof of Theorem 3.3, we can prove part (i) by taking $\epsilon = \bar{C}n^{-\tau}$. The proof of parts (ii)-(iii) follows the arguments in the proof of Theorem 3.2 directly. Thus, we complete the entire proof. □